\documentclass[11pt]{article}
\usepackage{axodraw2,pix}
\usepackage{epsfig}
\usepackage{amsfonts}
\usepackage{amsmath}
\usepackage{bbm,bm}
\usepackage{cite}
\allowdisplaybreaks[1]

\usepackage{tikz}
\usetikzlibrary{shapes,arrows,shadows,automata,positioning}
\newdimen\nodeDist
\usepackage[
	colorlinks=true,
	linkcolor=black,
	citecolor=black,
	filecolor=black,
	urlcolor=black,
        breaklinks=true
        ]{hyperref}
\usepackage{multicol}
\newcounter{dummy}

\renewcommand{\vec}[1]{{\bf #1}}

\newcommand{\ko}{\omega} 
\newcommand\he[1]{#1^\dagger}

\newcommand{\gammaE}{{\gamma_\rmii{E}}}

\newcommand{\Lamd}{\Lambda_{\rmii{3d}}}
\newcommand{\LamD}{\Lambda}
\newcommand\MSbar{$\overline{\rm MS}$}

\newcommand{\nf}{n_{\rm f}}
\newcommand{\Nc}{N_{\rm c}}

\newcommand{\Tc}{T_{\rm c}}

\newcommand{\CF}{C_\rmii{F}}

\newcommand{\Yl}{Y_{\ell}}
\newcommand{\Yq}{Y_{q}}
\newcommand{\Ye}{Y_{e}}
\newcommand{\Yu}{Y_{u}}
\newcommand{\Yd}{Y_{d}}
\newcommand{\Yf}{Y_{\rmi{f}}}
\newcommand{\Ys}{Y_{\phi}}

\newcommand{\mD}{m_\rmii{D}}

\newcommand{\gY}{g_\rmii{$Y$}}

\newcommand{\gp}{g'}
\newcommand{\gs}{g_\rmi{s}}

\newcommand{\rmO}{{\mathcal{O}}}

\def\lsi{\raise0.3ex\hbox{$<$\kern-0.75em\raise-1.1ex\hbox{$\sim$}}}
\def\gsi{\raise0.3ex\hbox{$>$\kern-0.75em\raise-1.1ex\hbox{$\sim$}}}

\newcommand{\gsim}{\mathop{\gsi}}

\newcommand{\nn}{\nonumber \\}
\newcommand{\rmi}[1]{{\mbox{\scriptsize #1}}}
\newcommand{\rmii}[1]{{\mbox{\tiny\rm{#1}}}}
\newcommand{\hp}{\hphantom}

\newcommand{\Tint}[1]{{\hbox{$\sum$}\!\!\!\!\!\!\!\int\,}_{\!\!\!\!\raise-0.9ex\hbox{$\scriptstyle{#1}$}}}
\newcommand{\Tinti}[1]{{{\Sigma}\!\!\!\!\raise0.3ex\hbox{$\int$}_\rmii{${#1}$}}}
\newcommand{\Tintip}[1]{{{\Sigma'}\!\!\!\!\!\raise0.3ex\hbox{$\int$}_\rmii{${#1}$}}}

\newcommand{\ZZ}{{\mathbb{Z}}}

\newcommand{\deltabar}{\raise-0.02em\hbox{$\bar{}$}\hspace*{-0.8mm}{\delta}}

\def\scfc{0.7}  
\def\phgt{21}   
\def\pwc{21}    
\def\pwcb{31.5} 
\def\pwcc{42} 
\newcommand{\PIC}[4]{\;\parbox[c]{#2 pt}{\begin{picture}(#2,#3)(0,0)
\SetWidth{1.0}\SetScale{#4} #1 \end{picture}}\;}
\renewcommand{\pic}[1]{\PIC{#1}{\pwc}{\phgt}{\scfc}}
\renewcommand{\picb}[1]{\PIC{#1}{\pwcb}{\phgt}{\scfc}}
\renewcommand{\picc}[1]{\PIC{#1}{\pwcc}{\phgt}{\scfc}}

\makeatletter \@addtoreset{equation}{section} \makeatother
\renewcommand{\theequation}{\arabic{section}.\arabic{equation}}
\makeatletter
\renewcommand\section{\@startsection{section}{1}{\z@}%
  {-5.5ex \@plus -1ex \@minus -.2ex}
  {2.3ex \@plus.2ex}%
  {\normalfont\large\bfseries}}
\renewcommand\subsection{\@startsection{subsection}{2}{\z@}%
  {-3.25ex\@plus -1ex \@minus -.2ex}%
  {1.5ex \@plus .2ex}%
  {\normalfont\normalsize\bfseries}}
\renewcommand\thesection{\@arabic\c@section}
\renewcommand\thesubsection{\thesection.\@arabic\c@subsection}
\renewcommand{\@seccntformat}[1]{%
  \csname the#1\endcsname.\hspace{1.0em}}
\makeatother

\begin{document}

\flushbottom

\begin{titlepage}

\begin{flushright}
HIP-2021-5/TH\\ 
NORDITA 2021-008\\
\end{flushright}
\begin{centering}

\vfill

{\Large{\bf
Robust approach to thermal resummation:\\
Standard Model meets a singlet
}}

\vspace{0.8cm}

\renewcommand{\thefootnote}{\fnsymbol{footnote}}
Philipp M.~Schicho$^{\rm a,b,}$%
\footnote{philipp.schicho@helsinki.fi},
Tuomas V.~I.~Tenkanen$^{\rm b,c,d,}$%
\footnote{tuomas.tenkanen@su.se},
Juuso {\"O}sterman$^{\rm a,}$%
\footnote{juuso.s.osterman@helsinki.fi}

\vspace{0.8cm}

$^{\rm a}$%
{\em
Department of Physics and Helsinki Institute of Physics,
P.O.\ Box 64,\\
FI-00014 University of Helsinki,
Finland\\}
\vspace{0.3cm}

$^\rmi{b}$%
{\em
AEC, Institute for Theoretical Physics, University of Bern,\\
Sidlerstrasse 5,
CH-3012 Bern,
Switzerland\\}
\vspace{0.3cm}

$^\rmi{c}$%
{\em
Nordita,
KTH Royal Institute of Technology and Stockholm University,\\
Roslagstullsbacken 23,
SE-106 91 Stockholm,
Sweden\\}
\vspace{0.3cm}

$^\rmi{d}$%
{\em
Tsung Dao Lee Institute/Shanghai Jiao Tong University,
Shanghai 200240,
China\\}

\vspace*{0.8cm}

\mbox{\bf Abstract}

\end{centering}

\vspace*{0.3cm}

\noindent

Perturbation theory alone fails to describe thermodynamics of the electroweak phase transition.
We review a technique combining perturbative and non-perturbative methods to overcome this challenge.
Accordingly, the principal theme
is a tutorial of high-temperature dimensional reduction.
We present an explicit derivation with a real singlet scalar and compute
the thermal effective potential at two-loop order.
In particular, we detail the
dimensional reduction for a real-singlet extended Standard Model.
The resulting effective theory will impact future non-perturbative studies
based on lattice simulations as well as purely perturbative investigations.

\vfill
\end{titlepage}

\tableofcontents
\clearpage

\renewcommand{\thefootnote}{\arabic{footnote}}
\setcounter{footnote}{0}

%
\section{Introduction}
\label{se:intro}

A strong first-order cosmic phase transition (SFOPT)
is a violent process that can trigger the generation of
a primordial gravitational wave (GW) background (cf.~%
\cite{Apreda:2001us,Grojean:2006bp} and reviews~%
\cite{Weir:2017wfa,Caprini:2018mtu,Hindmarsh:2020hop}).
Gravitational waves from astrophysical sources have been detected by
Earth-based detectors LIGO and VIRGO for
binary black hole~\cite{Abbott:2016blz,Abbott:2016nmj,Abbott:2017vtc} and
neutron star mergers~\cite{TheLIGOScientific:2017qsa,GBM:2017lvd,Monitor:2017mdv}.
Their success and
the mission to probe evidence of relic gravitational waves from the early Universe
have sparked interest for space-based gravitational wave observatories such as
LISA~\cite{Audley:2017drz},
BBO~\cite{Harry:2006fi},
Taiji~\cite{Guo:2018npi}, and
DECIGO~\cite{Kawamura:2011zz}.
A detection of
such a relic GW background
could scope
the underlying theories of particle physics
complimentary to collider physics~%
\cite{Ashoorioon:2009nf,Alves:2018jsw,Mazumdar:2018dfl,Hashino:2018wee}.

The electroweak phase transition (EWPT) is
a smooth crossover in the minimal Higgs sector of
the Standard Model (SM)~%
\cite{Kajantie:1996mn,Kajantie:1996qd,Gurtler:1997hr,Csikor:1998eu,DOnofrio:2015gop}.
Therein, the observed Higgs mass of $125$~GeV~\cite{Aad:2012tfa,Chatrchyan:2012ufa}
exceeds the requirements for a SFOPT which precludes both the
production of a cosmic GW background and
electroweak baryogenesis~\cite{Kuzmin:1985mm}.
The latter is a mechanism to produce the baryon asymmetry
during
the electroweak phase transition~\cite{Trodden:1998ym,Morrissey:2012db}.
New beyond the Standard Model (BSM) physics can alter the character of
the electroweak symmetry breaking towards a SFOPT.
To this end, new particles need to be sufficiently
light in the vicinity of the electroweak (EW) scale and
strongly enough coupled to the Higgs.
This indicates that such BSM theories offer
theoretical targets to guide future high-energy collider experiments~\cite{Ramsey-Musolf:2019lsf}.

One promising class of BSM candidates are theories with non-minimal Higgs sectors
with distinctive collider phenomenology signatures; cf.\
refs.~\cite{%
  No:2013wsa,Dorsch:2014qja,Craig:2014lda,Kotwal:2016tex,Huang:2017jws,
  Chen:2017qcz,Alves:2018jsw,Bell:2020gug}.
These theories form a theoretical playground for the EWPT
with ample related literature.
The SM Higgs doublet can be accompanied for example by a
singlet~\cite{%
  Kondo:1991jz,Enqvist:1992va,Espinosa:1993bs,Choi:1993cv,Profumo:2007wc,Ahriche:2007jp,Espinosa:2011ax,Cline:2012hg,Cline:2013gha,Alanne:2014bra,Profumo:2014opa,Curtin:2014jma,
  Kakizaki:2015wua,Beniwal:2017eik,Kurup:2017dzf,Chiang:2017nmu,Chiang:2018gsn,
  Alanne:2019bsm,Chen:2019ebq,Alanne:2020jwx,Papaefstathiou:2020iag},
second doublet~\cite{%
  Funakubo:1993jg,Davies:1994id,Cline:1995dg,Fromme:2006cm,Cline:2011mm,
  Dorsch:2013wja,Dorsch:2016nrg,Basler:2016obg,Basler:2017uxn,Bernon:2017jgv},
triplet~\cite{Patel:2012pi,Chala:2018opy},
higher-order representations of SU(2) symmetry~\cite{AbdusSalam:2013eya},
combinations of these~\cite{Chala:2016ykx,Basler:2019iuu} or
higher dimensional operators~\cite{%
  Grojean:2004xa,Bodeker:2004ws,Delaunay:2007wb,Grinstein:2008qi,Huang:2016odd,
  Cai:2017tmh,deVries:2017ncy,Chala:2018ari,Postma:2020toi}.
Different models with non-minimal Higgs sector can accommodate SFOPT
specifically but not exclusively at
the EW scale around temperatures of $100$~GeV.
In addition, they could invoke sources
for CP violation --
the missing ingredient in the SM~\cite{%
  Shaposhnikov:1987pf,Farrar:1993sp,Farrar:1993hn,Gavela:1993ts,Gavela:1994ds,
  Brauner:2011vb,Brauner:2012gu} required for electroweak baryogenesis --
and potential dark matter candidates
via new neutral scalars.
Compellingly,
a non-minimal Higgs sector can exhibit
a rich pattern of phase transitions that progress in
multiple consecutive steps~\cite{Land:1992sm,Patel:2012pi,Inoue:2015pza,Blinov:2015sna,Croon:2018new}.
Phase transitions could have even occurred in a dark sector which
couples to the SM only gravitationally.
These transitions could potentially source a primordial GW background%
~\cite{Schwaller:2015tja,Croon:2018erz,Croon:2018kqn,Hall:2019ank,Croon:2019rqu,Croon:2019iuh}.

Thermal field theories are plagued by the infrared problem~\cite{Linde:1980ts}.
Their perturbative description of long distance modes
is invalidated at high temperatures due to high occupancies of bosonic modes.
Nevertheless, perturbation theory is still widely used when
reorganising the perturbative expansion by resummation, such as in
hard thermal loop perturbation theory~\cite{Braaten:1991gm} and
daisy resummation~\cite{Arnold:1992rz}.

A robust solution to the IR problem is achieved by
an effective theory formulation of resummation.
This allows to treat high-temperature heavy degrees of freedom perturbatively, while
non-perturbative, light degrees of freedom are analysed with
lattice gauge field theory techniques.
Concretely, the phase transition thermodynamics is determined by
Monte Carlo lattice simulations~\cite{Farakos:1994xh,Kajantie:1995kf}
of dimensionally reduced
high-temperature effective field theories (3d EFT).
Originally established for non-Abelian gauge theories~\cite{%
  Ginsparg:1980ef,Appelquist:1981vg,Nadkarni:1982kb,Landsman:1989be},
the formalism was generalised
in the mid-1990s~\cite{Kajantie:1995dw} and
simultaneously successful in hot QCD~\cite{%
  Braaten:1995cm,Braaten:1995jr,
  Kajantie:1997tt,Kajantie:1997pd,Andersen:1997hq,Kajantie:1998yc,Kajantie:2000iz,Kajantie:2002wa,
  Kajantie:2003ax,Laine:2003bd,Hietanen:2004ew,Vepsalainen:2007ke,Hietanen:2008tv}
as reviewed in~\cite{Ghiglieri:2020dpq}.

Considering the vastness of recent studies of the EWPT in BSM theories,
the 3d EFT approach has been used scarcely.
Seminal work in the 1990s for the Standard Model~\cite{Kajantie:1995dw,Kajantie:1995kf}
were continued in~\cite{%
  Kajantie:1996qd,Gurtler:1996wx,Kajantie:1997ky,Rummukainen:1998as,Gynther:2005dj,
  Gynther:2005av,Vepsalainen:2007ji},
and also extended to
SUSY models~\cite{%
  Losada:1996ju,Losada:1996rt,Farrar:1996cp,Cline:1996cr,Cline:1997bm,Bodeker:1996pc,Laine:1998vn,
  Laine:1998qk,Laine:1998wi,Laine:2000kv,Laine:2012jy},
the Two-Higgs Doublet Model (2HDM)~\cite{Losada:1996ju,Andersen:1998br,Laine:2000rm},
the Abelian Higgs Model~\cite{Karjalainen:1996rk,Kajantie:1997vc,Kajantie:1997hn,Andersen:1997ba},
SU(5) GUT~\cite{Rajantie:1997pr}
and pure scalar field theory~\cite{Jansen:1998rj}.
Recent studies reinvigorated the 3d approach for the SM accompanied by
a real singlet (xSM)~\cite{Brauner:2016fla,Gould:2019qek},
a real triplet ($\Sigma$SM)~\cite{Niemi:2018asa,Niemi:2020hto},
the 2HDM~\cite{Helset:2017esj,Andersen:2017ika,Gorda:2018hvi,Kainulainen:2019kyp},
the SM with one simple higher dimensional operator~\cite{Croon:2020cgk} and
a real scalar field theory~\cite{Gould:2021dzl}.

Dimensional reduction implements the required resummations automatically upon
perturbatively constructing the 3d EFT.
Nonetheless, it is customary to study the EWPT in terms of
the thermal effective potential~\cite{Dolan:1973qd,Carrington:1991hz}
computed directly with other resummation schemes~\cite{Arnold:1992rz,Parwani:1991gq}.
While improved two-loop computations exist~\cite{%
  Arnold:1992rz,Buchmuller:1993bq,Fodor:1994bs,Kripfganz:1995jx}
(cf.\ also refs.~\cite{Laine:2017hdk,Ekstedt:2020abj}),
it is typical for recent EWPT literature to implement
a daisy-resummed thermal effective potential only at one-loop level.
However, the infrared problem persists and
these fully perturbative studies of the EWPT are severely limited with
their setbacks often underestimated.
Even their qualitative description can -- and often will -- fail.
In contrast to lattice studies,
transitions are often realised as (weak) first order
since a crossover character is incompatible with perturbation theory.

Describing the EWPT thermodynamics all the way by
a non-perturbative simulation poses a formidable task.
This roots in
analytical challenges related to the construction of required 3d EFTs and foremost
excessive computational cost of simulations.
Still, many perturbative studies of the EWPT could be significantly improved by
employing
{\em perturbation theory within the dimensionally reduced 3d EFT}.
While this approach is entirely perturbative and hence incapable of solving
the IR problem,
it allows for systematic resummation and straightforward computations at
two-loop level~\cite{Farakos:1994kx,Laine:1994bf}.
Thereby it supersedes the one-loop daisy-resummed thermal effective potential;
see refs.~\cite{Kainulainen:2019kyp,Niemi:2020hto,Croon:2020cgk}
for recent direct comparisons.
Indeed, this approach to perturbation theory was advocated already in ref.~\cite{Bodeker:1996pc}. 

The dimensional reduction can be largely automated and
the careful matching to multiple individual BSM theories streamlined.
The task has been tackled recently~\cite{Schicho:2020xaf,Croon:2020cgk} and in this work at hand.
As a consequence one can exploit the universality of the resulting dimensionally reduced EFTs
to efficiently examine the parameter space of different BSM theories.
This article combines these recent developments to
extend previous work~\cite{Brauner:2016fla} for
the xSM
-- a flagship model that is attractive for particle cosmology due to its minimal nature.
Based on the construction of the 3d EFT of the xSM,
its applications~\cite{Niemi:2021xxx,Gould:2021xxx}
chart a course of a state-of-the-art analysis of the EWPT thermodynamics.
Thereby, perturbative scans that utilise a 3d EFT approach
can guide non-perturbative simulations that finally solve the IR problem.

This article is organised as follows.
Section~\ref{se:thermo}
reviews the computation of thermodynamics in generic scalar driven phase transitions,
in particular focusing on the use of dimensionally reduced effective theories.
Section~\ref{se:tutorial}
is a pedagogic tutorial to the construction of such a 3d EFT
and computes the thermal effective potential at two-loop order,
in the simplest case of a real singlet scalar.
In sec.~\ref{se:dr:xsm},
for the first time, we generalise the dimensional reduction to the xSM.
Finally, sec.~\ref{se:discussion} discusses our results and outlook,
while several technical details relevant to our computation are collected in
the appendices.

%
\section{Thermal phase transitions}
\label{se:thermo}

The focus of this paper is to
take steps towards a state-of-art determination of the
cosmic phase transition thermodynamics for individual BSM theories with non-minimal Higgs sector.
Direct ab-initio lattice calculations are, however, not feasible
for a completely realistic 4d description of
the thermodynamics of electroweak phase transitions~\cite{Luscher:2000hn}
due to problems related to chiral fermions.%
\footnote{
  However, 4d simulations of purely bosonic theories are feasible to study;
  see ref.~\cite{Laine:1996nz} and references therein.
}
One alternative approach are
non-perturbative simulations of dimensionally reduced effective theories.
Following this idea, we survey
the required technology on a generic level in the following section.

%
\subsection{Down the pipeline}

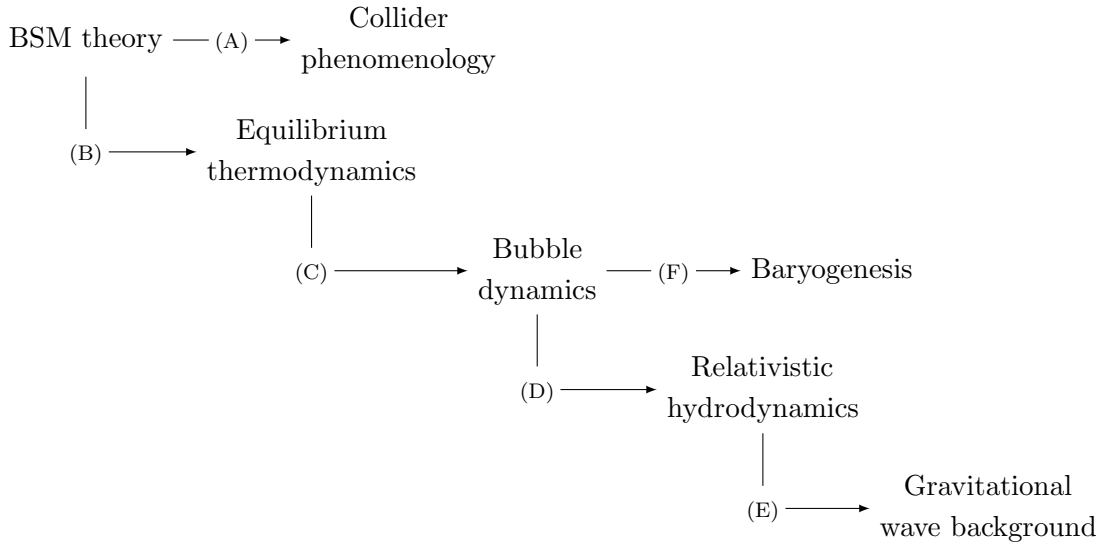
\begin{figure}
  \centering
\tikzstyle{zeroT}=[fill=white!20,minimum width=3.5em,
    align=center,minimum height=2.5em]
\def\blockw{3}
\def\blockh{1}
\newcommand\circled[1]{%
  \tikz[baseline=(X.base)]
    \node (X) [shape=circle,inner sep=-1pt,fill=white,text=black] {\strut\scriptsize #1};%
}
\begin{tikzpicture}
    \node(bsm) [zeroT]  {BSM theory};
    \path (bsm.east)+(\blockw,0) node (col) [zeroT] {Collider\\ phenomenology};
    \path (bsm.south)+(\blockw,-\blockh) node (thermo) [zeroT] {Equilibrium\\ thermodynamics};
    \path (thermo.south)+(\blockw,-\blockh) node (bubble) [zeroT] {Bubble\\ dynamics};
    \path (bubble.south)+(\blockw,-\blockh) node (hydro) [zeroT] {Relativistic\\ hydrodynamics};
    \path (bubble.east)+(\blockw,0) node (baryo) [zeroT] {Baryogenesis};
    \path (hydro.south)+(\blockw,-\blockh) node (GW) [zeroT] {Gravitational\\ wave background};

    \draw[->,>=latex] (bsm.east) -- node[midway] {\circled{(A)}} (col.west);
    \draw[->,>=latex] (bsm.south) |- node {\circled{(B)}} (thermo.west);
    \draw[->,>=latex] (thermo.south) |- node {\circled{(C)}} (bubble.west);
    \draw[->,>=latex] (bubble.south) |- node {\circled{(D)}} (hydro.west);
    \draw[->,>=latex] (bubble.east) -- node[midway] {\circled{(F)}} (baryo.west);
    \draw[->,>=latex] (hydro.south) |- node {\circled{(E)}} (GW.west);
\end{tikzpicture}
  \caption{
    A pipeline between
    collider phenomenology of BSM theory and
    stochastic gravitational wave background signature.
    Later we focus on step (B), as even relatively small uncertainties
    in this step can propagate all the way down the pipeline
    and cause significant errors in the end.
  }
  \label{fig:pipeline}
\end{figure}

Several steps have to be considered to accurately predict
gravitational waves from cosmological phase transitions.
To this end, we illustrate a ``pipeline'' ranging from
the collider phenomenology of BSM particle physics models to
a primordial, stochastic gravitational wave background.
Following a comprehensive ref.~\cite{Caprini:2019egz}, we
display different steps of this pipeline in fig.~\ref{fig:pipeline}
(ibid.\ ref.~\cite{Caprini:2019egz}).

From a theoretical standpoint,
it is natural to start by defining the Lagrangian of the corresponding BSM theory.
In our case of interest,
the
BSM
field content enters as a non-minimal Higgs sector which
contains one or more scalar fields.
In general, the scalar fields can occur in any representation of
the SU(2) symmetry and possess other symmetries and
couplings to new gauge field or fermion content in a dark sector.
However, several alternatives are conceivable (see refs.~\cite{Caprini:2019egz,Schmitz:2020rag}), such as
models for holographic phase transition~\cite{Dillon:2017ctw,Megias:2018sxv,Ares:2020lbt}
or Composite Higgs scenarios~\cite{Espinosa:2011eu,Bruggisser:2018mrt,Bian:2019kmg}.
The pipeline constitutes the following steps:
\begin{itemize}
\item[{\bf Step (A)}:]
  Relating collider signatures and BSM theory Lagrangian parameters.
  The Lagrangian (running) parameters are related to physical observables
  such as
  pole masses and mixing angles in zero-temperature perturbation theory.
  Then, actual
  collider signatures include production cross-sections and a relative shift in
  the Higgs couplings from their SM predicted values,%
  \footnote{
  For a concrete example, in the case of xSM
  the proposed experimental signatures include
  the $h_2 h_2$-production cross-section~\cite{Curtin:2014jma,Craig:2014lda,Chen:2017qcz} and
  a modification to
  the $h_1 ZZ$-coupling~\cite{Profumo:2007wc,Profumo:2014opa,Katz:2014bha}
    ($h_1$ and $h_2$ are
    `mostly Higgs-like' and
    `mostly singlet-like' scalar eigenstates).
  }
  and can constrain the available parameter space for phase transitions.
  The relation of
  the EWPT and collider physics is further discussed in ref.~\cite{Ramsey-Musolf:2019lsf}.

\item[{\bf Step (B)}:]
  Equilibrium thermodynamic properties as a function of BSM theory parameters.
  The former include the character of transition (crossover, first-order etc.),
  the critical temperature ($\Tc$) and
  latent heat ($L/\Tc^4$).
  They are encoded in the free energy of the system which is associated
  with the thermal effective potential in perturbation theory.
  Due to IR sensitivities at high-$T$, this step requires
  non-trivial resummations compared to perturbation theory at zero temperature, and eventually
  non-perturbative techniques.
  Step (B) is the main focus of the remaining sections of this article.
\item[{\bf Step (C)}:]
  If the phase transition is of first order, it proceeds by
  nucleation and expansion of bubbles of the broken phase in the presence of
  a surrounding plasma~\cite{%
      Langer:1967ax,Langer:1969bc,Coleman:1977py,Linde:1981zj}.
  The bubble nucleation rate can be computed in a semi-classical approximation from
  the effective action which includes quantum and thermal corrections.
  The relevant quantities~\cite{%
    Kosowsky:1992rz,Kamionkowski:1993fg,Ignatius:1993qn,KurkiSuonio:1995pp,Huber:2008hg}
  are the
  Hubble parameter ($H_*$) or
  temperature ($T_*$) when the phase transitions completes, its
  inverse duration ($\beta$),
  strength $(\alpha)$ at $T_*$ and
  the bubble wall velocity ($v_w$).
  The exact definitions and derivation of these quantities are detailed in e.g.\
  refs.~\cite{Caprini:2019egz,Croon:2020cgk,Giese:2020rtr}, and
  in particular~\cite{%
    Bodeker:2009qy,Kozaczuk:2015owa,Bodeker:2017cim,Dorsch:2018pat,Hoeche:2020rsg}
  for the bubble equations of motion and $v_w$.
  Also non-perturbative methods for nucleation have been developed~\cite{Moore:2000jw,Moore:2001vf}
  as an alternative to perturbation theory.
\item[{\bf Step (D)}:]
  Numerical, large scale lattice simulations of relativistic hydrodynamics;
  cf.\ refs.~\cite{%
    Hindmarsh:2013xza,Hindmarsh:2015qta,Hindmarsh:2017gnf,Cutting:2018tjt,
    Cutting:2019zws,Cutting:2020nla}.
  The parameters that describe the phase transition dynamics, $(T_*,\alpha,\beta/H_*,v_w)$,
  are input to simulations of colliding bubbles,
  cosmic fluid and sound waves after the phase transition completes.
  These determine the GW power spectrum.
  In practice, the approximate, analytical power spectrum has been solved
  from such simulations in terms of a generic set of input parameters.
  For an application of this, ref.~\cite{Caprini:2019egz} has devised the online tool
  {\tt PTPlot}.
\item[{\bf Step (E)}:]
  A detectable GW background signature depends on
  the architecture of the detector in addition
  to the predicted stochastic GW power spectrum.
  The determination of the signal-to-noise ratio 
  for a predicted signal at LISA is specified in ref.~\cite{Caprini:2019egz}.
\item[{\bf Step (F)}:]
  A necessary condition for the EW baryogenesis~\cite{Kuzmin:1985mm} are
  first order phase transitions occurring via bubble nucleation.
  For reviews cf.~\cite{Trodden:1998ym,Morrissey:2012db,white-book}.
  The generation of a baryon asymmetry could be realised when
  new BSM sources of C and CP violation are invoked
  and
  baryon number violating sphaleron transitions are sufficiently suppressed
  in the broken phase.
  The latter can be associated with sufficiently strong transitions.
\end{itemize}

Next, we detail step (B) starting by
a brief summary of the technique of high-temperature dimensional reduction.

%
\subsection{Dimensional reduction for a high-temperature 3d effective theory}
\label{se:dr}

High-temperature dimensional reduction encodes
the IR physics of the high-temperature plasma
in an effective three-dimensional theory to describe long wavelength phenomena.
In the context of electroweak theories, classic references are~\cite{%
  Kajantie:1995dw,Braaten:1995cm,Braaten:1995jr}
but we also refer~\cite{%
  Laine:1997qm, Andersen:2004fp,Laine:2016hma}.

The equilibrium thermodynamics of a thermal field theory is described
by an evolution in imaginary time $(\tau)$.
Therein,
bosonic (fermionic) fields satisfy (anti-)periodic boundary conditions
with period $\tau = 1/T$ and can be decomposed
into bosonic and fermionic Matsubara~\cite{Matsubara:1955ws} modes
\begin{equation}
\label{eq:ko}
  \phi(\tau,\vec{k}) =
    T \sum_{n=-\infty}^{\infty}
    \phi_{n}(\vec{k}) e^{i\ko^{ }_{n}\tau}
  \;,\quad
  \ko_{n} =
    \begin{cases}
      \ko^{\rmii{B}}_{n} = 2n\pi T &(\text{bosons})\\
      \ko^{\rmii{F}}_{n} = (2n+1)\pi T &(\text{fermions})
    \end{cases}
  \;,\quad
  n\in\ZZ
  \;,
\end{equation}
where
$\vec{k}$ is a three-dimensional (3d) momentum.
In other words the resulting theory is a 3d one with an
infinite tower of modes each carrying a mass $\omega_n^2$ corresponding
to the Matsubara frequency of mode $n$.

This system can be studied in an effective theory formulation.
In that EFT the central degree of freedom is
the static bosonic 3d zero mode ($\ko^{\rmii{B}}_{n=0}$) of
the original four-dimensional (4d) field.
The remaining non-zero modes of scale $\sim\pi T$ can been integrated out.
This is the {\em dimensional reduction step} which
is based on the high-temperature scale hierarchy
\begin{equation}
\label{eq:scale:hierarchy}
  \pi T \gg
  g^{ }T \gg
  g^{2}T/\pi
  \;.
\end{equation}
In the scale hierarchy we introduced a power counting parameter $g$ defining
the
hard ($\pi T$),
soft ($g^{ }T$), and
ultrasoft ($g^{2}T/\pi$) scales.
While the scaling of the hard scale is a direct consequence of
the Matsubara decomposition,
the soft and ultrasoft scale are pertinent to collective plasma effects.
Based on this hierarchy one can invoke a high-temperature expansion
$m_{\psi}/T\ll 1$ for generic scalar fields $\psi$, whereas
gauge bosons and fermions are massless in the unbroken phase.

In hindsight of the ensuing studies of a real scalar field,
we establish the formal scaling
$\lambda\sim g^{2}$
for the scalar quartic coupling $\lambda$
which is based on their appearance at one-loop.
In gauge field theories this power counting parameter is often set to be the gauge coupling.
For a scalar field we assume
the original mass squared parameter ($\mu^2$) to behave as
$\mu^2 \sim \lambda T^2 \sim (g^{ }T)^2$
which implies that the mass of the 3d soft mode ($\mu_{3}^{2}$) is thermally corrected by
$\mu_{3}^{2}\simeq\mu^2 + (g^{ }T)^2$ at leading order.

Phase transition physics can often be studied at
the ultrasoft scale by a simplified 3d EFT, where the soft scale has been integrated out.
In fact, the transition point resides near a vanishing
$\mu_{3}^{2}$
where thermal loop corrections cancel the tree-level part.
At this point the 3d mass scale is formally of the next natural order
which is the ultrasoft one
$\mu_{3}^{2}\sim (g^{2}T)^2$
where soft modes are screened.
The corresponding soft degrees of freedom are
the temporal (adjoint) scalar fields which are remnants of the zero components of gauge fields and
induced by the broken Lorentz symmetry from the heat bath.
They remain soft in the vicinity of the transition point.
For this {\em second step} of dimensional reduction, see ref.~\cite{Kajantie:1995dw}.

The EFT is constructed by determining the operator coefficients of the effective Lagrangian.
In practice, these parameters follow from matching correlation functions of both
the fundamental 4d theory and
effective 3d theory.
The generic rules of this procedure were established in
refs.~\cite{Kajantie:1995dw,Braaten:1995cm,Braaten:1995jr} and
applied recently~\cite{Croon:2020cgk,Gould:2021dzl}.

This construction of the 3d EFT by dimensional reduction is completely infrared-{\em safe}.
In the matching of correlators, the IR and 3d contributions cancel each other and
only the hard scale (non-zero Matsubara modes) contributes.
The corresponding sum-integrals over non-zero modes are IR-regulated by
non-vanishing Matsubara frequencies at high temperature.
Hence, the dimensional reduction defers
the IR problem of high-temperature bosonic perturbation theory to the 3d EFT.

At next-to-leading order (NLO) dimensional reduction, couplings are matched at one-loop and masses at two-loop level.
This ensures a $\mathcal{O}(g^4)$ accuracy in the established power counting.
To fully match this accuracy, the running parameters have to be related to
physical observables at one-loop order in zero temperature perturbation theory.
In 3d perturbation theory the effective potential is computed at two-loop order.
Notably, the frequently used 4d daisy-resummed thermal effective potential at one-loop includes some --
but crucially not all -- $\mathcal{O}(g^4)$ contributions~\cite{Arnold:1992rz}.

Instead of detailing the generic procedure in later sections,
we choose an alternative approach.
In an explicit hands-on demonstration, we dimensionally reduce
a scalar field theory
in sec.~\ref{se:tutorial}, and generalise it to the xSM in
sec.~\ref{se:dr:xsm} and
appendix~\ref{se:dr:xsm:details}.

%
\subsection{Approaches to thermodynamics of thermal phase transitions}
\label{se:thermo:approaches}

Let us assess the main approaches to access the thermodynamics of
the thermal phase transitions in electroweak theories.
See also similar summaries in
sec.~2 of ref.~\cite{Bodeker:1996pc},
sec.~2.2 of ref.~\cite{Laine:2012jy} and
sec.~1 of ref.~\cite{Gould:2021dzl}.
The following approaches are illustrated in fig.~\ref{fig:pipeline-thermo}:
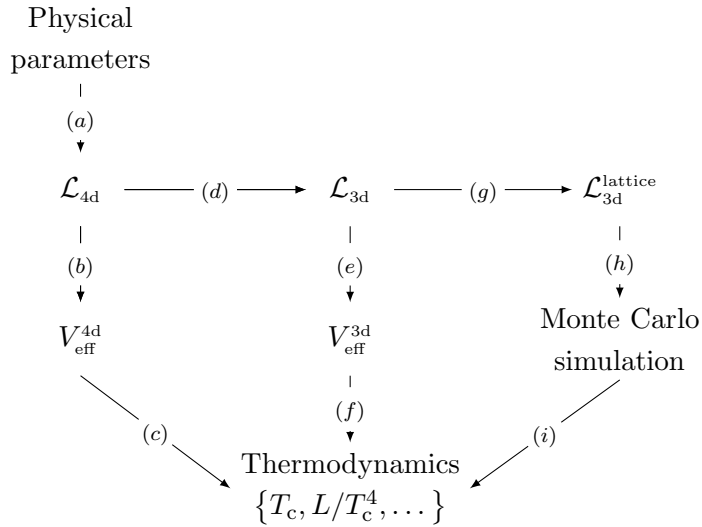
\begin{figure}
\centering
\tikzstyle{zeroT}=[fill=white!20,minimum width=3.0em,
    align=center,minimum height=2.5em]
\def\blockw{3}
\def\blockh{1.5}
\newcommand\circled[1]{%
  \tikz[baseline=(X.base)]
    \node (X) [shape=circle,inner sep=-1pt,fill=white,text=black] {\strut\scriptsize #1};%
  }
\begin{tikzpicture}
  \node (lag4d) [zeroT]  {$\mathcal{L}_{\rmii{4d}}$};
  \path (lag4d.north)+(0,\blockh) node (phys) [zeroT] {Physical\\ parameters};
  \path (lag4d.east)+(\blockw,0) node (lag3d) [zeroT] {$\mathcal{L}_{\rmii{3d}}$};
  \path (lag3d.south)+(0,-\blockh) node (veff3d) [zeroT] {$V^{\rmii{3d}}_{\rmii{eff}}$};
  \path (veff3d.south)+(0,-\blockh) node (thermo) [zeroT] {Thermodynamics \\$\bigl\{\Tc,L/\Tc^4,\dots\bigr\}$};
  \path (lag4d.south)+(0,-\blockh) node (veff4d) [zeroT] {$V^{\rmii{4d}}_{\rmii{eff}}$};
  \path (lag3d.east)+(\blockw,0) node (lag3dlattice) [zeroT] {$\mathcal{L}^\rmii{lattice}_{\rmii{3d}}$};
  \path (lag3dlattice.south)+(0,-\blockh) node (MC) [zeroT] {Monte Carlo\\ simulation};

  \draw[<-,>=latex] (lag4d.north) -- node[midway] {\hyperref[it:th:a]{\circled{$(a)$}}} (phys.south);
  \draw[->,>=latex] (lag4d.south) -- node[midway] {\hyperref[it:th:b]{\circled{$(b)$}}} (veff4d.north);
  \draw[->,>=latex] (lag4d) -- node[midway] {\hyperref[it:th:d]{\circled{$(d)$}}} (lag3d);
  \draw[->,>=latex] (lag3d.south) -- node[midway] {\hyperref[it:th:e]{\circled{$(e)$}}} (veff3d.north);
  \draw[->,>=latex] (veff3d.south) -- node[midway] {\hyperref[it:th:f]{\circled{$(f)$}}} (thermo.north);
  \draw[->,>=latex] (lag4d.south) --
      node[midway] {\circled{$(b)$}}
      (veff4d.north);
  \draw[->,>=latex] (lag4d.east) --
      node[midway,text=blue] {\circled{$(d)$}}
      (lag3d.west);
  \draw[->,>=latex] (veff4d.south) -- node[midway] {\hyperref[it:th:c]{\circled{$(c)$}}} (thermo.west);
  \draw[->,>=latex] (lag3d.east) -- node[midway] {\hyperref[it:th:g]{\circled{$(g)$}}} (lag3dlattice.west);
  \draw[->,>=latex] (lag3dlattice.south) -- node[midway] {\hyperref[it:th:h]{\circled{$(h)$}}} (MC.north);
  \draw[->,>=latex] (MC.south) -- node[midway] {\hyperref[it:th:i]{\circled{$(i)$}}} (thermo.east);
\end{tikzpicture}
\caption{
  Three different approaches towards the thermodynamics of the electroweak phase transition.
  We focus on the purely perturbative 3d approach with steps $(d)$ and $(e)$ for
  a real scalar theory and
  the Standard model supplemented by a real scalar singlet.
  }
  \label{fig:pipeline-thermo}
\end{figure}
\begin{itemize}
\item
  {\em ``4d approach''}\;
  $(a)\to(b)\to(c)$:\\
  Perturbative effective potential with daisy resummation.
\item
  {\em ``Perturbative 3d approach''}\;
  $(a)\to(d)\to(e)\to(f)$:\\
  Perturbative effective potential in 3d EFT.
\item
  {\em ``Non-perturbative 3d approach''}\;
  $(a)\to(d)\to(g)\to(h)\to(i)$:\\
  Non-perturbative lattice simulation of 3d EFT.
  Robust approach combining
  perturbative dimensional reduction and
  non-perturbative (Monte Carlo) methods.
\end{itemize}
The individual steps encompass:
\begin{itemize}
\item[$(a)$]\label{it:th:a}
  Relating physical parameters (such as pole masses) and Lagrangian (running) parameters
  at zero temperature.
  Often the ``4d approach'' uses only tree-level relations
  (e.g.\ refs.~\cite{Dorsch:2016nrg,Basler:2018cwe}), but
  in order to match the accuracy of dimensional reduction at NLO $\mathcal{O}(g^4)$,
  one-loop vacuum renormalisation is required~\cite{Kajantie:1995dw,Laine:2017hdk,Kainulainen:2019kyp}.
\item[$(b)$]\label{it:th:b}
  Perturbative computation of the thermal effective potential~\cite{Dolan:1973qd}.
  Frequently performed at one-loop, with leading order daisy resummation~\cite{Arnold:1992rz,Parwani:1991gq}.
  Two-loop computations are discussed for e.g. in refs.~\cite{Buchmuller:1993bq,Fodor:1994bs,Laine:2017hdk,Ekstedt:2020abj}.
  This computation suffers from the IR problem the most, and additionally can contain
  a dramatic artificial RG scale dependence if
  two-loop thermal masses are unaccounted~\cite{Gould:2021xxx}.
\item[$(c)$]\label{it:th:c}
  Computation of thermodynamics.
  At the critical temperature the minima of the effective potential are degenerate and
  thermodynamic quantities are obtained by differentiation with respect to temperature,
  {\em viz.}\ latent heat.
  Model-independent tools to locate degenerate minima
  have been implemented numerically in software including
  {\tt CosmoTransitions}~\cite{Wainwright:2011kj},
  {\tt BSMPT}~\cite{Basler:2018cwe}, and
  {\tt PhaseTracer}~\cite{Athron:2020sbe}.
  It is worth noting that location of minima of the effective potential are not gauge invariant.
  Thus, this computation frequently introduces unphysical estimates for thermodynamics
  as discussed in
  refs.~\cite{Patel:2011th,Ekstedt:2020abj,Croon:2020cgk} and also
  ref.~\cite{Laine:1994zq} (in 3d EFT context).
\item[$(d)$]\label{it:th:d}
  Dimensional reduction to a 3d EFT. See
  refs.~\cite{Kajantie:1995dw,Braaten:1995cm,Braaten:1995jr} and also recent
  refs.~\cite{Croon:2020cgk,Gould:2021dzl}.
  It is perturbative and IR-safe, since only the hard scale is integrated out, and
  systematically implements all required resummations.
  Furthermore, dimensional reduction at NLO is analytically independent of
  the 4d renormalisation scale up to that order,
  see ref.~\cite{Gould:2021xxx,Kajantie:1995dw}.
  This decreases the theoretical uncertainty in perturbation theory.
  A concrete computation is displayed in
  secs.~\ref{se:tutorial},~\ref{se:dr:xsm} and
  appendix~\ref{se:dr:xsm:details}.
\item[$(e)$]\label{it:th:e}
  Computation of 3d effective potential, see refs.~\cite{Farakos:1994kx,Laine:1994bf,Laine:1994zq}.
  The computation in the 3d EFT simplifies significantly compared to 4d because
  sum-integrals are replaced by vacuum integrals in $d=3-2\epsilon$ spatial dimensions.
  Hence, it straightforwardly extends to two-loop order, cf.\ sec.~\ref{se:tutorial:veff}.
  For recent applications, see
  refs.~\cite{Niemi:2020hto,Croon:2020cgk}.
  Even the three-loop effective potential has been computed for
  a pure scalar theory~\cite{Rajantie:1996np} and applied recently~\cite{Gould:2021dzl}.
\item[$(f)$]\label{it:th:f}
  Computation of thermodynamics from 3d effective potential.
  Again a pathological gauge-dependent analysis can be based on
  degenerate minima at the transition point.
  However, also a gauge invariant treatment is possible,
  in terms of gauge invariant condensates~\cite{Farakos:1994xh,Croon:2020cgk} or
  the pressure in $\hbar$-expansion.
  However, IR divergences arise at two-loop order for a radiatively generated transition~\cite{Kajantie:1995kf},
  compromising the analysis~\cite{Laine:1994zq}.
  On the other hand, these IR singularities are avoided in presence of a barrier
  at tree-level, and
  a manifestly gauge invariant treatment for
  the thermodynamics can be obtained in perturbation theory; see ref.~\cite{Croon:2020cgk}.
\item[$(g)$]\label{it:th:g}
  Lattice-continuum relations; see refs.~\cite{Farakos:1994xh,Kajantie:1995kf,Laine:1995np,Laine:1997dy,Gould:2021dzl}.
  The Lagrangian parameters of the lattice discretisation need to be related to those of
  the continuum theory.
  Thus, the results of Monte Carlo simulations can be associated
  with the 3d continuum theory
  and via dimensional reduction to temperature and physical parameters.
  This can be done by computing and equating effective potentials in both discretisations,
  to two-loop order.
  In super-renormalisable theories without higher dimensional operators,
  all divergences arise at finite loop order and hence
  relations between continuum and lattice are exact.
  However, this aggravates in the presence of higher dimensional operators
  as the 3d theory retains renormalisability but loses super-renormalisability.
  It remains a future challenge to overcome this technical issue.
  Note that in lattice gauge theories there is no need to fix the gauge, and
  the treatment is automatically gauge invariant by construction~\cite{Rothe:1992nt}.
\item[$(h)$]\label{it:th:h}
  Monte Carlo lattice simulations of spatial 3d EFT on finite volume and lattice spacing.%
  \footnote{
    See refs.~\cite{%
    Farakos:1994xh,Kajantie:1995kf,Kajantie:1996qd,Gurtler:1996wx,Kajantie:1997tt,Kajantie:1997pd,
    Kajantie:1997vc,Laine:1998vn,Laine:1998qk,Laine:2000rm,Hietanen:2008tv,
    Laine:2012jy,DOnofrio:2015gop,Kainulainen:2019kyp,Niemi:2020hto,Gould:2021dzl}.
  }
  Arbitrary field configurations are evolved --
  usually by a colourful cocktail of update algorithms --
  to form a Markov chain converging to a Boltzmann probability distribution.
  Thereof, physical quantities can be measured such as
  scalar and gauge condensates, and
  correlation lengths.
  Many autocorrelation times are measured to ensure that statistical errors remain small.
  At the transition point, the system is equally likely to occur in any of the phases.
  Multi-canonical methods in first order transitions ensure that
  the system can efficiently sample all phases while not getting stuck in one.
\item[$(i)$]\label{it:th:i}
  Extrapolate simulations of finite volume and fixed lattice spacing to the continuum.
  This corresponds to infinite volume and vanishing lattice spacing,
  thermodynamic and continuum limits, respectively.
  In practice, several lattice spacings are needed, each with several different volumes.
  This rapidly becomes computationally expensive and
  even a single parameter space point requires a large number of individual simulations.
  Furthermore, oftentimes manual effort is required to fit a proper continuum extrapolation to
  the data instead of an elephant.
\end{itemize}
This article focuses specifically on
steps $(d)$ and $(e)$
which are detailed for
a real scalar theory in
sec.~\ref{se:tutorial} and
a real singlet scalar coupled to the SM in
sec.~\ref{se:dr:xsm} and appendix~\ref{se:dr:xsm:details}.
The full non-perturbative path of the real scalar theory is presented in ref.~\cite{Gould:2021dzl},
where the corresponding results are compared with three-loop 3d perturbation theory.

Finally, let us summarise the different approaches and describe some of their merits.

%
\subsubsection{4d approach}
\label{se:4d}

The 4d approach is the accustomed ``bread and butter'' approach with the advantage
of its conceptual simplicity.
At one-loop order, a closed form expression for the effective potential is available in terms of
mass squared eigenvalues and
one-loop thermal mass corrections,
which straightforwardly automates to different models.
In addition,
numerical tools for thermodynamics (e.g.\ minimisation of potential)
have been developed~\cite{Wainwright:2011kj,Basler:2018cwe,Athron:2019nbd,Athron:2020sbe} and
can scan large regions parameter space of BSM models.

However, the 4d approach suffers from the IR problem of perturbation theory~\cite{Linde:1980ts}
and is often plagued with
large inaccuracies and
theoretical uncertainties~\cite{Farakos:1994xh,Kajantie:1995kf,Kajantie:1996qd,Kainulainen:2019kyp,Croon:2020cgk}.
In particular, weak transitions are poorly described by perturbation theory
and are sometimes even qualitatively mistaken.
Especially, crossover transitions are not predicted at all and
it is not expected to determine the critical temperature accurately
since it is highly IR-sensitive.
Conversely, large couplings are often required for strong transition
and can compromise the perturbative expansion, even at zero temperature~\cite{Kainulainen:2019kyp}.
Consistent ($\hbar$-)expansions leading to gauge invariant results are oftentimes unavailable,
since they require the knowledge of higher order contributions.
Furthermore, a truncation of the computation already at one-loop order
omits important thermal mass contributions at two-loop order.
In turn, this causes a large leftover renormalisation group (RG) scale dependence, see ref.~\cite{Gould:2021xxx}.

%
\subsubsection{Perturbative 3d approach}
\label{se:3d:pert}

Also this method still suffers from the IR problem of perturbation theory.
We emphasise that for the perturbative effective potential itself,
there is no real quantitative difference between 4d and 3d approaches,
provided that in both cases the computation is performed to the same order in
both coupling expansion and high-$T$ expansion.
However, dimensional reduction systematically accesses higher order resummations
and it is customary to include a consistent $\mathcal{O}(g^4)$ accuracy by
a two-loop level computation, which yields
a reduced RG scale dependence.
A gauge dependence of the analysis can still be a theoretical blemish,
but a gauge invariant treatment is possible by
employing a $\hbar$-expansion and
computing gauge invariant condensates~\cite{Farakos:1994xh,Croon:2020cgk}.
Although radiatively induced transitions suffer from IR divergences
at $\mathcal{O}(\hbar^2)$~\cite{Laine:1994zq}.

As a downside, the perturbative 3d approach is
harder to automate and streamline compared to 4d approach due to additional steps.
Although the computation of the 3d effective potential of a 3d EFT
can be related to known
topologies arising at two-loop order,
the automation of dimensional reduction and
the construction of the 3d EFT are still not common standard.
For developments, see~\cite{Nishimura:2012ee,Laine:2018lgj,Schicho:2020xaf,Croon:2020cgk}.
In the future, automated dimensional reduction for multiple BSM theories could permit
a perturbative 3d approach to be implemented to software that currently relies on
the 4d approach.

%
\subsubsection{Non-perturbative 3d approach}
\label{se:3d:nonpert}

This method solves the IR problem, by treating
perturbative (hard and soft) modes perturbatively while
non-perturbative ultrasoft modes are analysed by lattice simulations.
Furthermore, lattice simulations provide manifestly gauge invariant results.
While this approach is very technical and
computationally slow and demanding, it is still straightforward compared to
direct 4d simulations (see ref.~\cite{Laine:1996nz} and references therein).
Recent attempts~\cite{Kainulainen:2019kyp,Niemi:2020hto,Gould:2021dzl}
simulate phase transitions in
a limited number of BSM setups and benchmark points.
The hope is to expose general trends
regarding accuracy and reliability of simpler tools in perturbation theory.
However, model-independent or conclusive results are unavailable so far and
similar investigations are actively continued in the future. Finally, we highlight that
the 3d EFT approach is also an applicable and attractive framework for
non-equilibrium physics of phase transition, such as
bubble nucleation and sphaleron rate; see refs.~\cite{Moore:2000jw,Moore:2001vf,Moore:1998swa}.

%
\section{Dimensional reduction with a real scalar: A tutorial}
\label{se:tutorial}

The following tutorial constructs the dimensionally reduced 3d EFT of
a single real scalar field.%
\footnote{
  \label{footnote:Gould}
  An independent computation~\cite{Gould:2021dzl}
  treats masses and tadpole as interactions in
  strict perturbation theory and in the unbroken phase.
  The dictionary
  $\{
  \mu_1 \leftrightarrow \sigma,
  \mu_{\sigma}^{2} \leftrightarrow m^2,
  \mu_3 \leftrightarrow \frac{g}{2},
  \lambda_\sigma \leftrightarrow \frac{\lambda}{6}
  \}$
  gives a direct comparison,
  wherein $g$ denotes the cubic coupling of the real scalar
  not to be confused with
  a formal power counting parameter or
  the ${\rm SU}(2)$ gauge coupling.
}
The machinery presented
builds upon classic literature~\cite{Farakos:1994kx,Kajantie:1995dw,Braaten:1995cm,Braaten:1995jr}
and generalises straightforwardly to more
complicated BSM theories with non-minimal Higgs sector.
To guide upcoming generalisations of complicated models, we
detail step-by-step derivations that can be used for future crosschecks.
Furthermore, based on advances of automation in thermal field theories~\cite{%
  Nishimura:2012ee,Laine:2018lgj,Schicho:2020xaf,Croon:2020cgk},
we implemented in-house software in
{\tt FORM}~\cite{Ruijl:2017dtg} and applied
{\tt qgraph}~\cite{Nogueira:1991ex} for diagram generation.
Integration-by-parts reductions (IBP) follow a standard Laporta algorithm~\cite{Laporta:2001dd}
adapted for thermal integrals~\cite{Nishimura:2012ee}.
This algorithmic perturbative treatment fully automates
the computation of correlation functions within the unbroken phase and
their matching.
This software can tame
the ever increasing complexity of computations in future models
with multiple interacting new BSM fields.

The real singlet scalar model demonstrates all details of dimensional reduction.
Without coupling it to the SM with
the Higgs doublet, gauge fields, and fermions,
this model poses an ideal starting point.
The following computations employ
explicit resummation to cancel delicate soft/hard mixing contributions at two-loop order.
Practically,
these IR contributions are trivially dropped~\cite{Gould:2021dzl}
in {\em strict perturbation theory}.
As an instructive crosscheck, we perform the computation both in
the broken phase using the effective potential, the generator of correlators, as well as
the unbroken phase computing correlators directly.

Section~\ref{se:dr:xsm} focusses on full xSM -- where a real singlet scalar is coupled to the SM Higgs -- and
presents the definition of the EFT including results.
Details of this full computation are relegated to
appendix~\ref{se:dr:xsm:details} and
results of appearing (sum-)integrals are collected in
appendix~\ref{se:integrals}.
For the derivation of such integrals, we refer
refs.~\cite{Arnold:1994eb,Nishimura:2012ee,Osterman:2019xx}.
Our notation follows ref.~\cite{Brauner:2016fla} in which dimensional reduction for
the xSM was initially discussed.
This reference deferred
the case of a light (``soft'') singlet which remains dynamical in the 3d EFT
to our computation.

%
\subsection{Model and parameter matching}
\label{se:tutorial-model}

Consider the theory of
a single real scalar field $\sigma$,
given by bare 4d Lagrangian (with Euclidean metric) in the imaginary time
formalism
\begin{align}
\label{eq:lag:singlet:4d}
\mathcal{L} =
    \frac{1}{2} (\partial_\mu^{ } \sigma_{(b)}^{ })^2
  + \frac{1}{2} \mu_{\sigma(b)}^{2} \sigma_{(b)}^{2}
  + \mu_{1(b)}^{ } \sigma_{(b)}^{ }
  + \frac{1}{3} \mu_{3(b)}^{ } \sigma_{(b)}^{3}
  + \frac{1}{4} \lambda_{\sigma(b)}^{ } \sigma_{(b)}^{4}
  \;.
\end{align}
Definitions of bare quantities in terms of their renormalised versions and
counterterms in renormalised perturbation theory are found in sec.~2.1 of ref.~\cite{Brauner:2016fla}.
Note that we choose a general renormalisable theory with linear and cubic terms without
$Z_2$-symmetry $\sigma\to -\sigma$.
Conveniently, in the simple case of a real scalar (without gauge fields)
the 3d EFT and 4d parent theory bear the same form.
With the exception that after dimensional reduction
the couplings and field live in a spatial 3d theory.
We organise our perturbative expansion by establishing
the following formal power counting
\begin{equation}
\label{eq:scaling:singlet}
\mu_{1}^{ }\sim g^{ }T^{3}\;,\quad
\mu_{\sigma}^{2}\sim g^{2}T^{2}\;,\quad
\mu_{3}{ }\sim g^{ }T\;,\quad
\lambda_{\sigma}^{ }\sim g^{2}
\;,
\end{equation}
where $g$ is a formal power counting parameter that corresponds to
the weak coupling at zero temperature.
Once this theory couples to the SM in sec.~\ref{se:dr:xsm},
the formal power counting parameter $g$ is identified as the SU(2) gauge coupling.
Within the power counting~\eqref{eq:scaling:singlet},
we aim for a dimensional reduction at NLO with $\mathcal{O}(g^4)$ accuracy.
At loop-level, this requires
one-loop accuracy for cubic and quartic couplings, and
two-loop order for tadpole and mass parameter.

We emphasise that the above formal choice for the scaling of the cubic coupling
leads to a peculiarity illustrated along
the tree-level quartic interactions induced by the cubic coupling.
Its contribution
\begin{align}
\label{eq:scale:a}
\VtxvS(\Lsr1,\Lsr1,\Lsr1,\Lsr1,\Lsr1) &\simeq
  \frac{\mu_{3}^{2}}{\mu_{\sigma}^{2}} \sim
  \mathcal{O}(1) \gg
  \lambda_\sigma
  \;
\end{align}
parametrically dominates over the corresponding quartic coupling, and
could even compromise perturbativity at zero temperature.
In practice for dimensional reduction, this causes no complications
as the above interaction is 1-particle reducible.
Hence, it is absent in Green's functions that are matched during dimensional reduction for
the ultrasoft (light) field in 3d EFT.
By enforcing a different scaling, namely
$\mu_{3}\sim g^{2}T$,
the contribution~\eqref{eq:scale:a} formally scales as the corresponding quartic coupling.
However, this suppresses almost all contributions of
$\mu_{3}$ in the matching relations at $\mathcal{O}(g^4)$.

Hence, our strategy is the following:
for generality we indeed install a scaling of
$\mu_{3}\sim g^{ }T$ and
include all contributions of cubic couplings in our matching relations.
These contributions can always be trivially dropped if an extra suppression is assumed.
This choice allows us to more widely illustrate different aspects of
the dimensional reduction procedure, such as effects from field normalisation.
Indeed, a non-$Z_2$-symmetric theory can demonstrate
the high-temperature screening of the fields by the hard scale via
ring topology diagrams such as in eq.~\eqref{eq:fn-res}.
These are absent in a $Z_2$-symmetric case.
The motivation to include these contributions is the presence of similar diagrams
in theories with
gauge fields and fermions, even if the scalar sector is $Z_2$-symmetric.

Figure~\ref{fig:matching} illustrates the NLO matching of the parameters.
For references with explicit matching examples,
see refs.~\cite{Kajantie:1995dw,Braaten:1995cm,Braaten:1995jr,Brauner:2016fla,Croon:2020cgk,Gould:2021dzl}.
\begin{figure}
\centering
\begin{align*}
(\sigma^2)_{\rmii{3d}} &=
  \frac{1}{T} (\sigma^2)_{\rmii{4d}} \Big(1 + \hat{\Pi}'_{\sigma^2} \Big)
  \\
  &= \frac{1}{T} (\sigma^2)_{\rmii{4d}}
  \Bigl(
    1
    + \tfrac{{\rm d}}{{\rm d} K^2}
    \!\TopSi(\Lsr1,1)
  \Bigr)
  \;, \\[2mm]
\Vtxtn(\Lsr1,\Lsr1,\sigma)
\Bigr|_{\rmii{3d}} &=
\Bigl\{
  \Bigl(
      \Vtxt(\Lsr1,\Lsr1)
    + \TopSi(\Lsr1,1)
  \Bigr)
  \Bigl(
    1
    + \tfrac{{\rm d}}{{\rm d} K^2}\!\TopSi(\Lsr1,1)
  \Bigr)
  + \TopSi(\Lsr1,2)
  \Bigr\}_{\rmii{4d}} \quad
  \;,\\[2mm]
\Vtxvn(\Lsr1,\Lsr1,\Lsr1,\Lsr1,\sigma,\sigma,\sigma,\sigma)
\Bigr|_{\rmii{3d}} &= T
\Bigl\{
  \Vtxv(\Lsr1,\Lsr1,\Lsr1,\Lsr1)
  + \TopVi(\Lsr1,\Lsr1,\Lsr1,\Lsr1,1)
  + \Vtxv(\Lsr1,\Lsr1,\Lsr1,\Lsr1)
  \Bigl(\tfrac{{\rm d}}{{\rm d} K^2}\!\!\TopSi(\Lsr1,1)
  \Bigr)
\Bigr\}_{\rmii{4d}}
\end{align*}
  \caption{
    Illustration of a $\rmO(g^4)$ NLO matching of correlators between
    3d and 4d theories with $Z_2$-symmetry.
    A full non-$Z_2$-symmetric case is analogous.
    Blobs present the sum of hard contributions
    to one- and two-loop diagrams in perturbation theory,
    and the differentiation (prime) acts upon
    the external soft momentum $K=(0,\vec k)$.
  }
  \label{fig:matching}
\end{figure}
Loop corrections from the 3d side and
soft contributions match exactly.
Hence both drop out trivially in the matching,
leaving only hard contributions.
At two-loop order mixed soft/hard terms are cancelled by
counterterm-like resummation interactions at one-loop~\cite{Kajantie:1995dw,Gorda:2018hvi}.
We demonstrate this in sec.~\ref{se:hard-soft:cancel}.

For example, the matching of correlation functions for
the quartic self-interaction in both theories yields
\begin{align}
  \underbrace{T \Big(
    - 6 \lambda_{\sigma,3}
    - \langle \sigma^4 \rangle^{\rmii{3d}}_{\rmii{1loop}} \Big)}_{\rm 3d} =
    \underbrace{\vphantom{\Big(}
    - 6 \lambda_\sigma
    - \langle \sigma^4 \rangle^{\rmii{soft}}_{\rmii{1loop}}
    - \langle \sigma^4 \rangle^{\rmii{hard}}_{\rmii{1loop}}}_{\rm 4d}
  \;.
\end{align}
Note that here the correlators equal minus the sum of
the tree-level vertex and Feynman diagrams at higher orders.
For a consistent matching,
the resummation of parameters (see details in eq.~\eqref{eq:resum})
for zero modes allows to identify
3d loop corrections with
soft terms in the 4d computation,
and these two (IR) terms cancel.
Furthermore, resummation ensures a cancellation of all mixed soft/hard mode contributions
in one- and two-point correlation functions at two-loop order.
For details, see eqs.~\eqref{eq:veff-resum} and
\eqref{eq:resum:e}--\eqref{eq:resum:d}.
The effective vertices in both theories read
\begin{align}
  T \lambda_{\sigma,3^{ }}\,\sigma^4_{\rmii{3d}} = \Big(
    \lambda_\sigma
  + \frac{1}{6} \langle \sigma^4 \rangle^{\rmii{hard}}_{\rmii{1loop}} \Big) \sigma^4_{\rmii{4d}}
  \;,
\end{align}
and relating 3d and 4d fields by
the first line of fig.~\ref{fig:matching} leads to a $\mathcal{O}(g^4)$ result
\begin{align}
  T \lambda_{\sigma,3}^{ } = T \Big(
    \lambda_\sigma^{ }
  + \frac{1}{6} \langle \sigma^4 \rangle^{\rmii{hard}}_{\rmii{1loop}}
  - 2 \lambda_\sigma^{ } \Pi_2' \Big)
  \;.
\end{align}
Matching relations of other effective parameters are obtained analogously.

The 4d and 3d fields
at one-loop level are related by a computation of the ring topology diagram
at non-zero external static momentum $K=(0,\vec{k})$ with soft
$|\vec{k}| = k\sim g T$.
Denoting sum-integrals according to appendix~\ref{se:integrals},
a series expansion to quadratic order yields
\begin{align}
\label{eq:fn-res}
  \Pi_2' &\equiv
    \frac{{\rm d}}{{\rm d}k^2} \Pi_2(k)
 = \frac{{\rm d}}{{\rm d}k^2} \TopoSB(\Lsr1,\Asr1,\Asr1) \quad
  \nn &=
    \frac{{\rm d}}{{\rm d}k^2} \bigg( -2\mu_{3}^{2} \Tint{P}' \frac{1}{P^2(P+K)^2} \bigg)
  \nn &=
  \frac{{\rm d}}{{\rm d}k^2} \bigg(
    -2\mu_{3}^{2} \Tint{P}' \frac{1}{P^2} \Big(
        \frac{1}{P^2}
      - 2\frac{\vec k \cdot \vec p}{P^4}
      + 4 \frac{(\vec k \cdot \vec p)^2}{P^6}
      - k^2 \frac{1}{P^4}
      + \rmO(k^3)
  \Big) \bigg)
  \nn &=
  2 \mu_{3}^{2} \Tint{P}' \bigg(\frac{d-4}{d}\frac{1}{P^6} + \frac{4}{d} \frac{P^2_0}{P^8} \bigg)
  \nn &=
  \frac{2}{3} \mu_{3}^{2} \Tint{P}' \frac{1}{P^6}
  =
  2 \mu_{3}^2 \frac{1}{(4\pi)^4} \frac{2}{3} \frac{\zeta_{3}}{T^2}
  \;,
\end{align}
where we interchangeably denote
$\Pi_n = \langle \sigma^n \rangle$ as correlation functions
and utilise
the intact $d$-dimensional rotational symmetry
$(\vec{k}\cdot\vec{p})^2 \to \frac{1}{d} k^2 p^2$,
trivial manipulation $p^2 = P^2 - P^2_0$ and
integration-by-parts relations
$
\Tinti{P} \frac{(P^2_0)^{\beta+1}}{[P^2]^{\alpha+1}}
=\big(1-\frac{d}{2\alpha} \big)
\Tinti{P}\frac{(P^2_0)^\beta}{[P^2]^\alpha}
$.

The effective potential generates all correlation functions at zero external momenta.
In cases where an explicit momentum dependence exceeds the accuracy of the computation,
accessing correlators simplifies greatly by starting from the effective potential.
Shifting the scalar field
$\sigma \rightarrow \sigma + s$ with a real background field $s$,
the effective potential reads
\begin{align}
\label{eq:veff-s-expansion}
  V_{\rmii{eff}} = \sum_{n=1}^{\infty} \frac{\langle \sigma^n \rangle}{n!} s^n \equiv
    V_1 s
    + \frac{1}{2} V_2 s^2
    + \frac{1}{3} V_3 s^3
    + \frac{1}{4} V_4 s^4
    + \ldots
  \;,
\end{align}
where the ellipsis truncates
potential higher dimensional correlators that lead to
$\mathcal{O}(g^5)$ marginal operators in the EFT.

The complete matching relations that define the dimensionally reduced 3d EFT,
encode the thermodynamics of
the original 4d theory of eq.~\eqref{eq:lag:singlet:4d} at $\mathcal{O}(g^4)$.
They read
\begin{align}
\label{eq:matching-1}
\mu_{1,3}^{ } &=
  T^{-\frac{1}{2}}\Big(
    V^{\rmii{2loop}}_1
  - \frac{1}{2} V^{\rmii{1loop}}_1 \Pi_2'
  \Big)
  \;, \\
\mu_{\sigma,3}^{2} &=
    V^{\rmii{2loop}}_2
  - V^{\rmii{1loop}}_2 \Pi_2'
  \;, \\
\mu_{3,3}^{ } &=
  T^{\frac{1}{2}} \Big(
    V^{\rmii{1loop}}_3
    - \frac{3}{2} \mu_{3}^{ } \Pi_2'
  \Big)
  \;, \\
\label{eq:matching-4}
\lambda_{\sigma,3}^{ } &= T \Big(
    V^{\rmii{1loop}}_4
  - 2 \lambda_\sigma^{ } \Pi_2'
  \Big)
  \;,
\end{align}
where we indicated the required loop order for a $\mathcal{O}(g^4)$ accuracy.
The coefficients $V$ merely contain hard contributions
and correspond to correlators via eq.~\eqref{eq:veff-s-expansion}.

%
\subsection{Computation of correlators}
\label{se:corelators:singlet}

The computation of $n$-point correlation functions is expounded in two different ways.
In the following, we discuss their merits.
In the broken phase, the computation uses the mass eigenstate basis and
we employ the effective potential which is
the generator of the correlation functions.
Therein, the scalar field is shifted by a classical background field.
In the unbroken phase, in the gauge eigenstate basis,
we compute all correlators directly diagram-by-diagram.

These two approaches give rise to an equivalent result.
Technically, for a real scalar theory
the broken and unbroken phase computations vary marginally.
However, subtleties of these two approaches become more prominent
for gauge field theories with (multiple) scalars
in different representations of the underlying gauge symmetry group.

%
\subsubsection{Broken phase: Correlators from the two-loop effective potential}

\begin{figure}
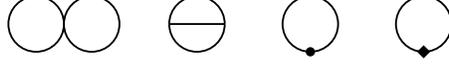

\centering
  \begin{align*}
   \ToptVE(\Asa3,\Asa3)\,\quad
   \ToptVS(\Asa3,\Asa3,\Lsa3)\,\quad
   \TopoVRo(\Asa3)\,\quad
   \TopoVRd(\Asa3)
  \end{align*}
  \caption{
    Two-loop level diagrams for the effective potential in 4d perturbation theory.
    The last two diagrams comprise renormalisation and resummation counterterms
  with quadratic interactions, respectively.
  }
  \label{fig:2loop-diagrams}
\end{figure}

Within the broken phase computation,
the Feynman rules for vertices read
\begin{align}
  V_{\sigma^3} &= -3! \Big( \frac{\mu_3}{3} + s \lambda_\sigma \Big)
  \;, \\
  V_{\sigma^4} &= -4! \Big( \frac{\lambda_\sigma}{4}\Big)
  \;.
\end{align}
Denoting the four-momentum by
$P=(P_0,\vec{p})$, where
$P_0 = 2\pi n T$ for each bosonic Matsubara mode,
the free scalar propagator is
\begin{align}
\langle \sigma(P) \sigma(Q) \rangle =
\frac{\deltabar(P+Q)}{P^2+m^2}
\;,
\end{align}
and employs the notation
$\,\deltabar(K)\equiv T^{-1}\delta_{K_{0},0}(2\pi)^{d}\delta^{(d)}(\vec{k})$
where
$\delta_{P_0}\equiv\delta_{P_0,0}$
denotes the Kronecker delta
for vanishing zero mode.
The broken phase mass parameter
$m^{2} =
  \mu_{\sigma}^{2}
+ 2 s \mu_{3}^{ }
+ 3 \lambda_{\sigma}^{ } s^2$
appearing in the propagator
corresponds to the squared mass eigenvalue of field $s$.
The resummation of the zero mode $\sigma_0$ writes%
\footnote{
  This (order-by-order) resummation identifies IR contributions by relating
  soft 4d loop contributions with 3d ones.
  The soft/hard mixing terms cancel explicitly at two-loop order.
  For gauge fields this procedure becomes technically complicated
  (cf.\ ref.~\cite{Gorda:2018hvi}) and
  ref.~\cite{Kajantie:1995dw} indicates that such an explicit resummation is
  somewhat cosmetical.
  While IR contributions in the matching must be identified,
  their specific expressions are obsolete.
  Therefore, soft/hard mixing terms never appear
  in the matching of strict perturbation theory~\cite{Gould:2021dzl},
  following refs.~\cite{Braaten:1995jr,Braaten:1995cm}.
}
\begin{align}
\label{eq:resum}
  \mathcal{L} &= \Big(
    \mathcal{L}_{\rmii{free}}
  + \frac{1}{2}\Pi^{\rmii{1loop}}_{2} \sigma^2_0
  + \frac{1}{3}\Pi^{\rmii{1loop}}_{3} \sigma^3_0
  + \frac{1}{4}\Pi^{\rmii{1loop}}_{4} \sigma^4_0
  \Big)
  \nn & + \Big(
    \mathcal{L}_{\rmii{int}}
  - \frac{1}{2}\Pi^{\rmii{1loop}}_{2} \sigma^2_0
  - \frac{1}{3}\Pi^{\rmii{1loop}}_{3} \sigma^3_0
  - \frac{1}{4}\Pi^{\rmii{1loop}}_{4} \sigma^4_0
  \Big)
  \;,
\end{align}
where $\Pi_{n}$ contain hard mode corrections.
Terms with
plus signs resum the zero mode mass and
terms with
minus sign act as interactions.
In particular, we have a quadratic resummation interaction
\begin{align}
\label{eq:resummation-interaction}
  V_{\sigma^2_0} =
    \Pi_s \equiv \Pi_{2}
    + 2 s^{ } \Pi_{3}
    + 3 s^{2} \Pi_{4}
    \;,
\end{align}
for the zero modes.
We also have a UV counterterm interaction for all modes
\begin{align}
  V_{\sigma^2} = -(\delta m^2 + P^2 \delta Z_{\sigma})
  \;,
\end{align}
where
$\delta m^2 =
  \delta \mu_{\sigma}^{2}
  + 2 s \delta \mu_3
  + 3 \delta \lambda_\sigma s^2
$
and
$\delta Z_{\sigma} = 0$ at one-loop level.
After resummation the propagator reads
\begin{align}
\langle \sigma(P) \sigma(Q) \rangle =
\frac{\deltabar(P+Q)}{P^2+m^2 + \delta_{P_0} \Pi_s}
  \;.
\end{align}
Perturbation theory is organised order-by-order.
Thus, thermal corrections $\Pi_{n}$ at one-loop are needed
explicitly for resummation at two-loop level.
Therefore, we already quote the result
\begin{align}
\label{eq:Pi:2}
\Pi_{2} &= T^2 \frac{\lambda_\sigma}{4} - 2 \mu_{3}^{2} \frac{L_b}{(4\pi)^2}
\;, \\
\Pi_{3} &= - 9 \mu_3 \lambda_\sigma \frac{L_b}{(4\pi)^2}
\;, \\
\Pi_{4} &= - 9 \lambda_{\sigma}^{2} \frac{L_b}{(4\pi)^2}
\;.
\end{align}
The effective potential including two-loop level reads
\begin{align}
  V^{\rmii{4d}}_{\rmii{eff}} =
    V_{\rmii{tree}}^{ }
  + V_{\rmii{CT}}^{ }
  + V_{\rmii{1loop}}^{ }
  + V_{\rmii{2loop}}^{ }
  \;,
\end{align}
and even though counterterm and resummation diagrams are
one-loop topologies they contribute at equal order as
two-loop topologies.
Figure~\ref{fig:2loop-diagrams} illustrates the corresponding two-loop level diagrams.
The separate terms in the potential yield
\begin{align}
V_{\rmii{tree}} &=
    \frac{1}{2} \mu_{\sigma}^{ } s^2
  + \mu_1 s
  + \frac{1}{3} \mu_{3}^{ } s^3
  + \frac{1}{4} \lambda_{\sigma}^{ } s^4
  \;, \\
V_{\rmii{CT}} &=
\frac{1}{2} \delta \mu_{\sigma}^{2} s^2
  + \delta\mu_1 s
  + \frac{1}{3} \delta\mu_{3}^{ } s^3
  + \frac{1}{4} \delta\lambda_{\sigma}^{ } s^4
  \;, \\
V_{\rmii{1loop}} &=
J_{\rmii{soft}}(m_{\rmii{3d}})
  + J_{\rmii{hard}}(m)
  \;, \\
\label{eq:veff-resum}
V_{\rmii{2loop}} &=
    -\Big( \frac{1}{8} V^{ }_{\sigma^4} (-1) \mathcal{D}_{SS}(m,m)
  + \frac{1}{12} V^{2}_{\sigma^3} \mathcal{D}_{SSS}(m,m,m)
\nn &\hp{=-\Big(}
  + \frac{1}{2}(-1) \mathcal{D}_{S}(m)
    + \frac{1}{2} \Pi_s I^3_1(m_{\rmii{3d}})
  \Big)
  \;,
\end{align}
where $m^2_{\rmii{3d}}$ corresponds to the mass eigenvalue in the 3d theory.
Therein, all {\em master} integrals are defined in appendix~\ref{se:integrals}
in the high-$T$ expansion and
in dimensional regularisation utilising the \MSbar-scheme.
On the UV side, all $T^2$-independent
$1/\epsilon$ and
$1/\epsilon^2$ poles cancel in dimensional regularisation.
On the IR-sensitive side,
non-analytic, mixed soft/hard terms
$\propto\sqrt{m^2_{\rmii{3d}}}$ cancel due
to resummation.

Expanding the effective potential in
$\epsilon$ and
the background field $s$ (cf.\ eq.~\eqref{eq:veff-s-expansion})
results in
\begin{align}
\label{eq:res-1}
V_1 &= \frac{1}{(4\pi)^2} \frac{T^2}{\epsilon} \frac{1}{2} \lambda_{\sigma}^{ }\mu_{3}^{ }
  + \mu_{1}^{ }(\Lambda)
  + \bigg[
    \frac{1}{12} T^2 \mu_{3}^{ }(\Lambda)
    - \frac{L_b}{(4\pi)^2} \mu_{3}^{ }\mu_{\sigma}^{ }(\Lambda) \bigg]_{\rmii{1loop}}
  \nn &
  + \bigg[
      \frac{L_b}{(4\pi)^2} \frac{3}{4} T^2 \lambda_\sigma^{ } \mu_3^{ }
    + \frac{1}{(4\pi)^4} \mu^3_3 \Big(3+2 L_{b}^{ } + L_{b}^{2} \Big)
    - \frac{1}{(4\pi)^2} 2 \lambda_{\sigma} \mu_{3} \Big( c + \ln \Big( \frac{3T}{\Lambda} \Big)\Big)
  \bigg]_{\rmii{2loop}}
  \;, \\
V_2 &= \frac{1}{(4\pi)^2} \frac{T^2}{\epsilon} \frac{3}{2} \lambda_{\sigma}^{2} + \mu_{\sigma}^{2}(\Lambda)
  \nn &
  + \bigg[
      \frac{1}{4} T^2 \lambda_\sigma(\Lambda)
      - \frac{L_b}{(4\pi)^2} \Big( 2 \mu_{3}^{2}(\Lambda) + 3 \lambda_{\sigma}^{ } \mu_{\sigma}^{ }(\Lambda) \Big)
    + \frac{\zeta_{3}}{(4\pi)^4 T^2} 8 \mu_3^2 \mu_{\sigma}^{2}(\Lambda)
  \bigg]_{\rmii{1loop}}
  \nn &
  + \bigg[
    - T^2 \frac{L_b}{(4\pi)^2} \frac{9}{4} \lambda_{\sigma}^{2}
    + \frac{1}{(4\pi)^4} \lambda_{\sigma}^{ } \mu_{3}^{2} \Big( 45 + 3 L_b (10+L_b) + 2 \zeta_{3} \Big)
  \nn &\hp{{}+\Big[}
    - \frac{\zeta_{3}}{(4\pi)^6 T^2} 8 (3+2 L_b) \mu_{3}^{4}
    - \frac{1}{(4\pi)^2} 6 \lambda^2_{\sigma} \Big( c + \ln \Big( \frac{3T}{\Lambda} \Big)\Big)
  \bigg]_{\rmii{2loop}}
  \;, \\
V_3 &= \mu_3
  - \frac{L_b}{(4\pi)^2} 9 \lambda_{\sigma}^{ }\mu_{3}^{ }
  + \frac{\zeta_{3}}{(4\pi)^4 T^2} 8 \mu_{3}^{3}
  \;, \\
\label{eq:res-4}
V_4 &= \lambda_{\sigma}
  - \frac{L_b}{(4\pi)^2} 9 \lambda_{\sigma}^{2}
  + \frac{\zeta_{3}}{(4\pi)^4 T^2} 48 \lambda_{\sigma}^{ }\mu_{3}^{2}
  - \frac{\zeta_{5}}{(4\pi)^6 T^4} 32 \mu_{3}^{4}
\;,
\end{align}
employing abbreviations for thermal logarithms
\begin{align}
c &=
  \frac{1}{2} \bigg(
    \ln \Big( \frac{8\pi}{9} \Big)
    + \frac{\zeta_{2}'}{\zeta_{2}}
  - 2 \gammaE \bigg)
\;, \\
\label{eq:Lbf}
L_{b} &\equiv
    2 \ln\Big( \frac{\Lambda}{T} \Big)
  - 2 \Big( \ln(4\pi) - \gammaE \Big)
    \;, \quad
L_{f} \equiv L_{b} + 4\ln2
    \;,
\end{align}
in which
$\Lambda$ is the 4d renormalisation scale and
$\gammaE$ the Euler-Mascheroni constant.
The uncancelled $T^2$-dependent divergences
correspond to the two-loop 3d counterterms
in eqs.~%
\eqref{eq:dmu13} and
\eqref{eq:dmus3}.

One hallmark of the broken phase computation is its diagrammatic simplicity:
the combinatorics of permuting external legs is intrinsic in the effective potential.
As a drawback one has to evaluate
{\em massive} sum-integrals at two-loop level
to generate the dependence on the background field.
Even though reaching $\mathcal{O}(g^4)$
the mass parameter $\mu_{\sigma}^{2}$ itself will not appear within two-loop pieces of
the matching relations.
This detail facilitates the unbroken phase computation in the next section.
As another drawback, in models with multiple scalars,
multiple background fields appear and it can be tedious to obtain
an analytic series expansion for the effective potential in these background fields.
This poses a complication, since an expansion in background fields,
analogous to eq.~\eqref{eq:veff-s-expansion}, is needed to extract the correlators.

%
\subsubsection{Unbroken phase: Correlators from the diagrammatic approach}
\label{se:cor:singlet}

An alternative approach computes the correlation functions directly diagram-by-diagram
(cf.\ footnote~\ref{footnote:Gould}).
The downside of this approach is its large number of diagrams
with several permutations of external legs.
Conversely, its extension to more complicated models is conceptually straightforward and
even multiple
gauge fields and scalars coupling to them can be handled algorithmically.
This poses an advantage compared to
the aforementioned complications in the broken phase where
series expansions in (multiple) background fields were needed.
In turn, at two-loop level one can set all propagators massless
for the NLO dimensional reduction at $\rmO(g^4)$
in analogy to strict perturbation theory (cf.\ ref.~\cite{Gould:2021dzl}).

Within the unbroken phase computation,
the Feynman rules for vertices read
\begin{align}
V_{\sigma^3} &= -3! \Big( \frac{\mu_3 + \delta \mu_3}{3} \Big)
  \;, \\
V_{\sigma^4} &= -4! \Big( \frac{\lambda_\sigma + \delta \lambda_\sigma}{4}\Big)
  \;, \\
V_{\sigma^2} &= -(\delta \mu_{\sigma}^{2} + P^2 \delta Z_\sigma)
  \;.
\end{align}
Note that the tadpole $\mu_1$ never contributes to 1PI diagrams required for the matching.
The scalar propagator reads
\begin{align}
\langle \sigma(P) \sigma(Q) \rangle =
  \frac{\deltabar(P+Q)}{P^2+\mu_{\sigma}^{2}}
  \;.
\end{align}
As mentioned, aiming for $\mathcal{O}(g^4)$ accuracy allows to treat propagators
inside two-loop diagrams as massless.
This provides the correct hard mode parts and non-analytic IR sensitive
contributions vanish trivially in dimensional regularisation due to
an absent mass scale.

The 3-point and 4-point correlator consist of
the following diagrams including their results
in terms of master integrals (cf.\ appendix~\ref{se:integrals})
\setcounter{dummy}{\value{equation}}
\begin{multicols}{2}
\noindent
\setcounter{equation}{0}
\renewcommand{\theequation}{a.\arabic{equation}}
\begin{align}
\Vtxr(\Lsr1,\Lsr1,\Lsr1)
&= -2\delta \mu_3
\;,\\[1mm]
\TopoTBl(\Lsr1,\Asr1,\Asr1)
&= \frac{1}{2} \times 3 \times 12 \lambda_{\sigma}^{ }\mu_{3}^{ } I^{4b}_2
\;,\\[1mm]
\TopoTC(fex(\Lsr1,\Lsr1,\Lsr1),\Asr1,\Asr1,\Asr1)
&= 1 \times 3 \times (-8) \mu_3^{2} I^{4b}_3
\;,
\end{align}
\columnbreak
\setcounter{equation}{0}
\renewcommand{\theequation}{b.\arabic{equation}}
\begin{align}
\Vtxv(\Lsr1,\Lsr1,\Lsr1,\Lsr1)
&= -6\delta \lambda_\sigma
\;,\\[1mm]
\TopoVBlr(fex(\Lsr1,\Lsr1,\Lsr1,\Lsr1),\Asr1,\Asr1)
&= \frac{1}{2} \times 3 \times 36 \lambda_{\sigma}^{2} I^{4b}_2
\;,\\[1mm]
\TopoVC(fex(\Lsr1,\Lsr1,\Lsr1,\Lsr1),\Asr1,\Asr1,\Asr1)
&= 1 \times 6 \times (-24) \mu_{3}^{2}\lambda_{\sigma}^{ } I^{4b}_3
\;,\\[1mm]
\TopoVD(fex(\Lsr1,\Lsr1,\Lsr1,\Lsr1),\Asr1,\Asr1,\Asr1,\Asr1)
&= 1 \times 3 \times 16 \mu_{3}^{4} I^{4b}_4
\;,
\end{align}
\end{multicols}
\noindent
where we indicated
symmetry factors and
combinatorial factors
related to permutations of external legs.
The tadpole (1-point) correlator up to two-loop level yields
\setcounter{equation}{0}
\renewcommand{\theequation}{c.\arabic{equation}}
\begin{multicols}{2}
\noindent
\begin{align}
\label{eq:sig:1pt:2l:a}
\Vtxo(\Lsr1)
&= -\delta \mu_1
\;,\\[2mm]
\TopoOT(\Lsr1,\Asr1)
&= -\mu_3 \Big( I^{4b}_1 - \mu_{\sigma}^{2} I^{4b}_2 \Big)
\;,\\[2mm]
\ToptOE(\Lsr1,\Asr1,\Asr1,\Asr1)
&= 3 \mu_{3}^{ } \lambda_\sigma^{ } I^{4b}_1 I^{4b}_2
\;,\\[2mm]
\ToptOS(\Lsr1,\Asr1,\Asr1,\Lsr1)
&= 2 \mu_3 \lambda_\sigma S_3
\;,\\[2mm]
\ToptOM(\Lsr1,\Asr1,\Asr1,\Asr1,\Lsr1)
&= -2 \mu^3_3 S_4
\;,\\[2mm]
\TopoOTa(\Lsr1,\Asr1)
&= - \delta \mu_3 I^{4b}_1
\;,\\[2mm]
\label{eq:sig:1pt:2l:g}
\TopoOTx(\Lsr1,\Asr1)
&= \mu_3 \Big( \delta Z_{\sigma} I^{4b}_1 + \delta \mu_{\sigma}^{2} I^{4b}_2 \Big)
\;.
\end{align}
\end{multicols}%
\noindent
Finally, the diagrammatic expressions of the self-energy (2-point correlator)
up to two-loop level read
\setcounter{equation}{0}
\renewcommand{\theequation}{d.\arabic{equation}}
\begin{multicols}{2}
\noindent
\begin{align}
\label{eq:sig:2pt:2l:a}
\Vtxt(\Lsr1,\Lsr1)
&= -\delta \mu_{\sigma}^{2}
\;,\\[1mm]
\TopoST(\Lsr1,\Asr1)
&= -3 \lambda_{\sigma}^{ } \Big( I^{4b}_1 - \mu_{\sigma}^{2} I^{4b}_2 \Big)
\;,\\[3mm]
\TopoSB(\Lsr1,\Asr1,\Asr1)
&= 2 \mu_{3}^{2} \Big( I^{4b}_2 - 2 \mu_{\sigma}^{2} I^{4b}_3 \Big)
\;,\\[4mm]
\label{eq:sig:2pt:2l:d}
\ToptSTT(\Lsr1,\Asr1,\Asr1,\Asr1)
&= 9 \lambda_\sigma^{2} I^{4b}_1 I^{4b}_2
\;,\\[3mm]
\label{eq:sig:2pt:2l:e}
\ToptSS(\Lsr1,\Asr1,\Asr1,\Lsr1)
&= 6 \lambda_\sigma^{2} S_3
\;,\\[3mm]
\ToptSE(\Lsr1,\Asr1,\Asr1,\Asr1,\Asr1)
&= -6 \mu_{3}^{2} \lambda_\sigma I^{4b}_2 I^{4b}_2
\;,\\[3mm]
\ToptSBT(\Lsr1,\Asr1,\Asr1,\Asr1,\Asr1)
&= -12 \mu_{3}^{2} \lambda_\sigma I^{4b}_1 I^{4b}_3
\;,\\[3mm]
\ToptSal(\Lsr1,\Asr1,\Asr1,\Asr1,\Asr1)
&= -24 \mu_{3}^{2} \lambda_\sigma S_4
\;,\\[3mm]
\ToptSTB(\Lsr1,\Asr1,\Asr1,\Asr1,\Asr1)
&= -6 \mu_{3}^{2} \lambda_\sigma S_4
\;,\\[3mm]
\ToptSBB(\Lsr1,\Asr1,\Asr1,\Asr1,\Asr1,\Asr1)
&= 8 \mu_{3}^{4} S_6
\;,\\[3mm]
\ToptSM(\Lsr1,\Asr1,\Asr1,\Asr1,\Asr1,\Lsr1)
&= 8 \mu_{3}^{4} S_5
\;,\\[3mm]
\TopoSTc(\Lsr1,\Asr1)
&= - 3 \delta \lambda_\sigma I^{4b}_1
\;,\\[3mm]
\TopoSBa(\Lsr1,\Asr1,\Asr1)
&= 4 \mu_3 \delta \mu_3 I^{4b}_1
\;,\\[3mm]
\TopoSTx(\Lsr1,\Asr1)
&= 3 \lambda_\sigma \Big( \delta Z_{\sigma} I^{4b}_1 + \delta \mu_{\sigma}^{2} I^{4b}_2 \Big)
\;,\\[3mm]
\label{eq:sig:2pt:2l:o}
\TopoSBx(\Lsr1,\Asr1,\Asr1)
&= -4 \mu_{3}^{2} \Big( \delta Z_{\sigma} I^{4b}_2 + \delta \mu_{\sigma}^{2} I^{4b}_3 \Big)
\;.
\end{align}
\end{multicols}%
\renewcommand{\theequation}{\arabic{section}.\arabic{equation}}
\setcounter{equation}{\value{dummy}}
\noindent
After summing individual diagrams for each correlator,
we apply integrals of appendix~\ref{se:integrals} and
recover the correlators of
eqs.~%
\eqref{eq:res-1}--%
\eqref{eq:res-4}.
Recall that the correlator itself is minus the sum of diagrams and
within our convention
$V_n = \langle \sigma^n \rangle/(n-1)!$ given in eq.~\eqref{eq:veff-s-expansion}.

%
\subsubsection{Cancellation of mixed hard/soft terms}
\label{se:hard-soft:cancel}

As mentioned earlier, explicit resummation is obsolete since all propagators at
two-loop diagrams are treated massless.
To this end, we demonstrate how resummation unfolds
as a cancellation of IR sensitive mixed hard/soft terms while keeping sum-integrals massive.
For simplicity, we discuss the $Z_2$-symmetric case and
employ the resummation
\begin{align}
\mathcal{L} &= \Big(
  \mathcal{L}_{\rmii{free}}
  + \frac{1}{2} \Pi^{\rmii{1loop}}_{2} \sigma^2_0
  + \frac{1}{4} \Pi^{\rmii{1loop}}_{4} \sigma^4_0
  \Big)
  \nn &
  + \Big(
  \mathcal{L}_{\rmii{int}}
  - \frac{1}{2} \Pi^{\rmii{1loop}}_{2} \sigma^2_0
  - \frac{1}{4} \Pi^{\rmii{1loop}}_{4} \sigma^4_0
  \Big)
  \;.
\end{align}
From the quadratic part we can read off the resummed propagator
\begin{align}
\label{eq:prop:res:dr}
\langle \sigma(P) \sigma(Q) \rangle =
\frac{\deltabar(P+Q)}{P^2+\mu_{\sigma}^{2} + \delta{ }_{P_0} \Pi^{\rmii{1loop}}_{2}}
  \;,
\end{align}
and the resummed vertex for the pure zero modes becomes
$
\lambda_\sigma^{ } \to
\lambda_\sigma^{ } + \Pi_{4}^{ }
$.
Also resummation interaction terms are introduced for the zero mode
\begin{align}
  V_{\sigma^2_0} &= \Pi_{\sigma}
  \;, \\
  V_{\sigma^4_0} &= 6 \Pi_{\sigma}
  \;.
\end{align}

\begin{figure}[t]
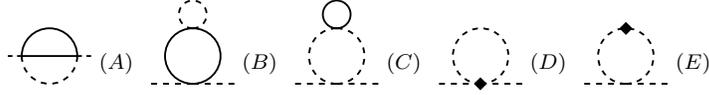

  \centering
  \begin{align*}
    \ToptSS(\Lxx,\Asa1,\Axx,\Lsa1)_{(A)}\;
    \ToptSTT(\Lxx,\Asa1,\Asa1,\Axx)_{(B)}\;
    \ToptSTT(\Lxx,\Axx,\Axx,\Asa1)_{(C)}\;
    \TopoSTco(\Lxx,\Axx,\diamond)_{(D)}\;
    \TopoSTxo(\Lxx,\Axx,\diamond)_{(E)}\;
  \end{align*}
  \caption{
    Resummation in the unbroken phase computation with $Z_2$ symmetry.
    Solid (dashed) lines denote scalar propagators with hard (soft) momenta.
    These IR-sensitive non-analytic mixed soft/hard contributions are
    compensated by resummation interaction diagrams (diamond).
    Alternatively, massless propagators in two-loop diagrams simplify the computation
    and retain
    the same result for the hard mode contribution relevant to matching.
  }
  \label{fig:resum}
\end{figure}

At two-loop order, the two diagrams that contribute are
\eqref{eq:sig:2pt:2l:d} and
\eqref{eq:sig:2pt:2l:e}.
By expanding the latter, massive sunset integral in~\eqref{eq:sig:2pt:2l:d},
at high-$T$
\begin{align}
\label{eq:resum:e}
\Tint{P,Q} \frac{1}{[P^2+m^2][Q^2+m^2][(P+Q)^2+m^2]} &=
  T^2 \int_{p,q} \frac{1}{[p^2+m^2][q^2+m^2][(p+q)^2+m^2]}
  \nn &
  + 3 T \int_p \frac{1}{[p^2+\mu^2_{\sigma,3}]} \Tint{Q}'\frac{1}{Q^4}
  + \mathcal{O}(\mu_{\sigma}^{2})
  \nn &\to
  6 \lambda_{\sigma}^{2} \Big( \underbrace{3 T I^3_1(\mu_{\sigma,3}^{ }) I^{4b}_2}_{(A)} \Big)
  \;,
\end{align}
one observes that masses
are expanded in the pure hard terms but
are kept in the mixed soft/hard terms.
In fact, the resummed 3d mass equals the one-loop dimensionally reduced mass parameter.
The last line above reintroduced numerical factors and the scalar self-coupling.
Similarly, the bubble integral~\eqref{eq:sig:2pt:2l:d} yields
\begin{align}
  \Tint{P} \frac{1}{[P^2+\mu_{\sigma}^{2}]}
  \Tint{Q} \frac{1}{[Q^2+\mu_{\sigma}^{2}]^2}
  \nn &\hspace{-2cm}=
  \bigg( T\int_{p}\frac{1}{[p^2+\mu^2_{\sigma,3}]^{ }}+\Tint{P}' \frac{1}{P^2} \bigg)
  \bigg( T\int_{q}\frac{1}{[q^2+\mu^2_{\sigma,3}]^{2}}+\Tint{Q}' \frac{1}{Q^4} \bigg)
  + \mathcal{O}(\mu_{\sigma}^{2})
  \\ &\hspace{-2cm}
  \simeq 9 \lambda_{\sigma}^{2} \bigg(
      T \int_{p}\frac{1}{[p^2+\mu^2_{\sigma,3}]^{ }} \Tint{Q}' \frac{1}{Q^4}
    + T \int_{q}\frac{1}{[q^2+\mu^2_{\sigma,3}]^{2}} \Tint{P}' \frac{1}{P^2} \bigg)
  \\ &\hspace{-2cm}
  \to 9 \lambda_\sigma \Big(
    \underbrace{\lambda_{\sigma} T I^3_1(\mu_{\sigma,3}^{ }) I^{4b}_2}_{(B)}
  + \underbrace{\lambda_{\sigma,3} I^3_2(\mu_{\sigma,3}^{ }) I^{4b}_1}_{(C)} \Big)
  \;,
\end{align}
where the pure 3d vertex is resummed and corresponds to the 3d effective one.
The mixed mode contributions
$(A)$, $(B)$ and $(C)$ are illustrated in fig.~\ref{fig:resum},
together with counterterm interaction diagrams that read
\begin{align}
\label{eq:resum:d}
\underbrace{\frac{1}{2}\times 6 \Pi_{\sigma,4} T I^3_1(\mu_{\sigma,3})}_{(D)} -
\underbrace{\frac{1}{2}\times 6 \lambda_{\sigma,3} \Pi_{\sigma,2} T I^3_2(\mu_{\sigma,3})}_{(E)}
\;.
\end{align}
Consequently, all non-analytic mixed soft/hard contributions vanish in resummation
since
$(A)+\dots+(E) = 0$.
In particular,
the IR-sensitive contributions
$(C)$ and $(E)$ are $\mathcal{O}(g^3)$ (instead of $\mathcal{O}(g^4)$) and
IR-divergent in the limit of vanishing $\mu_{\sigma,3}^{2}\to 0$.
Since dimensional reduction is IR-safe these problematic contributions were
expected to vanish.

In practice, the unbroken computation can be conducted by dealing
a zero-mass to propagators in two-loop diagrams
even though resummation is conceptually indispensable.
As a result, IR-divergent contributions vanish in dimensional regularisation
which, in turn, obscures the need for explicit resummation.

%
\subsection{Matching relations for 3d parameters}
\label{se:dr:singlet}

The ensuing matching relations define
the dimensionally reduced 3d EFT based on
the parameters of the fundamental 4d theory and temperature.
These follow the explicit computations of the previous sections:
\begin{align}
\label{eq:mu:1:3}
\mu_{1,3}^{ }(\Lamd) &=
  T^{-\frac{1}{2}}
  \bigg(
    \mu_{1}^{ }(\Lambda)
    + \bigg[
        \frac{1}{12} T^2 \mu_{3}^{ }(\Lambda)
      - \frac{L_b}{(4\pi)^2} \mu_{3}^{ }\mu_{\sigma}^{2}(\Lambda)
    \bigg]_{\rmii{1loop}}
  \nn &\hp{=\frac{1}{\sqrt{T}}\bigg(}
  - \bigg[
    \Big( \frac{\zeta_{3}}{(4\pi)^4} \frac{2}{3} \frac{\mu_{3}^{2}}{T^2} \Big)
    \Big( \mu_{1} + \frac{1}{12} T^2 \mu_3 \Big)
  \bigg]_{\rmii{f.n.}}
  \nn &\hp{=\frac{1}{\sqrt{T}}\bigg(}
  + \bigg[
    \frac{L_b}{(4\pi)^2} \frac{3}{4} T^2 \lambda_{\sigma}^{ }\mu_{3}^{ }
    + \frac{1}{(4\pi)^4} \mu_{3}^{3} \Big(3+2 L_{b}^{ } + L_{b}^2 \Big)
  \bigg]_{\rmii{2loop}}
  \bigg)
  \nn &
  - \bigg[
      \frac{1}{(4\pi)^2} 2 \lambda_{\sigma,3} \mu_{3,3} \Big( c + \ln \Big( \frac{3T}{\Lamd} \Big)\Big)
    \bigg]_{\rmii{3d-running}}
  \;, \\[2mm]
\label{eq:mu:sig:3}
\mu^2_{\sigma,3}(\Lamd) &= \mu_{\sigma}^{2}(\Lambda)
  \nn &
  + \bigg[
      \frac{1}{4} T^2 \lambda_\sigma(\Lambda)
    - \frac{L_b}{(4\pi)^2} \Big( 2 \mu_{3}^{2}(\Lambda) + 3 \lambda_{\sigma}^{ }\mu_{\sigma}^{2}(\Lambda) \Big)
    + \frac{\zeta_{3}}{(4\pi)^{4}T^{2}} 8 \mu_{3}^{2}\mu_{\sigma}^{2}(\Lambda)
  \bigg]_{\rmii{1loop}}
  \nn &
  - \bigg[
      \frac{\zeta_{3}}{(4\pi)^4 T^{2}} \frac{4}{3} \mu_{3}^{2} \Big(
          \mu_{\sigma}^{2}(\Lambda)
        + \frac{1}{4} T^2 \lambda_{\sigma}^{ }
        - \frac{L_b}{(4\pi)^2} 2 \mu_{3}^{2}
      \Big)
    \bigg]_{\rmii{f.n.}}
  \nn &
  + \bigg[
    -T^2 \frac{L_b}{(4\pi)^2} \frac{9}{4} \lambda_{\sigma}^{2}
    + \frac{1}{(4\pi)^4} \lambda_\sigma \mu_{3}^{2} \Big( 45 + 3 L_b (10+L_b) + 2 \zeta_{3} \Big)
  \nn &\hp{{}+\Big(}
    - \frac{\zeta_{3}}{(4\pi)^{6}T^{2}} 8 (3+2 L_b) \mu_{3}^{4}
  \bigg]_{\rmii{2loop}}
  - \bigg[ \frac{1}{(4\pi)^2} 6 \lambda^2_{\sigma,3} \Big( c + \ln \Big( \frac{3T}{\Lamd} \Big)\Big)
  \bigg]_{\rmii{3d-running}}
  \;, \\
\label{eq:mu:3:3}
    \mu_{3,3}^{ } &= T^{\frac{1}{2}}\Big(
    \mu_3^{ }(\Lambda)
  - \frac{ L_b}{(4\pi)^2} 9 \lambda_{\sigma}^{ } \mu_3^{ }
  + \frac{\zeta_{3}}{(4\pi)^4 T^{2}} 2 \mu_{3}^{3} (4 - \underbrace{1}_{\text{f.n.}})
  \Big)
  \;, \\
\label{eq:lambda:sig:3}
\lambda_{\sigma,3}^{ } &= T\Big(
    \lambda_{\sigma}^{ }(\Lambda)
  - \frac{L_b}{(4\pi)^2} 9 \lambda_{\sigma}^{2}
  + \frac{\zeta_{3}}{(4\pi)^{4}T^{2}} \lambda_{\sigma}^{ }\mu_{3}^{2}
    \Big(48 - \underbrace{\frac{8}{3}}_{\text{f.n.}}\Big)
  - \frac{\zeta_{5}}{(4\pi)^{6}T^{4}} 32 \mu_{3}^{4}
  \Big)
  \;,
\end{align}
indicating contributions originating from
field normalisation (f.n.),
one-loop, and
two-loop
level.
Note that the high-$T$ expansion gives rise to NLO terms at one-loop which
are $\mu_{\sigma}^{2}$-proportional.
Importantly, we explicitly denoted the 4d scale dependence ($\Lambda$) in terms of which
the running of LO terms produce contributions at NLO ($\mathcal{O}(g^4)$).
In addition, we indicated that the 3d
tadpole $\mu_{1,3}$ and
mass parameter $\mu_{\sigma,3}^{2}$ run
with the 3d renormalisation scale $\Lamd$.
An {\em exact} dependence on $\Lamd$ is presented even though
it includes higher contributions than $\mathcal{O}(g^4)$.
This exact dependence can be solved due to the super-renormalisability of the 3d EFT;
see sec.~\ref{se:tutorial:veff}.

By applying $\beta$-functions of appendix~\ref{se:4d:ct:beta} and ref.~\cite{Brauner:2016fla},
we immediately observe that
{\em all}
3d parameters are independent of the 4d renormalisation scale
$\Lambda$ at $\mathcal{O}(g^4)$:
\begin{align}
\label{eq:scale:cancel:singlet}
\Lambda \frac{{\rm d}}{{\rm d}\Lambda} \mu_{1,3} = 0\;, \quad
\Lambda \frac{{\rm d}}{{\rm d}\Lambda} \mu^2_{\sigma,3} = 0\;, \quad
\Lambda \frac{{\rm d}}{{\rm d}\Lambda} \mu_{3,3} = 0\;, \quad
\Lambda \frac{{\rm d}}{{\rm d}\Lambda} \lambda_{\sigma,3} = 0\;.
\end{align}
For example, the temperature-dependent scale dependence of
the 3d mass parameter $\mu_{\sigma,3}^{2}$ in eq.~\eqref{eq:mu:sig:3}
arises via
its one-loop running contribution
$\frac{1}{4}T^{2}\lambda_\sigma(\Lambda)$ and cancels upon
its two-loop logarithmic term $\propto T^{2}L_b$.
As a general feature for other scale-dependent terms,
this renormalisation scale dependence is discussed in
ref.~\cite{Gould:2021xxx}.
It is worth to point out that -- as depicted in
eqs.~\eqref{eq:mu:1:3}--\eqref{eq:lambda:sig:3} --
the tadpole and mass parameter are running in terms of
the 3d renormalisation scale $\Lamd$ whereas couplings do not.
This dependence of 3d RG scale cancels in computations within the EFT,
as we illustrate in the next section.

%
\subsection{Two-loop effective potential in 3d EFT}
\label{se:tutorial:veff}

To conclude this section, we illustrate the computation of
the two-loop thermal effective potential, within
dimensionally reduced 3d perturbation theory.
This corresponds to step $(e)$ of sec.~\ref{se:thermo:approaches}.
At two-loop level, the 3d effective potential composes of
\begin{align}
    V^{\rmii{3d}}_{\rmii{eff}}
  = V^{\rmii{3d}}_{\rmii{tree}}
  + V^{\rmii{3d}}_{\rmii{CT}}
  + V^{\rmii{3d}}_{\rmii{1loop}}
  + V^{\rmii{3d}}_{\rmii{2loop}}
  \;.
\end{align}
The 3d mass is given by the mass eigenvalue:
$m^2_{\rmii{3d}} =
    \mu_{\sigma,3}^{2}
  + 2 \mu_{3,3}^{ } s_{3}^{ }
  + 3 \lambda_{\sigma,3}^{ } s_{3}^{2}
$,
where $s_3$ is a background field of the 3d theory.
The corresponding vertices read
\begin{align}
V_{\sigma^3,\rmii{3d}} &= -3! \Big( \frac{\mu_{3,3}}{3} + s_3 \lambda_{\sigma,3} \Big)
\;, \\
V_{\sigma^4,\rmii{3d}} &= -4! \Big( \frac{\lambda_{\sigma,3}}{4}\Big)
\;.
\end{align}
The individual pieces of the potential read
\begin{align}
V^{\rmii{3d}}_{\rmii{tree}} &=
\frac{1}{2} \mu_{\sigma,3}^{2}(\Lamd) s_{3}^{2}
+ \mu_{1,3}^{ } (\Lamd) s_{3}^{2}
  + \frac{1}{3} \mu_{3,3} s^3_3
  + \frac{1}{4} \lambda_{\sigma,3}^{ } s_{3}^{4}
  \;, \\
V^{\rmii{3d}}_{\rmii{CT}} &=
    \delta V^{\rmii{3d}}_0
  + \frac{1}{2} \delta\mu_{\sigma,3}^{2} s_{3}^{2}
  + \delta\mu_{1,3}^{ } s_{3}^{ }
  \;, \\
V^{\rmii{3d}}_{\rmii{1loop}} &= J_{\rmii{soft}}(m_{\rmii{3d}})
  \;, \\
V^{\rmii{3d}}_{\rmii{2loop}} &=
    -\bigg( \frac{1}{8} V_{\sigma^4,\rmii{3d}}(-1) \mathcal{D}^{\rmii{3d}}_{SS}(m_{\rmii{3d}},m_{\rmii{3d}})
  + \frac{1}{12} (V_{\sigma^3,\rmii{3d}})^2 \mathcal{D}^{\rmii{3d}}_{SSS}(m_{\rmii{3d}},m_{\rmii{3d}},m_{\rmii{3d}}) \bigg)
  \;,
\end{align}
with the respective master (loop) integrals collected in appendix~\ref{se:integrals}.
In the 3d EFT divergences stemming from the field dependence appear only
at two-loop level wherefore
one-loop diagrams with counterterms contribute at three-loop level.
The only parameters in need of renormalisation are
the mass, tadpole, and field-independent vacuum counterterm.
Their counterterms are
\begin{align}
\label{eq:dmu13}
\delta\mu_{1,3} &= \frac{1}{(4\pi)^2} \frac{1}{4\epsilon} 2\lambda_{\sigma,3}\mu_{3,3}
\;, \\
\label{eq:dmus3}
\delta\mu_{\sigma,3}^{2} &= \frac{1}{(4\pi)^2} \frac{1}{4\epsilon} 6\lambda_{\sigma,3}^{2}
\;, \\
\delta V^{\rmii{3d}}_0 &= \frac{1}{(4\pi)^2} \frac{1}{4\epsilon} \frac{1}{3} \mu_{3,3}^{2}
\;.
\end{align}
The upper two of these counterterms are exact due to super-renormalisability of the 3d EFT;
new divergences are absent for tadpole and scalar mass at higher loop orders.
The field-independent vacuum counterterm gets contributions up to four-loop order~\cite{Gould:2021dzl}.
Hence, we can solve an
exact renormalisation scale ($\Lamd$) dependence of the 3d parameters
by requiring that the bare parameters are scale-invariant:
\begin{align}
\Lamd \frac{{\rm d}}{{\rm d}\Lamd} \mu_{1,3(b)} &=
\Lamd \frac{{\rm d}}{{\rm d}\Lamd} \bigg( \Lambda^{4\epsilon}_{\rmii{3d}}
  \Big( \mu_{1,3}(\Lamd) + \delta \mu_{1,3}\Big) \bigg) = 0
\;, \\
\Lamd \frac{{\rm d}}{{\rm d}\Lamd} \mu^2_{\sigma,3(b)} &=
\Lamd \frac{{\rm d}}{{\rm d}\Lamd} \bigg( \Lambda^{4\epsilon}_{\rmii{3d}}
  \Big( \mu^2_{\sigma,3}(\Lamd) + \delta \mu^2_{\sigma,3}\Big) \bigg) = 0
\;.
\end{align}
Their solution yields
\begin{align}
\mu_{1,3}(\Lamd) &=
  - \frac{1}{(4\pi)^2} 2 \lambda_{\sigma,3} \mu_{3,3} \ln \Big( \frac{\Lambda_0}{\Lamd} \Big)
  + \mathcal{C}_1
  \;, \\
\mu^2_{\sigma,3}(\Lamd) &=
  - \frac{1}{(4\pi)^2} 6 \lambda^2_{\sigma,3} \ln \Big( \frac{\Lambda_0}{\Lamd} \Big)
  + \mathcal{C}_2
  \;,
\end{align}
where temperature-dependent initial conditions are parametrised by
$\Lambda_0 \equiv 3 T e^{c}$ and
coefficients $\mathcal{C}_{1,2}$ are fixed to reproduce
the $\mathcal{O}(g^4)$ hard mode contributions of the matching relations at
$\Lamd = \Lambda_{\rmii{4d}}$.
The above renormalised parameters
$\mu_{3,3}$ in eq.~\eqref{eq:mu:3:3} and
$\lambda_{\sigma,3}$ in eq.~\eqref{eq:lambda:sig:3} are RG-invariant.
In total, the effective potential at two-loop has a compact result
\begin{align}
V^{\rmii{3d}}_{\rmii{eff}} &=
\frac{1}{2} \mu^2_{\sigma,3}(\Lamd) s_{3}^{2}
  + \mu_{1,3}^{ } (\Lamd) s_{3}^{ }
  + \frac{1}{3} \mu_{3,3}^{ } s_{3}^{3}
  + \frac{1}{4} \lambda_{\sigma,3}^{ } s_{3}^{4}
  - \frac{(m^2_{\rmii{3d}})^{\frac{3}{2}}}{12\pi}
  \nn &
    + \frac{1}{(4\pi)^2} \bigg(
     \frac{3}{4} \lambda_{\sigma,3}^{ } m^2_{\rmii{3d}}
    - \frac{1}{6} (3 \lambda_{\sigma,3} s_{\rmii{3d}} + \mu_{3,3})^2
      \Big[1+ 2 \ln \Big( \frac{\Lamd}{3 m_{\rmii{3d}}} \Big) \Big]
  \bigg)
  \;.
\end{align}
We observe that
the tree-level running of the 3d parameters compensates
the scale dependence in the two-loop logarithmic terms.
This {\em RG-improved} effective potential (and even the three-loop effective potential)
is compared to non-perturbative lattice simulations in ref.~\cite{Gould:2021dzl}.
The latter agrees surprisingly well even with large expansion parameters
in perturbation theory.

This concludes our instructions to the dimensional reduction of
the real scalar field theory.

%
\section{Dimensional reduction of the real-singlet extended Standard Model}
\label{se:dr:xsm}

This section
details the dimensionally reduced 3d EFT for the SM coupled to a real scalar singlet at NLO.
This is the novel result of this article.
Applications of this 3d EFT to study of electroweak phase transition in the xSM
are presented in refs.~\cite{Niemi:2021xxx,Gould:2021xxx}.

When extending the Standard Model by a real scalar singlet,
the position of that scalar in the high-temperature hierarchy
is {\em a priori} undetermined.
If the scalar singlet assumes a hard (or ``superheavy'') scale it is integrated out entirely
during the dimensional reduction~\cite{Brauner:2016fla}.%
\footnote{
  However, ref.~\cite{Brauner:2016fla}
  lacks full $\mathcal{O}(g^4)$ accuracy
  since the Higgs 3d mass parameter at two-loop order is not computed.
}
The resulting version of the SM 3d EFT encodes effects of the singlet merely in its
matching relations.

The following analysis relaxes this assumption and performs the dimensional reduction
with a soft (or ``heavy'') singlet.
The singlet remains a dynamical field in the 3d EFT and
can eventually become ultrasoft (or ``light'').
Such a configuration allows for dynamical transitions with two consecutive steps
which are a viable candidate for EWPT with SFOPT in this model (cf.\ singlet refs.\ in sec.~\ref{se:intro}).
During such a dynamical two-step transition,
first the singlet acquires a non-zero vacuum expectation value (vev) at high temperatures
which is followed by a
SFOPT in Higgs-direction once temperature is lowered further.

The scalar sector of the 4d Lagrangian reads
\begin{align}
\mathcal{L}^{\rmii{4d}}_\rmii{scalar} &=
  (D_{\mu}^{ }\phi)^\dagger (D_{\mu}^{ }\phi)
  + \mu_{h}^2\phi^\dagger \phi
  + \lambda_{h}^{ }(\phi^\dagger \phi)^2
  \nn &
  + \frac{1}{2} (\partial_{\mu}^{ }\sigma)^2
  + \frac{1}{2} \mu_{\sigma}^2\sigma^2
  + \mu_{1}^{ }\sigma
  + \frac{1}{3} \mu_{3}^{ }\sigma^3
  + \frac{1}{4} \lambda_{\sigma}^{ }\sigma^4
  \nn &
  + \frac{1}{2} \mu_{m}^{ }\sigma\phi^\dagger \phi
  + \frac{1}{2} \lambda_{m}^{ }\sigma^2\phi^\dagger \phi
\;,
\end{align}
which notationally aligns with ref.~\cite{Brauner:2016fla} (see sec.~2 ibid.)
except the {\em opposite} sign convention for the 4d Higgs mass parameter $\mu_{h}^{2}$.

We assume the following formal power counting,
or scaling in powers of the ${\rm SU}(2)$ gauge coupling $g$:
\begin{equation}
\gp,\gs,\gY\sim g^{ }\;,\quad
\lambda_{h},\lambda_{\sigma},\lambda_{m}\sim g^{2}\;,\quad
\mu_{h}^{2},\mu_{\sigma}^{2}\sim(g^{ }T)^2
\;.
\end{equation}
Since both mass parameters are soft, this leads to an EFT with
two dynamical light scalars $\phi_{3}$ and $\sigma_{3}$.
The scaling of dimensionful couplings is more delicate
(cf.\ sec.~2.1.2 in ref.~\cite{Brauner:2016fla}), where
for the tadpole and cubic couplings we assume
\begin{equation}
\label{eq:scaling:xsm:cubic}
\mu_{1}\sim g^{ }T^{3}\;,\quad
\mu_{m},\mu_{3}\sim g^{ }T
\;.
\end{equation}
In analogy to eq.~\eqref{eq:scale:a}
this formal choice leads to similar peculiarities.
Contributions
\begin{align}
\label{eq:scale:b}
\VtxvS(\Lsc1,\Lsc1,\Lcs1,\Lcs1,\Lsr1) &\simeq
  \frac{\mu_{m}^{2}}{\mu_{\sigma}^{2}} \sim
  \mathcal{O}(1) \gg
  \lambda_h
  \;, \\
\label{eq:scale:c}
\VtxvS(\Lsr1,\Lsc1,\Lcs1,\Lsr1,\Lsr1) &\simeq
  \frac{\mu_m \mu_3}{\mu_{\sigma}^{2}} \sim
  \mathcal{O}(1) \gg
  \lambda_m
  \;,
\end{align}
parametrically dominate over the corresponding quartic couplings.
We reiterate the strategy below eq.~\eqref{eq:scale:a}:
for generality we install
$\mu_{m},\mu_{3}\sim g^{ }T$ and
include all contributions of cubic couplings in our matching relations.
Note, that contributions proportional to
$\mu^2_3$ and $\mu^4_3$ are further numerically
suppressed by extra powers of $1/(4\pi)$.
In the chosen formal scaling,
the one-loop contributions of the tadpole correlator $\sim\mu_{m,3}\times T^{2}$
are formally of
the same order as the tree-level tadpole.
Therefore,
the $\beta$-functions of tadpole and mass parameters (cf.\ appendix~\ref{se:4d:ct:beta}),
are partly needed at two-loop level for a $\mathcal{O}(g^4)$ accuracy.

%
\subsection{Effective 3d theories}
\label{ss:xsm:dr:matching}

The corresponding effective 3d Lagrangian is
\begin{align}
\label{eq:lag:soft}
\mathcal{L}^{\rmii{3d}}_\rmii{scalar} &=
  (D_{r}^{ }\phi)^\dagger (D_{r}^{ }\phi)
  + \mu_{h,3}^2\phi^\dagger \phi
  + \lambda_{h,3}^{ }(\phi^\dagger \phi)^2
  \nn &
  + \frac{1}{2} (\partial_{r}^{ }\sigma)^2
  + \frac{1}{2} \mu_{\sigma,3}^{2}\sigma^2
  + \mu_{1,3}^{ }\sigma
  + \frac{1}{3} \mu_{3,3}^{ }\sigma^3
  + \frac{1}{4} \lambda_{\sigma,3}^{ }\sigma^4
  \nn &
  + \frac{1}{2} \mu_{m,3}^{ }\sigma\phi^\dagger \phi
  + \frac{1}{2} \lambda_{m,3}^{ }\sigma^2\phi^\dagger \phi
  \;,
\end{align}
with $r\in\{1,...,d\}$ and
the 3-subscript denoting parameters in the dimensionally reduced 3d EFT.%
\footnote{
  For simplicity, we keep the notation for 3-dimensional fields identical as in
  the fundamental 4d theory.
}
In addition, we include the following (non-kinetic) pure scalar marginal operators
\begin{align}
\label{eq:lag:marginal}
\mathcal{L}^{\rmii{3d}}_\rmii{marginal} &=
    c_{0,5}\sigma^5
  + c_{2,3}(\phi^\dagger \phi)^{ }\sigma^3
  + c_{4,1}(\phi^\dagger \phi)^{2}\sigma
  \nn &
  + c_{6,0}(\phi^\dagger \phi)^{3}
  + c_{0,6}\sigma^6
  + c_{4,2}(\phi^\dagger \phi)^{2}\sigma^2
  + c_{2,4}(\phi^\dagger \phi)^{ }\sigma^4
  \;.
\end{align}
The nomenclature of the effective field theory~\cite{Jenkins:2013zja,Jenkins:2013wua,Alonso:2013hga}
classifies these operators as $S^6$.
We omit classes with higher dimensional kinetic operators such as
$D^2 S^4$ and
$D^4 S^2$
where $D$ formally presents a derivative operator and
classes with gauge fields such as
$F^3$
where $F$ presents a field strength tensor.
This choice is purely practical:
The matching of class $S^6$ is straightforward since corresponding correlators
can be computed at zero external momenta both from
the effective potential in the broken phase and even
diagrammatically directly in unbroken phase.
On the other hand,
the derivative structure of kinetic operators requires a computation
with explicit external momenta dependence.
We defer this challenge to a future comprehensive analysis of
the numerical relevance of different higher dimensional operators.
However, in the presence of large portal couplings it is natural to expect
the class $S^6$ to numerically dominate over other classes that are always
suppressed by $g^2$.

The breaking of Lorentz symmetry by the heat bath induces temporal scalars.
These are remnants of the temporal gauge field components~\cite{Laine:2016hma}
and obtain Debye screening masses at the soft scale.
In analogy with ``electrostatic'' QCD~\cite{Braaten:1995jr},
the Lagrangian composes of
\begin{align}
\label{eq:L:temp:soft}
\mathcal{L}^{\rmii{3d}}_\rmii{temporal}&=
    \frac{1}{2}(D_{r}^{ }A^a_0)^2
  + \frac{1}{2}\mD^{2}\,A^a_0A^a_0
  + \frac{1}{2}(\partial_{r}^{ }B_{0}^{ })^2
  + \frac{1}{2}\mD'\,^{2}B_{0}^{2}
  + \frac{1}{2}(D_{r}^{ }C^\alpha_0)^2
  + \frac{1}{2}\mD''^{2}\,C^\alpha_0C^\alpha_0
  \nn &
  + \frac{1}{4}\kappa_{3}^{ }\,(A^a_0A^a_0)^2
  + \frac{1}{4}\kappa_3'\,B_0^4
  + \frac{1}{4}\kappa_3''\,A^a_0A^a_0B_{0}^{2}
  \nn &
  + h_3^{ }\,\he\phi\phi A^a_0A^a_0
  + h_3'\,\he\phi\phi B_0^2
  + h_3''\,B_{0}^{ }\he\phi A_{0}^{a}\tau^{a} \phi
  + \delta_{3}^{ }\,\he\phi\phi C^\alpha_0 C^\alpha_0
  \nn[2mm] &
  + x_{3}^{ }\,\sigma A^a_0A^a_0
  + x_{3}'\,\sigma B_0^2
  + y_{3}^{ }\,\sigma^2 A^a_0A^a_0
  + y_{3}'\,\sigma^{2}B_0^2
  \;,
\end{align}
wherein
$\tau^{a}$ denote the Pauli matrices and
the covariant derivatives act on the adjoint scalars as
$D_{r}^{ }A_{0}^{a} =
    \partial_{r}^{ } A^a_0
  + g_{3}^{ } \epsilon^a_{\phantom{a}bc}A^b_rA^c_0$ and
$D_{r}^{ }C_{0}^{\alpha} =
    \partial_{r}^{ } C^\alpha_0
  + g_{\rmi{s},3}^{ } f^\alpha_{\phantom{\alpha}\beta\rho}C^\beta_rC^\rho_0$.

Several interaction terms among adjoint scalars were omitted in the temporal Lagrangian
since they are of secondary interest in our computation~\cite{Brauner:2016fla}.
Among these omissions are operators with an odd number of temporal fields
such as
$\sigma\he\phi A_{0}^{a}\tau^{a}\phi$.
These only appear in the presence of a finite chemical potential due to
the breaking of parity~\cite{Kajantie:1997ky}.
Exceeding the accuracy of our analysis,
we exclude higher dimensional operators involving temporal scalars
since they are numerically suppressed compared to large scalar
portal couplings~\cite{Kajantie:1995dw}.
Besides, the effect of the temporal sector is numerically subdominant
in strong phase transitions driven by ultrasoft (light) scalar fields.
This suppression is often empirically observed in BSM theories since physically
the temporal scalars are screened at length scales
much shorter than those relevant to the phase transition dynamics.

At the ultrasoft scale, the dynamics of the soft (heavy) temporal scalars
$A_{0}^{a}$,
$B_{0}^{ }$ and
$C_{0}^\alpha$ has been integrated out by
the second step of dimensional reduction~\cite{Kajantie:1995dw}.
Remember that their masses are at the soft scale and are not dynamical
in the vicinity of the transition.
The resulting Lagrangian resembles
eqs.~\eqref{eq:lag:soft} and \eqref{eq:lag:marginal}
but all parameters are denoted with a bar.

%
\subsection{Integrating out the hard scale}
\label{se:hard:soft}

The first step of the dimensional reduction occurs from hard to soft scale by
integrating out all hard, non-zero Matsubara modes.
Here, the dimensional reduction is performed at NLO ($\mathcal{O}(g^4)$), which means at
one-loop in the couplings,
two-loop in the tadpole and masses, and
one-loop in the field renormalisations.
This chapter merely quotes final results and matching relations,
casting details of the computation of correlators to
appendix~\ref{se:dr:xsm:details}.
We employ a general covariant gauge with gauge parameters
$\xi_1$ for ${\rm U}(1)_{\rmii{Y}}$, and
$\xi_2$ for ${\rm SU}(2)$
opposed to Landau gauge in ref.~\cite{Brauner:2016fla}.

%
\subsubsection*{Normalisation of fields}

The relations between 4d and 3d scalar fields are (for generic field $\psi$)
\begin{align}
(\psi^2)_{\rmii{3d}} &=
  \frac{(\psi^2)_{\rmii{4d}}}{T}
  \Big[1 + \hat{\Pi}_{\psi^2}' \Big]
\;,
\end{align}
where primed correlators are differentiated with respect to the external momentum squared.
The hat denotes the correlator in renormalised perturbation theory
with implicit counterterms $\delta Z_{\psi}$ in appendix~\ref{se:4d:ct:beta}.
The corresponding renormalised 2-point correlation functions yield
\begin{align}
\hat{\Pi}'_{A_{0}^{a}A_{0}^{b}} &=
  \frac{g^{2}}{(4\pi)^2} \bigg(3 + \frac{(\Nc+1)}{3}\nf(L_f-1) + \Big(\xi_2 - \frac{25}{6} \Big) L_b - 2 \xi_2 \bigg)
    \;, \\
\hat{\Pi}'_{A_{r}^{a}A_{s}^{b}} &=
  \frac{g^{2}}{(4\pi)^2} \bigg( -\frac{2}{3} + \frac{(\Nc+1)}{3} \nf L_f + \Big(\xi_2 - \frac{25}{6} \Big) L_b \bigg)
    \;, \\
\hat{\Pi}'_{B_{0}B_{0}} &=
\frac{\gp^{2}}{(4\pi)^2} \frac{1}{6}\Big(
    \Ys^{2}(L_{b}+2)
    + Y_{\rmi{2f}}^{ }\,\nf^{ }(L_f-1)
  \Big)
  \;, \\
\hat{\Pi}'_{B_{r}B_{s}} &=
\frac{\gp^{2}}{(4\pi)^2}\frac{1}{6}\Big(
      \Ys^{2}L_b
      + Y_{\rmi{2f}}^{ }\,\nf^{ } L_f
  \Big)
  \;, \\
\label{eq:phi:2pt:1l:1d}
\hat{\Pi}'_{\phi^\dagger \phi} &=
    \frac{1}{(4\pi)^2} \bigg(
      - \frac{L_b}{4} \Big( 3(3-\xi_2)g^{2} + (3-\xi_1)\gp^{2} \Big)
      + \Nc L_f \gY^{2} \bigg)
      + \frac{\zeta_{3}}{(4\pi)^4} \frac{1}{6} \frac{\mu_{m}^{2}}{T^2}
    \;, \\
\hat{\Pi}'_{\sigma\sigma} &=
  \frac{\zeta_{3}}{(4\pi)^4 T^2}\frac{1}{3} \Big(
      4 \mu_{3}^{2}
    + \mu_{m}^{2}
  \Big)
\;,
\end{align}
where
$\nf=3$ is the number of quark and lepton families and
$\Nc=3$ the number of colours.
The ${\rm U}(1)_{\rmii{Y}}$ hypercharges are
\begin{equation}
\label{eq:hypercharge}
\Yl = -1\;,\quad
\Ye = -2\;,\quad
\Yq = \frac{1}{3}\;,\quad
\Yu = \frac{4}{3}\;,\quad
\Yd = -\frac{2}{3}\;,\quad
\Ys = 1
\;,
\end{equation}
for which we abbreviate recurring sums as
\begin{align}
  \sum_{f} \Yf^{4} \equiv
  Y_{\rmi{4f}}^{ } &=
    \Bigl[ (\Ye^{4} + 2\Yl^{4}) + \Nc(\Yu^{4} + \Yd^{4} + 2\Yq^{4})\Bigr]
  = \frac{2}{81}(729 + 137\Nc)
  = \frac{760}{27}
  \;,\\
  \sum_{f} \Yf^{2} \equiv
  Y_{\rmi{2f}}^{ } &=
    \Bigl[ (\Ye^{2} + 2\Yl^{2}) + \Nc(\Yu^{2} + \Yd^{2} + 2\Yq^{2})\Bigr]
  = \frac{2}{9}(27 + 11\Nc)
  = \frac{40}{3}
  \;.
\end{align}

The matching between the 4d and effective 3d theory
relates their couplings at one-loop order
\begin{align}
\label{eq:match:lh}
\lambda_{h,3}^{ } &=
  T \Big( \lambda_h^{ }
    + \frac{1}{2} \hat{\Gamma}_{(\phi^\dagger \phi)^2}
    - 2\lambda_{h}^{ }\hat{\Pi}_{\phi^\dagger \phi}' \Big)
  \;, \\
\mu_{3,3}^{ } &=
  T^{\frac{1}{2}}
  \Big( \mu_3^{ }
    + \frac{1}{2} \hat{\Gamma}_{\sigma^3}
  - \frac{3}{2} \mu_{3}^{ } \hat{\Pi}_{\sigma\sigma}' \Big)
  \;, \\
\lambda_{\sigma,3}^{ } &=
  T \Big( \lambda_\sigma^{ }
    + \frac{1}{6} \hat{\Gamma}_{\sigma^4}
    - 2 \lambda_\sigma^{ } \hat{\Pi}_{\phi^\dagger \phi}' \Big)
  \, \\
\mu_{m,3}^{ } &=
  T^{\frac{1}{2}}
  \Big( \mu_m^{ }
    + 2 \hat\Gamma_{\phi^\dagger \phi \sigma}
    - \mu_m^{ } ( \hat\Pi_{\phi^\dagger \phi}'
    + \frac{1}{2}\hat\Pi_{\sigma\sigma}') \Big)
  \;, \\
\lambda_{m,3}^{ } &=
  T \Big( \lambda_m^{ }
    + \hat\Gamma_{\phi^\dagger \phi \sigma^2}
    - \lambda_m^{ } ( \hat\Pi_{\phi^\dagger \phi}'
    + \hat\Pi_{\sigma\sigma}' ) \Big)
  \;, \\
x_3^{ } &=
  \frac{1}{2} T\,\hat\Gamma_{A^a_0 A^b_0 \sigma}
  \;, \\
x_3' &=
  \frac{1}{2} T\,\hat\Gamma_{B_0 B_0 \sigma}
  \;, \\
y_3^{ } &=
  \frac{1}{4} T\,\hat\Gamma_{A^a_0 A^b_0 \sigma^2}
  \;, \\
y_3' &=
  \frac{1}{4} T\,\hat\Gamma_{B_0 B_0 \sigma^2}
  \;,
\end{align}
where $\hat\Gamma$ represents a $n$-point correlation function with subscript
corresponding to external fields.
The mass parameters and tadpole at two-loop match according to
\begin{align}
\mu_{h,3}^{2} &= \mu_{h}^{2}
  + \hat\Pi^{\rmii{1loop}}_{\phi^\dagger \phi}
  - (\mu_{h}^{2} + \hat\Pi^{\rmii{1loop}}_{\phi^\dagger \phi})\hat\Pi_{\phi^\dagger \phi}'
  + \hat\Pi^{\rmii{2loop}}_{\phi^\dagger \phi}
  \;, \\
\mu_{\sigma,3}^{2} &= \mu_{\sigma}^{2}
  + \hat\Pi^{\rmii{1loop}}_{\sigma\sigma}
  - (\mu_{\sigma}^{2} + \hat\Pi^{\rmii{1loop}}_{\sigma\sigma})\hat\Pi_{\sigma\sigma}'
  + \hat\Pi^{\rmii{2loop}}_{\sigma\sigma}
  \;, \\
\label{eq:match:mu1}
\mu_{1,3}^{ } &= \mu_{1}^{ }
  + \hat\Gamma^{\rmii{1loop}}_{\sigma}
  - \frac{1}{2} (\mu_{1}^{ } + \hat\Gamma^{\rmii{1loop}}_{\sigma})\hat\Pi_{\sigma\sigma}'
  + \hat\Gamma^{\rmii{2loop}}_{\sigma}
  \;.
\end{align}
The matching of marginal operators~\eqref{eq:lag:marginal} is analogous and
we do not explicate their formulas.%
\footnote{
  In fact, matching relations of marginal operators are simpler since for them
  field normalisations are of even higher order and thus absent.
}
We showcase the computation of the required correlators in
appendix~\ref{se:cor:2l}.
As an example, for the quartic Higgs coupling, we obtain
\begin{align}
  \lambda_{h,3}^{ } &= T\bigg[
    \underbrace{\lambda_h^{ }(\Lambda)}_{\rmO(g^4)\,{\rm running}}
    + \frac{1}{(4\pi)^2} \bigg(
      \frac{2-3L_b}{16}\Big(
        3g^4 + 2g^2\gp^2 + \gp^4
      \Big)
      + \Nc\gY^{2} L_f \Big( \gY^{2}
      - \underbrace{2 \lambda_{h}^{ }}_{\text{f.n.}} \Big)
    \nn &\hp{=T\bigg[}
  + L_b \Big(
      \underbrace{\frac{3}{2}(3g^2 + \gp^2) \lambda_{h}^{ }}_{\text{f.n.}}
    - 12 \lambda_{h}^{2}
    - \frac{1}{4} \lambda_{m}^{ }
    \Big)
  \bigg)
  - \frac{1}{2} (\underbrace{1}_{\text{f.n.}} -1) L_b \lambda_{h}^{ } (3g^3 \xi_2 + \gp^2 \xi_1)
  \nn &\hp{=T\bigg[}
  + \frac{\zeta_{3}}{(4\pi)^4 T^2} \Big(
      \Big(3 -\underbrace{\frac{1}{3}}_{\text{f.n.}}\Big) \lambda_{h}^{ }
    + \frac{1}{2} \lambda_{m}^{ } \Big) \mu_{m}^{2}
  - \frac{\zeta_{5}}{(4\pi)^6 T^4} \frac{\mu_{m}^{4}}{4}
  \bigg]\;,
\end{align}
where we indicated the origin of individual contributions,
displaying cancellations
between correlators and field normalisation (f.n.) for the
{\em RG-scale} and
{\em gauge dependence}.
To witness a cancellation of the RG-scale $\Lambda$ at $\mathcal{O}(g^4)$ order,
we include the one-loop $\beta$-function~\eqref{eq:ct:dlamh}
for the tree-level piece,
which cancels the logarithmic scale dependence in $L_{b,f}$.
In these cancellations, the contribution from field normalisation
is essential.
Note that the coefficient of
$\ln(\Lambda/T)$ (in $L_{b,f}$) matches the $\beta$-function for the tree-level parameter.

The matching of other parameters yields analogously
\begin{align}
\mu_{3,3}^{ } &=
  T^{\frac{1}{2}}
  \bigg[
    \mu_3^{ }(\Lambda)
  - \frac{3}{(4\pi)^2} L_b \Big(3 \lambda_\sigma^{ } \mu_3^{ } + \frac{1}{2}\lambda_m^{ } \mu_m^{ } \Big)
  + \frac{\zeta_{3}}{ (4\pi)^4 T^2} \Big(
      6 \mu_{3}^{3}
    - \frac{1}{2} \mu_{3}^{ } \mu_{m}^{2}
    + \frac{1}{2} \mu_{m}^{3} \Big) \bigg]
\;, \\
\lambda_{\sigma,3}^{ } &= T \bigg[
    \lambda_\sigma^{ }(\Lambda)
  - \frac{1}{(4\pi)^2} L_b \Big(\lambda_{m}^{2} + 9 \lambda_{\sigma}^{2} \Big)
  + \frac{\zeta_{3}}{(4\pi)^4 T^2} \Big(
      \frac{136}{3} \lambda_\sigma^{ } \mu_{3}^{2}
      + 2 \lambda_{m}^{ }\mu_{m}^{2}
    - \frac{2}{3} \lambda_\sigma^{ }\mu_{m}^{2} \Big)
  \nn &\hp{=\bigg[}
  - \frac{\zeta_{5}}{(4\pi)^6 T^4}\Big(32 \mu_{3}^{4} + \frac{1}{2} \mu_{m}^{4} \Big)
  \bigg]
\;, \\
\mu_{m,3}^{ } &=
  T^{\frac{1}{2}}
  \bigg[
    \mu_m^{ }(\Lambda)
  + \frac{1}{(4\pi)^2} \bigg(
      L_b \Big(
        \Big( \frac{3}{4}(3g^2 + \gp^2) - 6 \lambda_h^{ } - 2 \lambda_m^{ } \Big) \mu_m^{ }
      - 2 \lambda_m^{ } \mu_3 ^{ }
    \Big)
    - \Nc L_f \gY^{2} \mu_m^{ }
  \bigg)
  \nn &\hp{=T^{\frac{1}{2}}\bigg[}
  + \frac{\zeta_{3}}{(4\pi)^4 T^2} \mu_m^{ } \Big(
    - \frac{2}{3} \mu_{3}^{2}
    + 2 \mu_{3}^{ } \mu_{m}^{ }
    + \frac{1}{6} \mu_{m}^{2}
  \Big)
  \bigg]
\;.
\end{align}
We observe an important -- and bluntly surprising -- exception for
the $Z_2$-symmetric Higgs-singlet portal coupling
\begin{align}
\label{eq:lamm}
\lambda_{m,3}^{ } &= T \bigg[
  \lambda_{m}^{ }(\Lambda)
  + \frac{\lambda_m^{ }}{(4\pi)^2} \bigg(
      L_b  \Big(
        \frac{3}{4}(3g^2 + \gp^2)
      - 6\lambda_h^{ }
      - 2\lambda_m^{ }
      - 3\lambda_\sigma^{ }
    \Big)
    - \Nc L_f \gY^{2}
  \bigg)
  \nn &\hp{= T \bigg[}
  + \frac{\zeta_{3}}{(4\pi)^4 T^2} \bigg(
      \lambda_m^{ } \Big(\frac{20}{3} \mu_{3}^{2} + 8 \mu_3^{ } \mu_m^{ } \Big)
    + (6 \lambda_h^{ } + 2 \lambda_m^{ } + 3 \lambda_\sigma^{ })\mu_{m}^{2}
  \bigg)
  \nn &\hp{= T \bigg[}
  - \frac{\zeta_{5}}{(4\pi)^6 T^4} \mu_{m}^{2} \Big( 8 \mu_{3}^{2} + 2 \mu_3^{ } \mu_m^{ } + \frac{1}{2} \mu_{m}^{2} \Big)
  - \frac{\zeta_{3}}{(4\pi)^4 T^2} \mu_{m}^{2} \frac{1}{4}
    \underbrace{\Big(3g^2 \xi_2 + \gp^2 \xi_1 \Big) }_{\xi-\text{dependent}}
  \bigg]
\;.
\end{align}
The remaining gauge dependence in $\lambda_{m,3}$ originates from the correlator
$\hat\Gamma_{\phi^\dagger \phi \sigma^2}$ (cf.\ appendix~\ref{se:marginal}) and
is uncancelled by the field renormalisation contribution
which is proportional to the portal coupling $\lambda_m$ instead.
Since the matched parameters are merely 3d effective Lagrangian parameters, they are not
directly associated with physical observables of the 3d theory and
may well depend on the gauge fixing of the 4d theory.
A similar discussion in ref.~\cite{Laine:2018lgj} addresses
the role of higher dimensional operators in hot QCD.
We defer the topic of gauge dependence
in matching of hot electroweak theories to future research.
Meanwhile, an immediate solution
(used in ref.~\cite{Niemi:2021xxx}) is provided by a stricter
power counting $\mu_{m}\sim g^{2}T$ for the cubic portal coupling.
Hence, by sticking to $\rmO(g^4)$,
all gauge-dependent contributions are cast to higher orders.

The singlet interacts with temporal scalars through the couplings
\begin{align}
x_{3}^{ } &= T \frac{1}{(4\pi)^2} g^2 \mu_m^{ }
\;, \\
x_{3}' &= T \frac{1}{(4\pi)^2} \gp^2 \mu_m^{ }
\;, \\
y_{3}^{ } &= T \Big(
    \frac{1}{(4\pi)^2} \frac{1}{2} g^2 \lambda_m^{ }
    - \frac{\zeta_{3}}{(4\pi)^{4}T^{2}} \frac{1}{2} g^{2}\mu_{m}^{2}
  \Big)
\;, \\
y_{3}' &= T \Big(
    \frac{1}{(4\pi)^2} \frac{1}{2} \gp^2 \lambda_m^{ }
    - \frac{\zeta_{3}}{(4\pi)^{4}T^{2}} \frac{1}{2} \gp^{2}\mu_{m}^{2}
  \Big)
\;.
\end{align}
The results for the two-loop mass parameters and tadpole coupling read
\begin{align}
\mu_{h,3}^{2} &=
  \big(\mu_{h,3}^{2}\big)_{\rmii{SM}}
  + \frac{T^2}{24} \lambda_{m}^{ }(\Lambda)
  \nn &
  - \frac{L_b}{(4\pi)^2} \Big(
      \frac{1}{4} \mu_{m}^{2}(\Lambda)
    + \frac{1}{2} \lambda_{m}^{ } \mu_{\sigma}^{2}(\Lambda)
  \Big)
  + \frac{\zeta_{3}}{(4\pi)^4} \frac{\mu_{m}^{2}}{T^2} \frac{1}{2} \Big(
      \mu_{h}^{2}(\Lambda)
    + \mu_{\sigma}^{2}(\Lambda) \Big)
  \nn &
  - \frac{\zeta_{3}}{6(4\pi)^4} \frac{\mu_{m}^{2}}{T^2} \bigg(
      \mu_{h}^{2}(\Lambda)
      + \frac{T^2}{12}\Big(
        \frac{3}{4}(3g^2 +\gp^2) + \Nc\gY^{2}
        + 6\lambda_h^{ }
        + \frac{1}{2}\lambda_m^{ }
      \Big)
    - \frac{1}{(4\pi)^2} \frac{1}{4} L_b \mu_{m}^{2}(\Lambda)
  \bigg)
  \nn &
  + \frac{1}{(4\pi)^2}
    \bigg(\frac{3}{4}(3g^2 + \gp^2) L_b - \Nc\gY^{2} L_f \bigg)
    \bigg(\frac{T^2}{24} \lambda_m^{ } - \frac{1}{(4\pi)^2} \frac{1}{4} L_b \mu_{m}^{2}(\Lambda) \bigg)
  \nn &
  + \frac{1}{(4\pi)^4} \bigg[
    \frac{(3+2L_b+L_b^2)}{2}\Big(
        \lambda_{m}^{ }(\mu_{3}^{2} +\mu_{3}^{ } \mu_{m}^{ })
        +\frac{9}{2} \mu_{m}^{2}\lambda_{h}
    \Big)
  \nn &\hp{+\frac{1}{(4\pi)^4}\bigg[}
    - L^2_b \mu_{m}^{2} \Big(
        \frac{3}{32}(3g^2 + \gp^2)
      - \frac{5}{8} \lambda_{m}^{ }
    \Big)
    + \Nc\,\mu_{m}^{2}\gY^{2} \frac{L_f}{4} \Big(L_b -\frac{1}{2}L_f \Big)
  \bigg]
  \nn &
  - \frac{T^2}{(4\pi)^2} L_b \lambda_{m}^{ } \Big(
      \frac{1}{4}\lambda_h^{ }
    + \frac{5}{24}\lambda_m^{ }
    + \frac{1}{8}\lambda_\sigma^{ }
  \Big)
  - \frac{1}{(4\pi)^2} \frac{1}{2} \lambda^2_{m,3} \Big(c + \ln\Big(\frac{3T}{\Lamd}\Big)\Big)
  \nn &
  - \frac{1}{(4\pi)^4} \frac{(3 + 2L_b)}{8}\mu_{m}^{2} \Big(
      \frac{1}{4}(3g^2 + \gp^2)
    - \Nc\gY^{2}
    - 3\lambda_{m}^{ }
  \Big)
  \nn &
  + \frac{\zeta_{3}}{(4\pi)^4} \mu_{m}^{2} \Big(
    \frac{1}{32}(3g^2 + \gp^2)
    + \frac{1}{4} \lambda_{h}^{ }
    + \frac{5}{48} \lambda_{m}^{ }
    + \frac{1}{8} \lambda_\sigma^{ }
    + \frac{\Nc}{24} \gY^{2}
  \Big)
  \nn &
  - \frac{\zeta_{3}}{(4\pi)^{6}T^{2}} \frac{(3 + 2 L_b)}{16} \mu_{m}^{2} (8 \mu_{3}^{2} + 3 \mu_{m}^{2})
  \;, \\[3mm]
\mu_{\sigma,3}^{2} &=
  \mu_{\sigma}^{2}(\Lambda)
  + T^2 \Big(
      \frac{1}{6}\lambda_m^{ }(\Lambda)
    + \frac{1}{4}\lambda_\sigma^{ }(\Lambda)
  \Big)
  + \frac{1}{(4\pi)^4} \frac{\zeta_{3}}{T^2} \Big(2 \mu_{m}^{2} \mu_{h}^{2}(\Lambda) + 8 \mu_{3}^{2} \mu_{\sigma}^{2}(\Lambda) \Big)
  \nn &
  - \frac{L_b}{(4\pi)^2}\Big(
      2\mu_{3}^{2}(\Lambda)
    + \frac{1}{2} \mu_{m}^{2}(\Lambda)
    + 2 \lambda_m^{ } \mu_{h}^{2}(\Lambda)
    + 3 \lambda_\sigma^{ } \mu_{\sigma}^{2}(\Lambda)
  \Big)
  \nn &
  - \frac{\zeta_{3}}{(4\pi)^4 T^{2}} \frac{2}{3}
    \Big( 2\mu_{3}^{2} + \frac{1}{2} \mu_{m}^{2} \Big)
    \bigg(
        \mu_{\sigma}^{2}(\Lambda)
      + T^2 \Big(\frac16 \lambda_m^{ } + \frac{1}{4} \lambda_\sigma^{ } \Big)
      - \frac{L_b}{(4\pi)^2} \Big( 2 \mu_{3}^{2} + \frac{1}{2} \mu_{m}^{2} \Big)
    \bigg)
  \nn &
  + \frac{1}{(4\pi)^4} \bigg[
    \frac{(3+2L_b)}{2}\bigg(
        30 \lambda_\sigma^{ } \mu_{3}^{2}
      + 4 \lambda_m^{ } \mu_3^{ } \mu_m^{ }
      - \mu_{m}^{2}\Big(
          (3g^2 + \gp^2)
        - \Nc\gY^{2}
      \Big)
    \bigg)
    \nn &\hp{+T^2}
    + \frac{(3 + 2L_b + L^2_b)}{4} \mu_{m}^{2} (5\lambda_m^{ }+ 3\lambda_\sigma^{ })
    + \Nc\gY^{2}\mu_{m}^{2}L_{f}\Big( L_{b} -\frac{1}{2}L_{f}\Big)
    \nn &\hp{+T^2}
    + L_{b}^{2} \Big(
        21 \lambda_\sigma^{ } \mu_{3}^{2}
      + 4 \lambda_m^{ } \mu_3^{ } \mu_m^{ }
      + 3 \lambda_h^{ }\mu_{m}^{2}
      - \frac{3}{8}(3g^2 + \gp^2)\mu_{m}^{2}
    \Big)
  \bigg]
  \nn &
  + \frac{1}{(4\pi)^2} \Big(
    (3g^2_3 + \gp^2_3) \lambda_{m,3}^{ }
    - 2\lambda_{m,3}^{2}
    - 6\lambda_{\sigma,3}^{2}
  \Big) \Big( c + \ln\Big(\frac{3T}{\Lamd} \Big) \Big)
  \nn &
  + \frac{T^2}{(4\pi)^2} \bigg[
      \frac{(2+3L_b)}{24}(3g^2 + \gp^2) \lambda_m^{ }
    - L_b \bigg(
      \Big(
          \lambda_h^{ }
        + \frac{7}{12}\lambda_m^{ }
        + \frac{1}{2} \lambda_\sigma^{ }
      \Big) \lambda_m^{ }
    + \frac{9}{4} \lambda_{\sigma}^{2} \bigg)
  \nn &\hp{-\frac{T^2}{(4\pi)^2}\bigg[}
  - \frac{\Nc}{12}(3L_b - L_f) \gY^{2}\lambda_{m}^{ }
  \bigg]
  \nn &
  + \frac{\zeta_{3}}{(4\pi)^4} \bigg(
    \Big(
        \frac{1}{8} (3g^2 + \gp^2)
      + \frac{\Nc}{6}\gY^{2}
      + \lambda_h
      + \frac{1}{12}\lambda_m^{ }
    \Big) \mu_{m}^{2}
  + \Big( \frac{4}{3} \lambda_m^{ } + 2 \lambda_\sigma^{ } \Big) \mu_{3}^{2}
  \bigg)
  \nn &
  - \frac{\zeta_{3}}{(4\pi)^{6}T^{2}} \frac{( 3 + 2 L_b)}{4} \Big(
      32 \mu_{3}^{4}
    + 8 \mu_{3}^{2} \mu_{m}^{2}
    + \mu_{m}^{4}
  \Big)
  \;,
\end{align}
and
\begin{align}
\mu_{1,3}^{ } &=
  T^{-\frac{1}{2}}\bigg\{
    \mu_{1}^{ }(\Lambda)
  + \frac{T^2}{12} \Big( \mu_3^{ }(\Lambda) + \mu_m^{ }(\Lambda) \Big)
  - \frac{L_b}{(4\pi)^2} \Big(
      \mu_m^{ } \mu_{h}^{2}(\Lambda)
    + \mu_3^{ } \mu_{\sigma}^{2}(\Lambda) \Big)
  \nn &\hp{-T\bigg\{}
  - \frac{\zeta_{3}}{(4\pi)^4 T^2} \frac{1}{6}(4 \mu_{3}^{2} + \mu_{m}^{2}) \Big(\mu_1^{ } + \frac{1}{12} T^2 (\mu_3^{ } + \mu_m^{ }) \Big)
  \nn &\hp{-T\bigg\{}
  + \frac{1}{(4\pi)^4} \frac18 (3 + 2 L_{b}^{ } + L_{b}^{2}) (8 \mu_{3}^{3} + 2 \mu_{3}^{ }\mu_{m}^{2} + \mu_{m}^{3})
  \nn &
  + \frac{T^2}{(4\pi)^2}\bigg[
    \frac{(2+3L_b)}{48}(3g^2 + \gp^2)\mu_{m}^{ }
    - \frac{L_b}{2}\bigg(
      \Big(
        \lambda_h
      + \frac{7}{12} \lambda_m^{ }
      \Big) \mu_m^{ }
    + \Big(
        \frac{1}{3} \lambda_m^{ }
      + \frac{3}{2}\lambda_\sigma^{ }
    \Big)\mu_3^{ }
  \bigg)
  \nn &\hp{-\frac{T^2}{(4\pi)^2}\bigg[}
    - \frac{\Nc}{24} \mu_m^{ }\gY^{2} \big(
        3L_b - L_f
  \big) \bigg]
\bigg \}
  \nn &
  - \frac{1}{(4\pi)^2} \Big(
      2 \lambda_{\sigma,3}^{ } \mu_{3,3}^{ }
    - \frac{1}{2} \mu_{m,3}^{ } \big(3g^2_3 + {\gp_3}^{2} - 2 \lambda_{m,3}^{ } \big)
  \Big)
  \Big( c + \ln\Big(\frac{3T}{\Lamd} \Big) \Big)
  \;.
\end{align}
We explicated the dependence of the 4d RG-scale $\LamD$ in terms where
the running can be verified to cancel the logarithmic scale dependence,
rendering the 3d parameters $\LamD$-independent.
The running is determined by the $\beta$-functions listed in appendix~\ref{se:4d:ct:beta}.
In the xSM the cancellation of the RG-scale is slightly more subtle than in the SM,
since the running of the masses starts at order
$\mathcal{O}(\mu_{m}^{2}, \mu_{3}^{2})$.
Therefore, the running of the one-loop mass corrections is crucial.
In the above formulas, the 3d running is in fact not exact, as we did not include
two-loop terms with singlet-temporal scalar couplings
$x_3,x_3',y_3$ and
$y_3'$.
However, this approximation is justified since temporal couplings only arise
at $\mathcal{O}(g^4)$ because the singlet couples only indirectly to gauge fields.
Hence, the omitted terms proportional to $x_{3}^{2},\dots$ are numerically negligible.
The SM result for the three-dimensional mass parameter reads
\clearpage
\begin{align}
\label{eq:muh:sm}
\big(\mu_{h,3}^{2}\big)_\rmii{SM} &=
  \mu_{h}^{2}(\Lambda)
  + \frac{T^2}{12}\bigg(
      \frac{3}{4}(3g^2(\Lambda) + \gp^2(\Lambda))
    + \Nc\gY^{2}(\Lambda)
    + 6\lambda_{h}^{ }(\Lambda)
  \bigg)
  \nn &
  + \frac{\mu_{h}^{2}(\Lambda)}{(4\pi)^2}\bigg(
      \Big(\frac{3}{4}(3g^2 + \gp^2) - 6 \lambda_h^{ } \Big)L_b
    - \Nc\gY^{2} L_f
  \bigg)
  \nn &
    + \frac{T^2}{(4\pi)^2} \bigg[
        \frac{167}{96}g^4
      + \frac{1}{288}\gp^4
      - \frac{3}{16}g^2\gp^2
      + \frac{(1+3L_b)}{4}\lambda_h^{ }(3g^2+\gp^2)
      \nn &\hp{+T^2\bigg(}
      + L_b \Big(
          \frac{17}{16}g^4
        - \frac{5}{48}\gp^4
        - \frac{3}{16}g^2\gp^2
        - 6 \lambda_{h}^{2}
      \Big)
      \nn &\hp{+T^2\bigg(}
      + \frac{1}{T^2}\Big( c + \ln\Big(\frac{3T}{\Lamd}\Big) \Big)\Big(
          \frac{39}{16}g^4_3
        + 12g^2_3 h_{3}^{ }
        - 6 h_{3}^{2}
        + 9 g_{3}^{2} \lambda_{2,3}^{ }
        - 12\lambda_{h,3}^{2}
      \nn &\hp{+T^2\bigg(+\frac{1}{T^2}\Big(c+\ln\Big(\frac{3T}{\Lamd}\Big)\Big)\Big(}
        - \frac{5}{16} \gp^4_3
        - \frac{9}{8} g^2_3 \gp^2_3
        - 2h'^2_{3}
        - 3h''^2_3
        + 3 \gp^2_3 \lambda_{h,3}^{ }
      \Big)
      \nn &\hp{+T^2\bigg(}
        - \frac{1}{96}\Big(9L_b - 3L_f -2\Big)
        \Big((\Nc+1)\,g^4 + \frac{1}{6}Y_{\rmi{2f}}^{ }\,\gp^4 \Big)\nf^{ }
      \nn &\hp{+T^2\bigg(}
        + \frac{\Nc}{32}\Big(7L_b -L_f - 2\Big)g^{2}\gY^{2}
        - \frac{\Nc}{4}\Big(3L_b + L_f\Big)\lambda_{h}^{ }\gY^{2}
      \nn &\hp{+T^2\bigg(}
        + \frac{\Nc}{96}\bigg(
            \Big(9(L_b - L_f) +4\Big)\Ys^{2}
          - 2\Big(L_b - 4L_f + 3\Big)(\Yq^{2} + \Yu^{2})
        \bigg)\gp^{2}\gY^{2}
      \nn &\hp{+T^2\bigg(}
        - \frac{\Nc\CF}{6}\Big(L_b - 4L_f +3\Big)\gs^{2}\gY^{2}
        + \frac{\Nc}{24}\Big(3L_b - 2(\Nc-3)L_f\Big)\gY^{4}
  \bigg]
  \;,
\end{align}
where
$\CF=\frac{\Nc^2-1}{2\Nc}=\frac{4}{3}$ is the fundamental quadratic Casimir of SU(3).
In the xSM the running of $\mu_{h}^{2}(\Lambda)$ starts at $\mathcal{O}(\mu_{m}^{2})$
which is apparent from the $\beta$-function~\eqref{eq:muh:beta}.
Therefore, also the one-loop mass correction in the second line of $\mu_{h,3}^{2}$
is affected by it which is required for the cancellation of the RG-scale.
This result can be found in refs.~\cite{Gorda:2018hvi,Kajantie:1995dw}
where the latter neglects the two-loop contributions involving $\gp$.
The matching relations of the marginal operator coefficients are listed in
appendix~\ref{se:marginal}.
Additionally also the couplings between
temporal scalars and
the Higgs doublet receive singlet-induced corrections:
\begin{align}
h_3&=\frac{g^2(\Lambda)T}{4}\bigg[
  1
  + \frac{1}{(4\pi)^2}\bigg(
      \Big(
        \frac{43}{6}L_b
      + \frac{17}{2}
      - \frac{(\Nc+1)\nf}{3}(L_f-1)
    \Big)g^2
    + \frac{\gp^2}{2}
    - 2\Nc\gY^{2}
    + 12\lambda_h\bigg)
  \nn &\hp{{}=\frac{g^2(\Lambda)T}{4}\bigg[1}
    - \frac{\zeta_{3}}{(4\pi)^{4}T^{2}}\frac{2}{3}\mu_{m}^{2}
  \bigg]
\;,\\
h'_3&=\frac{\gp^2(\Lambda)T}{4}\bigg[
  1
  + \frac{1}{(4\pi)^2}\bigg(
      \frac{3g^2}{2}
      - \frac{1}{6}\Big(
          (L_b-1)\Ys^{2}
        + (L_f-1)Y_{\rmi{2f}}^{ }\,\nf^{ }
      \Big)\gp^2
  \nn &\hp{{}=\frac{\gp^2(\Lambda)T}{4}\bigg[1}
      - 2(\Yq^{2}+\Yu^{2})\Nc\gY^{2}
      + 12\lambda_{h}^{ }
    \bigg)
    - \frac{\zeta_{3}}{(4\pi)^{4}T^{2}}\frac{2}{3}\mu_{m}^{2}
  \bigg]
\;,\\
h''_{3}&=\frac{g(\Lambda)\gp(\Lambda)T}{2}\bigg[
    1
    + \frac{1}{(4\pi)^2}\bigg(
        \Big(\frac{43}{12}L_b - 1\Big) g^2
      - \frac{\Ys^{2}}{3}\Big(\frac{1}{4}L_b - 1\Big)\gp^2
      + 4\lambda_{h}^{ }
      + \frac{2}{3}\Nc^{ }\,\gY^{2}
  \nn &\hp{{}=\frac{g(\Lambda)\gp(\Lambda)T}{2}\bigg[1}
      - (L_f-1)\Big(\frac{\Nc +1}{6}g^2 +\frac{Y_{\rmi{2f}}^{ }}{12}\gp^2\Big)\nf^{ }
    \bigg)
    - \frac{\zeta_{3}}{(4\pi)^{4}T^{2}}\frac{2}{3}\mu_{m}^{2}
\bigg]
\;.
\end{align}
Due to the absence of singlet contributions,
other 3d parameters in the dimensionally reduced Lagrangian~\eqref{eq:L:temp:soft}
agree with the Standard model.
For completeness, we collect these results below,
with
$\nf=3$ the number of fermion generations
\begin{align}
g_3^2 &=
  g^2(\Lambda)T\bigg[
    1
  + \frac{g^2}{(4\pi)^2}\Big(
      \frac{43}{6}L_b
    + \frac{2}{3}
    - \frac{(\Nc+1)\nf}{3}L_f
  \Big)\bigg]
\;,\\
\gp^{2}_3 &=
  \gp^2(\Lambda)T\bigg[
    1
    - \frac{\gp^2}{(4\pi)^2}\frac{1}{6}\Big(
        L_{b}\Ys^{2}
      + L_{f}Y_{\rmi{2f}}^{ }\,\nf^{ }
  \Big)\bigg]
\;,\\
\mD^{2}&=g^2T^2\frac{1}{3}\Big(
  \frac{5}{2}
  + \frac{(\Nc + 1)}{4}\nf
  \Big)
\;,\\
{\mD'}^{2}&=\gp^2T^2\frac{1}{24}\Big(
    4\Ys^2
  + Y_{\rmi{2f}}^{ }\,\nf^{ }
  \Big)
\;,\\
\label{eq:mD3}
{\mD''}^{2} &=
\gs^{2} T^2 \Big(\frac{\Nc}{3}+\frac{\nf}{3}\Big)
\;,\\
\kappa_3&=T\frac{g^4}{(4\pi)^2}\frac{1}{3}\Big(17-(\Nc + 1)\nf\Big)
\;,\\
\kappa_3'&=T\frac{\gp^4}{(4\pi)^2}\frac{1}{6}\Big(
    2 \Ys^{4}
  - Y_{\rmi{4f}}^{ }\,\nf^{ }\Big)
\;,\\
\kappa_3''&=T\frac{g^2\gp^2}{(4\pi)^2}
  2\Big(\Ys^{2}-(\Yl^{2} + \Nc^{ }\Yq^{2})\nf^{ }\Big)
\;.
\end{align}
We point out, that eq.~(3.87) in ref.~\cite{Brauner:2016fla} misprints
the gluon Debye mass ${\mD''}^{2}$, which we corrected above in
eq.~\eqref{eq:mD3}.

%
\subsubsection*{Two-loop electroweak Debye masses $\mD^{ }$ and $\mD'$}
\label{se:debye:2l}

Above, we quoted the one-loop electroweak Debye masses $\mD$ and $\mD'$,
which are standard, as their two-loop corrections are of higher order at final ultrasoft scale EFT.
To reach a consistent $\rmO(g^4)$ order at the soft scale,
these Debye masses should be computed at two-loop order, and at this order they receive
contributions from the singlet.
We point out, that the QCD Debye mass is independent of singlet contributions at two-loop order.
We omit its two-loop result here, since these are further suppressed at
the ultrasoft scale than two-loop contributions to the EW Debye masses because
the Higgs couples only indirectly to the gluon sector.

The singlet-induced two-loop corrections to the electroweak Debye masses read
\begin{align}
\label{eq:mD1}
{\mD'^{2}}
&=
      \frac{T^{2}}{(4\pi)^{2}} \frac{1}{12}\,\gp^{2}\,\lambda_{m}^{ }
    - \frac{1}{(4\pi)^{4}}\,\gp^{2}\,\mu_{m}^{2} \Big(
      1 + \frac{1}{2} L_{b}
    \Big)
    + (\text{SM terms})
  \;, \\[3mm]
\label{eq:mD2}
{\mD^{2}}
&=
      \frac{T^{2}}{(4\pi)^{2}} \frac{1}{12}\,g^{2}\,\lambda_{m}^{ }
    - \frac{1}{(4\pi)^{4}}\,g^{2}\,\mu_{m}^{2} \Big(
      1 + \frac{1}{2} L_{b}
    \Big)
    + (\text{SM terms})
  \;.
\end{align}
Their standard Model contributions have initially been computed in
refs.~\cite{Gynther:2005dj,Gynther:2005av} and
are reproduced in eqs.~\eqref{eq:mD1:SM} and \eqref{eq:mD2:SM}
in appendix~\ref{se:cor:2l}.

%
\subsection{Integrating out the soft scale}
\label{se:soft:ultrasoft}

The second step of dimensional reduction integrates out heavy temporal scalars
at the soft scale.
The resulting simplified 3d EFT at the ultrasoft scale assumes
light, dynamical doublet and singlet scalars.
Since the singlet couples to gauge fields only indirectly,
its couplings to temporal scalars
$A_{0}^{a}$ and
$B_{0}^{ }$ are suppressed already at leading order.
Hence, we include only one-loop effects of temporal scalars in
ultrasoft singlet parameters
instead of including two-loop corrections for tadpole and singlet mass parameter.
All correlators are then encoded in
the one-loop contribution of the effective potential
(see appendix~\ref{se:integrals} for the one-loop master integral)
\begin{align}
V^{\rmii{1loop}}_{\rmii{eff, soft}} \simeq
  3 J_\rmii{soft}(m_A)
  + J_\rmii{soft}(m_B)
  \;.
\end{align}
Denoting the background fields
$v_3$ for the doublet and
$s_3$ for the singlet,
the (3d) background field-dependent
mass eigenvalues read
\begin{align}
m^2_A &= \mD^{2}
  + h_{3}^{ } v_3^{2}
  + 2 s_{3}^{ } (x_3^{ } + y_3^{ } s_3^{ })
\;, \\
m^2_B &= \mD'^{2}
  + h_3' v^2_3
  + 2 s_3^{ } (x_3' + y_3' s_3^{ })
\;,
\end{align}
and give rise to the (one-loop) matching relations
\begin{align}
\bar{\mu}_{1,3} &= \mu_{1,3}
  - \frac{1}{4\pi} \Big(3 \mD x_3 + \mD' x'_3 \Big)
\;, \\
\bar{\mu}^2_{\sigma,3} &= \mu^2_{\sigma,3}
  - \frac{1}{2\pi} \Big( 3 \mD^{ } y_3^{ } + \mD' y'_3 \Big)
  - \frac{1}{4\pi} \Big( 3\frac{x^2_3}{\mD^{ }} + \frac{x'^2_3}{\mD'} \Big)
\;, \\
\bar{\mu}_{3,3} &= \mu_{3,3}
  - \frac{3}{4\pi} \Big( 3 \frac{x_3^{ } y_3^{ }}{\mD^{ }} + \frac{x'_3 y'_3}{\mD'} \Big)
  + \frac{1}{8\pi} \Big( 3 \frac{x^3_{3}}{\mD^3} + \frac{x'^3_3}{\mD'^3} \Big)
\;, \\
\bar{\lambda}_{\sigma,3} &= \lambda_{\sigma,3}
  - \frac{1}{8\pi} \Big( 3\frac{x^4_3}{\mD^5} + \frac{x'^4_3}{\mD'^5} \Big)
  + \frac{1}{2\pi} \Big( 3\frac{x^2_3 y_3^{ }}{\mD^3} + \frac{x'^2_3 y'_3}{\mD'^3} \Big)
  - \frac{1}{2\pi} \Big( 3\frac{y^2_3}{\mD^{ }} + \frac{y'^2_3}{\mD'} \Big)
\;, \\
\bar{\mu}_{m,3} &= \mu_{m,3}
  - \frac{1}{2\pi} \Big( 3\frac{h_3^{ } x_3^{ }}{\mD^{ }} + \frac{h'_3 x'_3}{\mD'} \Big)
\;, \\
\bar{\lambda}_{m,3} &= \lambda_{m,3}
  + \frac{1}{4\pi} \Big( 3\frac{h_3^{ } x^2_3}{\mD^3} + \frac{h'_3 x'^2_3}{\mD'^3} \Big)
  - \frac{1}{2\pi} \Big( 3\frac{h_3^{ } y_3^{ }}{\mD^{ }} + \frac{h'_3 y'_3}{\mD'} \Big)
\;.
\end{align}
Therein soft corrections stem only from correlators since
field normalisations contribute at a higher order
due to non-existing tree-level contributions.
In particular, all soft contributions at leading order are gauge-independent,
as there are no gauge field propagators involved at one-loop order.
The remaining parameters are referred from~\cite{Kajantie:1995dw,Croon:2020cgk,Niemi:2021xxx}
\begin{align}
\bar{g}^2_3 &= g_{3}^{2} \Big( 1 - \frac{g_{3}^{2}}{6 (4\pi) \mD} \Big)
  \;, \\
\bar{g}'^2_3 &= {g}'^2_{3}
  \;, \\
\bar{\mu}^2_{h,3} &= \mu_{h,3}^{2}
    - \frac{1}{4\pi}\Big(
      3 h_{3}^{ }\mD^{ }
      + {h_{3}'}^{ }\mD'
      + 2\Nc^{ }\CF^{ }\delta_{3}^{ }\mD''
    \Big)
    \nn &
    + \frac{1}{(4\pi)^2} \bigg(
        3 g_{3}^{2} h_{3}^{ }
      - 3 h_{3}^{2}
      - {h_{3}'}^{2}
      - \frac{3}{2} {h_{3}''}^{2}
    \nn &\hp{+\frac{1}{(4\pi)^2}\bigg(}
    - \Big(\frac{3}{4}g_{3}^{4} - 12g_{3}^{2}h_{3}^{ } \Big) \ln\Big(\frac{\Lamd}{2\mD} \Big)
    - 6 h_{3}^{2} \ln\Big(\frac{\Lamd}{2\mD} \Big)
    \nn &\hp{+\frac{1}{(4\pi)^2}\bigg(}
    - 2 {h_{3}'}^{2} \ln\Big(\frac{\Lamd}{2\mD'} \Big)
    - 3 {h_{3}''}^{2} \ln\Big(\frac{\Lamd}{\mD^{ }+\mD'} \Big)
    \bigg)
\;, \\
\bar{\lambda}_{3} &= \lambda_{3} - \frac{1}{2(4\pi)}\Big(
        \frac{3 h_{3}^{2}}{\mD^{ }}
      + \frac{{h_{3}'}^{2}}{\mD'}
      + \frac{{h_{3}''}^{2}}{\mD^{ }+\mD'}
    \Big)
    \;.
\end{align}
In general the ultrasoft Higgs self-energy $\bar{\mu}^2_{h,3}$ receives contributions
from interactions with singlet and temporal scalars.
Even though these are two-loop topologies, we discard them due to the suppression of
$x_3^{ },x'_3,y_3^{ },y'_3$
in analogy with discarding contributions with quartic self-interactions of temporal scalars.
This is apparent for
$
\kappa_3^{ },
\kappa'_3,
\kappa''_3
$
that lack
a tree-level contribution $\mathcal{O}(g^2)$ and consequently their
leading contribution is $\mathcal{O}(g^4)$.
For simplicity, we drop corrections from temporal scalars to marginal operators
due to their numerical insignificance.

These relations complete our construction of the high-$T$ 3d EFT of
the SM augmented with a real scalar singlet.
As an effective theory, it can be used to
examine the thermodynamics of the electroweak phase transition of the fundamental model
(cf.\ refs.~\cite{Niemi:2021xxx,Gould:2021xxx}).
In particular, ref.~\cite{Niemi:2021xxx} showcases
the computation of the two-loop thermal effective potential
in the 3d EFT of the xSM constructed in this section.
This is analogous to our section~\ref{se:tutorial:veff}.

%
\section{Discussion}
\label{se:discussion}

The pipeline between collider phenomenology of BSM theories and
their implications to early universe cosmology and
the potential birth of stochastic GW background convolves multiple complicated stages.
One goal of this article is to take steps towards that, on theoretical grounds,
uncertainties related to the prediction of
the thermodynamics are not the largest in this pipeline.

Concretely,
we gave a fresh qualitative review of
the thermodynamics of the electroweak phase transition and
focused on scalar extensions of the SM.
In particular, we concentrated on the framework of high-temperature dimensional reduction.
As an automatic all-order resummation scheme it
perturbatively defers the infrared problem of thermal field theory.
Thereafter, the IR sensitive physics is encoded in a dimensionally reduced EFT
that can be studied non-perturbatively on the lattice.
However, the constructed 3d EFT is powerful already in perturbative studies.
The tutorial-styled computations explicated in sec.~\ref{se:tutorial}
aim to make this technique more accessible with
emphasis on a scientific community studying the thermodynamics of the EWPT
for a wide variety of BSM theories.

The majority of studies of the electroweak phase transition in BSM setups are restricted to
perturbative computations of the thermal effective potential.
They are often limited to
one-loop order and
naive leading order resummation usually of Arnold-Espinosa type.
Therefrom, a straightforward extension to a non-perturbative treatment is less apparent
and the IR problem remains in its core.
On perturbative grounds, important contributions in the weak coupling expansion
are missed which roots in a misalignment of loop and coupling (power counting) expansions.
This omission causes a residual, artificial RG scale dependence which cannot be
compensated by RG-improvement at one-loop.
A recent study~\cite{Croon:2020cgk} concludes that
this kind of leftover artificial renormalisation scale dependence can pose
a dramatic two to three orders-of-magnitude theoretical uncertainty for subsequent analyses of
the cosmic gravitational wave background originating from cosmic phase transitions.
Such an uncertainty for thermodynamic parameters can compromise predictions for
e.g.\
the signal-to-noise ratio for LISA and
other future GW experiments.

Tools to automate dimensional reduction are much needed to handle
large numbers of Feynman diagrams that arise at multi-loop orders.
By adopting sophisticated tool from zero temperature,
developments towards such automation have been taken recently~\cite{Schicho:2020xaf,Croon:2020cgk}.
As a concrete application of dimensional reduction,
we derived for the first time the high-temperature 3d EFT
of the real-singlet extended Standard Model (with a dynamical singlet) --
one of the most widely studied BSM models in particle cosmology.
This poses the main technical part of our investigation displayed in sec.~\ref{se:dr:xsm}.
Perturbative studies of this 3d EFT to scrutinise the EWPT in this model
are presented in
refs.~\cite{Niemi:2021xxx,Gould:2021xxx}.

We conclude by envisioning specific but also model-independent future avenues:
\begin{itemize}
\item[(i)]
  The derived 3d EFT of the xSM is indispensable for subsequent studies.
  Lattice simulations can probe its equilibrium thermodynamics 
  and in particular expose the character of the phase transition and
  determine parameter regions that admit SFOPT.
  Additionally, out-of-equilibrium properties of the phase transition such
  as bubble nucleation rate can be
  investigated by non-perturbative studies of the 3d EFT.

  A leftover gauge dependence indicates an incomplete basis of
  higher dimensional operators in the dimensionally reduced theory.
  In general, it is interesting how their effects influence the IR dynamics of
  the system~\cite{Laine:2018lgj}.

\item[(ii)]
  The real singlet scalar model (not coupled to SM)
  offers a testing platform for different approaches.
  Implications could be drawn for dark sector phase transitions,
  by
  determining the mapping of 4d parameters and temperature to its 3d phase structure.
  The latter was comprehensively analysed in ref.~\cite{Gould:2021dzl}.

  Since a real scalar theory is purely bosonic,
  it evades problems of
  discretising chiral fermions on lattice.
  Hence, a comparison between 3d EFT and full 4d lattice simulations is feasible (see related ref.~\cite{Jansen:1998rj})
  also when including higher dimensional operators.
  In this context, its lattice-continuum relations with higher dimensional operators
  are needed.
  It remains to be determined which lattice measurements and extrapolations are required
  to extract the continuum physics from simulations in presence
  of higher dimensional operators.
\item[(iii)]
  It would be worth investigating how large parameter space scans
  of past EWPT studies (using a one-loop thermal potential) are affected
  when complete $\mathcal{O}(g^4)$ effects are included.
  We advocate pre-existing software to implement
  the perturbative dimensionally reduced 3d EFT approach.
  An example are parameter space scans using
  {\tt CosmoTransitions}~\cite{Wainwright:2011kj},
  {\tt BSMPT}~\cite{Basler:2018cwe},
  {\tt PhaseTracer}~\cite{Athron:2020sbe}
  to examine the phase structure of individual BSM models.
\end{itemize}

%
\section*{Acknowledgements}
Over the years of this work, the authors would like to thank
Jens O.~Andersen,
Tom\'a\v s Brauner,
Djuna Croon,
Daniel Cutting,
Ioan Ghi{\c s}oiu,
Tyler Gorda,
Oliver Gould,
Mark Hindmarsh,
Keijo Kajantie,
Thomas Konstandin,
Jonathan Kozaczuk,
Mikko Laine,
Lauri Niemi,
Jose Miguel No,
Hiren Patel,
Michael J.~Ramsey-Musolf,
Arttu Rajantie,
Kari Rummukainen,
York Schr{\"o}der,
Anders Tranberg,
Aleksi Vuorinen,
Jorinde van de Vis,
David J.~Weir
and
Graham White
for enlightening discussions.
TT thanks Sara T{\"a}htinen for crosschecking many individual Feynman diagrams,
and correcting multiple related symmetry and combinatorial factors.
This work was partly supported by
the Swiss National Science Foundation (SNF) under grant 200020B-188712.
PS has been supported
by the European Research Council, grant no.~725369, and
by the Academy of Finland, grant no.~1322507.
J{\"O} acknowledges financial support from
the Vilho, Yrj{\"o} and Kalle V{\"a}is{\"a}l{\"a} Foundation of
the Finnish Academy of Science and Letters.

%
\appendix
\renewcommand{\thesection}{\Alph{section}}
\renewcommand{\thesubsection}{\Alph{section}.\arabic{subsection}}
\renewcommand{\theequation}{\Alph{section}.\arabic{equation}}

%
\section{Detailed computation in the xSM}
\label{se:dr:xsm:details}

This appendix collects details of
the dimensional reduction computation in the xSM from sec.~\ref{se:dr:xsm} and
extends our results of the 3d parameters to
marginal operators defined in eq.~\eqref{eq:lag:marginal}.

%
\subsection{Counterterms and $\beta$-functions of the 4d theory}
\label{se:4d:ct:beta}

One-loop counterterms and $\beta$-functions are listed e.g.\
in sec.~3.2 in ref.~\cite{Brauner:2016fla}.
Using field renormalisations
$Z_\phi$ for the Higgs,
$Z_q$ the left handed quark doublet and
$Z_t$ the top quark,
we define the bare top Yukawa parameter
\begin{align}
  g_{\rmii{$Y$}(b)} \equiv
    Z^{-\frac{1}{2}}_{\phi}
    Z^{-\frac{1}{2}}_{q}
    Z^{-\frac{1}{2}}_{t}
    \Lambda^\epsilon
    (\gY + \delta\gY)
    \;.
\end{align}
This convention for $\gY$ and its counterterm $\delta\gY$ align with
eqs.~(C.22) and (C.29) in ref.~\cite{Gorda:2018hvi} in contrast to
eqs.~(2.20) and (3.38) of ref.~\cite{Brauner:2016fla}.
Since these references use Landau gauge, we merely display
the $\xi$-dependent counterterms in general covariant gauge:
\begin{align}
\delta Z_\phi & = \frac{1}{(4\pi)^2} \frac{1}{\epsilon} \Big(
    \frac{3}{4}(3-\xi_2)g^2
  + \frac{1}{4}(3-\xi_1)\gp^2
  - \Nc \gY^{2}
  \Big)\
  \;, \\
\delta Z_A & = \frac{1}{(4\pi)^2} \frac{1}{\epsilon} g^2 \Big(
    \frac{25}{6}
  - \frac{1}{3}\nf(1+\Nc)
  - \xi_2
  \Big)
  \;, \\
\delta Z_B & = \frac{1}{(4\pi)^2} \frac{1}{\epsilon} \gp^2 \frac{1}{6}\Big(
    \Ys^{2}
  + Y_{\rmi{2f}}^{ }\,\nf^{ }
  \Big)
  \;, \\
\delta Z_q & = -\frac{1}{(4\pi)^2} \frac{1}{\epsilon} \Big(
    \frac{1}{2} \gY^{2}
  + \frac{3}{4}g^2 \xi_2
  + \frac{1}{4}\gp^2 \Yq^2 \xi_1
  + \CF^{ } \gs^2 \xi_3
  \Big)
  \;, \\
\delta Z_u & = -\frac{1}{(4\pi)^2} \frac{1}{\epsilon} \Big(
    \gY^{2}
  + \frac{1}{4}\gp^2 \Yu^2 \xi_1
  + \CF^{ } \gs^{2}\xi_3
  \Big)
  \;, \\
\delta\gY^{ } & = -\frac{1}{(4\pi)^2} \frac{1}{\epsilon} \Big(
    \frac{3\Yq\Yu}{4}\gp^2
  + \frac{3}{4} g^2 \xi_2
  + \frac{\Yq\Yu-\Ys(\Yq-\Yu)}{4}\gp^2 \xi_1
  + \CF^{ } \gs^{2}(3 + \xi_{3})
  \Big)
  \;, \\
\label{eq:ct:dlamh}
\delta\lambda_h^{ } & = \frac{1}{(4\pi)^2} \frac{1}{\epsilon} \Big(
    \frac{3}{16} (3g^4 + 2 g^2 \gp^2 +\gp^4 )
  - 3 \gY^{4}
  + 12 \lambda_h^{ }
  + \frac{1}{4} \lambda_{m}^{ }
  - \frac{1}{2} \lambda_{h}^{ } (3g^2 \xi_2 + \gp^2 \xi_1)
  \Big)
  \;, \\
\delta\mu_m^{ } & = \frac{1}{(4\pi)^2} \frac{1}{\epsilon} \Big(
    6 \lambda_h^{ } \mu_m^{ }
  + 2 \lambda_m^{ } (\mu_m^{ } + \mu_3^{ })
  - \frac{1}{4} \mu_m^{ } (3g^2 \xi_2 + \gp^2 \xi_1)
  \Big)
  \;, \\
\delta\lambda_m^{ } & = \frac{1}{(4\pi)^2} \frac{1}{\epsilon} \lambda_m^{ } \Big(
    6 \lambda_h^{ }
  + 2 \lambda_m^{ }
  + 3 \lambda_\sigma^{ }
  - \frac{1}{4} (3g^2 \xi_2 + \gp^2 \xi_1)
  \Big)
  \;,
\end{align}
where hypercharges are defined in eq.~\eqref{eq:hypercharge}.
Essentially also the unphysical gauge fixing parameter receives renormalisation:
$\xi_{(b)} = \xi(1+\delta Z_\xi)$
with
$\delta Z_{\xi_1} = \delta Z_{B}$ and
$\delta Z_{\xi_2} = \delta Z_{A}$.

The two-loop computation of tadpole and
mass parameters receives contributions that require new counterterms:
\begin{align}
\delta\mu_{1}^{ } &=
    \frac{1}{(4\pi)^2} \frac{1}{\epsilon} \Big(
        \mu_{h}^{2}\mu_{m}^{ }
      + \mu_{\sigma}^{2}\mu_{3}^{ }
  \Big)
  + \frac{1}{(4\pi)^4} \Big(\frac{1}{\epsilon^2} - \frac{1}{\epsilon}\Big)
  \Big(
      \frac{1}{8} \mu_{m}^{3}
    + \mu_{3}^{3}
    + \frac{1}{4} \mu_{3}^{ }\mu_{m}^{2} \Big)
  \;,\\
\delta\mu_{h}^{2} &=
    \frac{1}{(4\pi)^2} \frac{1}{\epsilon} \frac{1}{4} \Big(
        24 \lambda_{h}^{ }\mu_{h}^{2}
      + \mu_{m}^{2}
      + 2\lambda_{m}^{ }\mu_{\sigma}^{2}
      - \mu_{h}^{2}(\gp^{2}\xi_{1} + 3 g^{2}\xi_{2})
    \Big)
  \nn &
  - \frac{1}{(4\pi)^4} \frac{1}{\epsilon} \bigg(
      \frac{1}{2} \lambda_{m}^{ } (\mu_{3}^{2} + \mu_{3}^{ }\mu_{m}^{ })
    - \mu_{m}^{2} \Big(
        \frac{1}{32}(3g^2 + \gp^2)
      - \frac{\Nc}{8} \gY^{2}
      - \frac{9}{4} \lambda_{h}^{ }
      - \frac{3}{8} \lambda_{m}^{ }
  \Big) \bigg)
  \nn &
  + \frac{1}{(4\pi)^4} \frac{1}{\epsilon^2} \bigg(
    \frac{1}{2} \lambda_{m}^{ }(\mu_{3}^{2} + \mu_{3}^{ }\mu_{m}^{ })
  \nn &\hp{+\frac{1}{(4\pi)^4} \frac{1}{\epsilon^2}\bigg(}
    - \mu_{m}^{2} \Big(
        \frac{1}{32}(3 + 2\xi_{1}) \gp^2
      + \frac{3}{32}(3 + 2\xi_{2}) g^2
      - \frac{\Nc}{8} \gY^{2}
      - \frac{9}{4} \lambda_{h}^{ }
      - \frac{5}{8} \lambda_{m}^{ }
  \Big) \bigg)
  \;, \\
\delta\mu_{\sigma}^{2} &=
    \frac{1}{(4\pi)^2} \frac{1}{\epsilon} \frac{1}{2} \Big(
        6 \lambda_{\sigma}^{ }\mu_{\sigma}^{2}
      + 4 \lambda_{m}^{ }\mu_{h}^{2}
      + 4 \mu_{3}^{2}
      + \mu_{m}^{2}
    \Big)
  \nn &
  - \frac{1}{(4\pi)^4} \frac{1}{\epsilon} \bigg(
      15 \lambda_{\sigma}^{ }\mu_{3}^{2}
    + 2 \lambda_{m}^{ }\mu_{3}^{ }\mu_{m}^{ }
    - \mu_{m}^{2} \Big(
        \frac{1}{2}(3g^2 + \gp^2)
      - \frac{\Nc}{2} \gY^{2}
      - \frac{3}{4} \lambda_{\sigma}^{ }
      - \frac{5}{4} \lambda_{m}^{ } \Big) \bigg)
    \nn &
  + \frac{1}{(4\pi)^4} \frac{1}{\epsilon^2} \bigg(
      21 \lambda_{\sigma}^{ }\mu_{3}^{2}
    + 4 \lambda_{m}^{ }\mu_{3}^{ }\mu_{m}^{ }
    - \mu_{m}^{2} \Big(
        \frac{3}{8}(3g^2 + \gp^2)
      - \frac{\Nc}{2} \gY^{2}
      - \frac{3}{4} \lambda_{\sigma}^{ }
      - \frac{5}{4} \lambda_{m}^{ }
      - 3 \lambda_{h}^{ }
  \Big) \bigg)
  \;,
\end{align}
and their corresponding $\beta$-functions are
\begin{align}
\Lambda\frac{{\rm d}}{{\rm d}\Lambda} \mu_{1}^{ } &= \frac{1}{(4\pi)^2} 2 \Big(
    \mu_{h}^{2} \mu_{m}^{ }
    + \mu_{\sigma}^{2} \mu_{3}^{ }
  \Big)
  - \frac{1}{(4\pi)^4} \Big(
      \frac{1}{2} \mu_{m}^{3}
      + 4 \mu_{3}^{3}
      + \mu_{3}^{ } \mu_{m}^{2}
  \Big)
  \;, \\
\label{eq:muh:beta}
\Lambda\frac{{\rm d}}{{\rm d}\Lambda} \mu_{h}^{2} &= \frac{1}{(4\pi)^2} 2 \bigg(
      \frac{1}{4}\mu_{m}^{2}
    + \frac{1}{2}\lambda_{m}^{ } \mu_{\sigma}^{2}
    - \mu_{h}^{2} \Big(
        \frac{3}{4}(3g^2 +\gp^2)
      - \Nc\gY^{2}
      - 6\lambda_{h}^{ }
    \Big)
  \bigg)
  \nn &
  - \frac{1}{(4\pi)^4}\bigg(
      2 \lambda_{m}^{ }\mu_{3}^{2}
    + 2 \lambda_{m}^{ }\mu_{3}^{ }\mu_{m}^{ }
    - \mu_{m}^{2} \Big(
        \frac{1}{8}(3g^2 + \gp^2)
      - \frac{\Nc}{2} \gY^{2}
      - 9 \lambda_{h}^{ }
      - \frac{3}{2} \lambda_{m}^{ } \Big) \bigg)
  \;, \\
\Lambda\frac{{\rm d}}{{\rm d}\Lambda} \mu_{\sigma}^{2} &= \frac{1}{(4\pi)^2} 2 \bigg(
    2 \mu_{3}^{2}
  + \frac{1}{2}\mu_{m}^{2}
  + 3 \lambda_{\sigma}^{ }\mu_{\sigma}^{2}
  + 2 \lambda_{m}^{ } \mu_{h}^{2}
  \bigg)
  \nn &
  - \frac{1}{(4\pi)^4}2\bigg(
      30 \lambda_{\sigma}^{ }\mu_{3}^{2}
    + 4 \lambda_{m}^{ }\mu_{3}^{ }\mu_{m}^{ }
    - \mu_{m}^{2} \Big(
        (3g^2 + \gp^2)
      - \Nc\gY^{2}
      - \frac{3}{2}\lambda_{\sigma}^{ }
      - \frac{5}{2}\lambda_{m}^{ }
  \Big) \bigg)
  \;,
\end{align}
which are necessarily gauge-independent.
The remaining $\beta$-functions are listed in
sec.~3.2 of ref.~\cite{Brauner:2016fla}.

%
\subsection{Correlators from the one-loop effective potential in general covariant gauge}

In the background field method, the scalar fields are shifted by
$\phi_i\to\phi_i + \delta_{i,2} v/\sqrt{2}$ for $i=1,2$ and
$\sigma\to\sigma + s$, around real background fields $v$ and $s$.
We can read the scalar correlators from
the effective potential expanded in these background fields
\begin{align}
\label{eq:bg:expansion}
V_\rmii{eff} &=
  V_{0,0}
  + \frac{1}{2} V_{2,0}v^{2}
  + \frac{1}{4} V_{4,0}v^{4}
  +V_{0,1}s^{ }
  +V_{0,2}s^{2}
  \nn &
  +V_{0,3}s^{3}
  +V_{0,4}s^{4}
  +\frac{1}{2} V_{2,1} v^{2}s^{ }
  +\frac{1}{2} V_{2,2} v^{2}s^{2}
  \nn &
  +V_{0,5}s^5
  +\frac{1}{2} V_{2,3} v^{2}s^{3}
  +\frac{1}{4} V_{4,1} v^{4}s^{ }
  \nn &
  +\frac{1}{8} V_{6,0} v^{3}
  +V_{0,6}s^6
  +\frac{1}{4} V_{4,2} v^{4}s^{2}
  +\frac{1}{2} V_{2,4} v^{2}s^{4}
  \;,
\end{align}
up to dimension-6 terms.
In our convention, the coefficients $V$ relate to
correlators appearing in the matching relations of
eqs.~\eqref{eq:match:lh}--\eqref{eq:match:mu1} as
$$
\Gamma_{(\phi^\dagger \phi)^2} = 8 V_{4,0}\;,\quad
\Gamma_{\sigma^4} = 24 V_{0,4}\;,\quad
\Gamma_{\phi^\dagger \phi \sigma^2} = 4 V_{2,2}\;,\quad
\Pi_{\phi^\dagger \phi \sigma} = 2 V_{2,1}\;,\quad
\Pi_{\sigma^3} = 6 V_{4,0}\;,
$$ and
similarly for 1- and 2-point correlators and marginal operators.

In Landau gauge,
the background-dependent mass eigenvalues can be solved from the mass matrix constructed from
the coefficients of the bilinear parts of the $\phi_i$- and $\sigma$-fields
that mix in the broken phase.
By employing the shorthand notation for the parameters of the shifted theory
\begin{align}
\tilde\mu_{h}^{2} &\equiv
    \mu_{h}^{2}
  + \lambda_{h}^{ } v
  + \frac{1}{2}\mu_{m}^{ } s
  + \frac{1}{2}\lambda_{m}^{ } s^2
  \;, \\
\tilde\mu_{\sigma}^{2} &\equiv
    \mu_{\sigma}^{2}
  + 2\mu_{3}^{ } s
  + 3\lambda_{\sigma}^{ } s^2
  + \frac{1}{2}\lambda_{m}^{ } v
  \;, \\
\tilde\mu_{3}^{ } &\equiv
    \mu_{3}^{ }
  + 3\lambda_{\sigma}^{ } s
  \;,\\
\tilde\mu_m &\equiv
    \mu_{m}^{ }
    + 2\lambda_{m}^{ } s
  \;,
\end{align}
the scalar mass eigenvalues read
\begin{align}
m^2_G &= \tilde\mu_h^2
  \;, \\
m^2_\pm &=
    \frac{1}{2}(\tilde\mu_h^2 + \tilde\mu_\sigma^2)
  + \lambda_{h}^{ } v^2
  \pm \sqrt{
    \Big( \frac{1}{2}(-\tilde\mu_h^2 + \tilde\mu_\sigma^2) - \lambda_{h}^{ } v^2 \Big)^2
  + v^2 \Big( \frac{1}{2} \mu_{m}^{ } + \lambda_{m}^{ } s \Big)^2}
  \;,
\end{align}
where the Goldstone mass eigenvalue $m^2_G$ is triple degenerate.
Since the singlet does not couple to the gauge fields or top quark,
their mass eigenvalues align with the SM
\begin{align}
m^2_{W^{\pm}} &= \frac{1}{4} g^2 v^2
  \;, &
m^2_{Z} &= \frac{1}{4} (g^2+\gp^2) v^2
  \;, \\
m^2_{\gamma} &= 0
  \;, &
m^2_{t} &= \frac{1}{2} \gY^{2} v^2
\;.
\end{align}
In general covariant gauge,
the three Goldstone mass eigenvalues are replaced by~\cite{Andreassen:2013hpa,Laine:1994bf}
\begin{align}
{m^\pm_1}^2 &= \frac{1}{2} \bigg(m^2_G \pm \sqrt{m^2_G \Big(m^2_G - \xi_2 g^2 v^2 - \xi_1 \gp^2 v^2 \Big) } \bigg)
\;, \\
{m^\pm_2}^2 &= \frac{1}{2} \bigg(m^2_G \pm \sqrt{m^2_G \Big(m^2_G - \xi_2 g^2 v^2 \Big) } \bigg)
\;,
\end{align}
where ${m^\pm_2}^2$ is double degenerate.
Based on these background field dependent mass eigenvalues,
the one-loop effective potential becomes
\begin{align}
V^{\rmii{1loop}}_\rmii{eff} &=
    J_b(m_{1}^{+})
  + J_b(m_{1}^{-})
  + 2J_b(m_{2}^{+})
  + 2J_b(m_{2}^{-})
  + J_b(m_{+}^{ })
  + J_b(m_{-}^{ })
  \nn &
  + d \Big( 2 J_b(m_{W}^{\pm}) + J_b(m_{Z}^{ }) \Big)
  - 4 \Nc J_f(m_{t}^{ })
  \;.
\end{align}
And by comparing to the expansion~\eqref{eq:bg:expansion},
we can solve for the desired correlators.
For a crosscheck, we also compute all one-loop correlators directly in the unbroken phase.

%
\subsection{Matching of marginal operators}
\label{se:marginal}

Marginal operators of eq.~\eqref{eq:lag:marginal} arise at
$\mathcal{O}(g^5)$ and $\mathcal{O}(g^6)$,
at which contributions from the field normalisations are absent in their matching.
We get
\begin{align}
\label{eq:c60}
c_{6,0}&=  T^2 \bigg[
    \frac{\zeta_{3}}{(4\pi)^{4}T^2} \Big(
        \frac{3}{8} g^6
      + \frac{3}{8} g^4 \gp^2
      + \frac{3}{8} g^2 \gp^4
      + \frac{1}{8}\gp^6
      - \frac{28}{3}\Nc\,\gY^6
      + 80 \lambda_{h}^{3}
      + \frac{1}{3} \lambda_{m}^{3}
    \nn &\hp{T^2 \bigg[\frac{\zeta_{3}}{(4\pi)^{4}T^2}\Big(}
    - 2\lambda_{h}^{2} (3g^2 \xi_2 + \gp^2 \xi_1)
  \Big)
  \nn &\hp{=T^2\bigg[}
  - \frac{\zeta_{5}}{(4\pi)^{6}T^4} \mu_{m}^{2} \Big(
      36 \lambda_{h}^{2}
    + 6 \lambda_{h}^{ }\lambda_{m}^{ }
    + \lambda_{m}^{2} \Big)
  \nn &\hp{=T^2\bigg[}
  + \frac{\zeta_{7}}{(4\pi)^{8 }T^6} \frac{5}{4}(6\lambda_{h}^{ } + \lambda_{m}^{ })\mu_{m}^{4}
  - \frac{\zeta_{9}}{(4\pi)^{10}T^8} \frac{7}{12}\mu_{m}^{6}
  \bigg]
  \;, \\[3mm]
c_{0,6}&= T^2 \bigg[
    \frac{\zeta_{3}}{(4\pi)^{4}T^2} \Big( \frac{1}{6} \lambda_{m}^{3} + 9 \lambda_{\sigma}^{3} \Big)
  - \frac{\zeta_{5}}{(4\pi)^{6}T^4} \frac{3}{4} \Big( 144 \lambda_{\sigma}^{2}\mu_{3}^{2} + \lambda_{m}^{2}\mu_{m}^{2} \Big)
  \nn &\hp{=T^2\bigg[}
  + \frac{\zeta_{7}}{(4\pi)^{8 }T^6} \Big( 240\lambda_{\sigma}\mu_{3}^{4} + \frac{5}{8}\lambda_{m}^{ }\mu_{m}^{4}\Big)
  - \frac{\zeta_{9}}{(4\pi)^{10}T^8} \frac{7}{48} \Big( 1024\mu_{3}^{6} + \mu_{m}^{6} \Big)
  \bigg]
  \;, \\[3mm]
c_{4,2}&= T^2 \bigg[
    \frac{\zeta_{3}}{(4\pi)^{4}T^2}  \lambda_m \bigg(
      24 \lambda_{h}^{2}
    + 2 \lambda_{m}^{2}
    + \lambda_{m}^{ } (12 \lambda_{h}^{ } + 3 \lambda_{\sigma}^{ })
    - \lambda_{h}^{ } (3g^2 \xi_2 + \gp^2 \xi_1)
  \bigg)
  \nn &\hp{=T^2\bigg[}
  - \frac{\zeta_{5}}{(4\pi)^{6}T^4} \bigg(
      6 \lambda_{m}^{ }\mu_{m}^{ } \Big(
        8\lambda_{h}^{ }\mu_{3}^{ }
      + 5\lambda_{h}^{ }\mu_{m}^{ }
      + \lambda_{\sigma}^{ }\mu_{m}^{ }
    \Big)
    + \lambda_{m}^{2} \Big( 12 \mu_{3}^{2} + 16 \mu_{3}^{ }\mu_{m}^{ } + \frac{17}{2} \mu_{m}^{2} \Big)
    \nn &\hp{=T^2\bigg[}
    + \mu_{m}^{2} \Big(
      36 \lambda_{h}^{2}
      + 18 \lambda_{h}^{ } \lambda_{\sigma}^{ }
      - \frac{3}{2}\lambda_{h}^{ } (3g^2 \xi_2 + \gp^2 \xi_1)
      + \frac{1}{16}(3 g^4 \xi^2_2 + 2 g^2 \gp^2 \xi_2 \xi_1 + \gp^4 \xi^2_1)
    \Big)\bigg)
  \nn &\hp{=T^2\bigg[}
  + \frac{\zeta_{7}}{(4\pi)^{8 }T^6} \frac{5}{4}\mu_{m}^{2}\Big(
      24(2\lambda_{h}^{ } + \lambda_{m}^{ })\mu_{3}^{2}
    + 4(6\lambda_{h}^{ } + 5\lambda_{m}^{ })\mu_{3}^{ }\mu_{m}^{ }
    + (9\lambda_{h}^{ } + 5\lambda_{m}^{ } + 3\lambda_{\sigma}^{ })\mu_{m}^{2}
  \Big)
  \nn &\hp{=T^2\bigg[}
  - \frac{\zeta_{9}}{(4\pi)^{10}T^8} \frac{7}{16}\mu_{m}^{4}\Big(
      48\mu_{3}^{2}
    + 16\mu_{3}^{ }\mu_{3}^{ }
    + 3\mu_{m}^{2}
  \Big)
  \bigg]
  \;, \\[3mm]
c_{2,4}&= \bigg[
    \frac{\zeta_{3}}{(4\pi)^{4}T^2} \Big(
        \lambda_{m}^{3}
      + 9 \lambda_{m}^{ } \lambda_{\sigma}^{2}
      + 3 \lambda_{m}^{2} (\lambda_{h}^{ } + 2\lambda_{\sigma}^{ })
      - \frac{1}{8} \lambda_{m}^{2} (3g^2 \xi_2 + \gp^2 \xi_1)
    \Big)
  \nn &\hp{\bigg[}
  - \frac{\zeta_{5}}{(4\pi)^{6}T^4} \Big(
      9 \lambda_{\sigma}^{2} \mu_{m}^{2}
    + \frac{1}{4} \lambda_{m}^{2} \Big( 64 \mu_{3}^{2} + 32 \mu_3^{ } \mu_m^{ } + 13 \mu_{m}^{2}
    \Big)
    \nn &\hp{\bigg[-\frac{\zeta_{5}}{(4\pi)^{6}T^4}\Big(}
    + \frac{3}{8} \lambda_m \Big(
        4 \lambda_\sigma^{ } (4 \mu_3^{ } + \mu_m^{ })(12 \mu_3^{ } + 5 \mu_m^{ })
      + 24 \mu_{m}^{2} \lambda_h^{ } \Big)
    - \frac{3}{8} \lambda_m^{ } (3g^2 \xi_2 + \gp^2 \xi_1)
  \Big)
  \nn &\hp{\bigg[}
  + \frac{\zeta_{7}}{(4\pi)^{8 }T^6} \frac{5}{32}\mu_{m}^{2}\Big(
      12\lambda_{\sigma}^{ }\mu_{m}^{2}(48\mu_{3}^{3} + 8\mu_{3}^{ }\mu_{m}^{ } + \mu_{m}^{2})
    \nn[2mm] &\hp{\bigg[+\frac{\zeta_{7}}{(4\pi)^{8}T^{6}}\frac{5}{32}\mu_{m}^{2}\Big(}
    + 2\lambda_{m}^{ }(
        256\mu_{3}^{4}
      + 256\mu_{3}^{3}\mu_{m}^{ }
      + 80\mu_{3}^{2}\mu_{m}^{2}
      + 24\mu_{3}^{ }\mu_{m}^{3}
      + 7\mu_{m}^{4}
      )
    \nn[2mm] &\hp{\bigg[+\frac{\zeta_{7}}{(4\pi)^{8}T^{6}}\frac{5}{32}\mu_{m}^{2}\bigg(}
    + 24\lambda_{h}\mu_{m}^{4}
    - \mu_{m}^{4}(\gp^{2}\xi_{1} + g^{2}\xi_{2})
  \Big)
  \nn &\hp{\bigg[}
  - \frac{\zeta_{9}}{(4\pi)^{10}T^8} \frac{7}{16}\mu_{m}^{2}\Big(
      256\mu_{3}^{4}
    + 64\mu_{3}^{3}\mu_{m}^{ }
    + 16\mu_{3}^{2}\mu_{m}^{2}
    + 4\mu_{3}^{ }\mu_{m}^{3}
    + \mu_{m}^{4}
  \Big)
  \bigg]
  \;, \\[3mm]
c_{0,5}&= T^{\frac{3}{2}} \bigg[
    \frac{\zeta_{3}}{(4\pi)^{4}T^2} \Big(18 \lambda_{\sigma}^{2}\mu_{3}^{ } + \frac{1}{2} \lambda_{m}^{2}\mu_{m}^{ } \Big)
  - \frac{\zeta_{5}}{(4\pi)^{6}T^4} \Big(48 \lambda_{\sigma}^{ }\mu_{3}^{3} + \frac{1}{2} \lambda_{m}^{ }\mu_{m}^{3} \Big)
  \nn &\hp{=T^{\frac{3}{2}}\bigg[}
  + \frac{\zeta_{7}}{(4\pi)^{8}T^6} \Big(32 \mu_{3}^{5} + \frac{1}{8}\mu_{m}^{5} \Big)
  \bigg]
  \;, \\[3mm]
c_{2,3}&= T^{\frac{3}{2}} \bigg[
    \frac{\zeta_{3}}{(4\pi)^4 T^2} \frac{\lambda_m}{4} \Big(
        8 (\lambda_m^{ }+3 \lambda_\sigma^{ }) (2 \mu_3^{ } + \mu_m^{ })
      + 24 \lambda_h^{ } \mu_m^{ }
      - \mu_m (3g^2 \xi_2 + \gp^2 \xi_1)
    \Big)
    \nn &\hp{=T^{\frac{3}{2}}\bigg[}
    - \frac{\zeta_{5}}{(4\pi)^6 T^4} \frac{1}{8} \Big(
        4 \lambda_m^{ } (4 \mu_3^{ } + 3 \mu_m^{ } ) (8 \mu_{3}^{2} + 2\mu_3^{ } \mu_m^{ } + \mu_{m}^{2})
    \nn &\hp{=T^{\frac{3}{2}}\bigg[-\frac{\zeta_{5}}{(4\pi)^6 T^4}\frac{1}{8}\Big(}
    + 12\mu_{m}^{2} \Big(
        \lambda_\sigma^{ } (8 \mu_3^{ } + \mu_m^{ })
      + 2\mu_m^{ } \lambda_h^{ }
    \Big)
    - \mu_{m}^{3} (3g^2 \xi_2 + \gp^2 \xi_1)
    \Big)
    \nn &\hp{=T^{\frac{3}{2}}\bigg[}
  + \frac{\zeta_{7}}{(4\pi)^8 T^6} \frac{5}{16} \mu_{m}^2 (4 \mu_3^{ } + \mu_m^{ } ) (16 \mu_{3}^{2} + \mu_{m}^2)
  \bigg]
  \;,\\[3mm]
c_{4,1}&= T^{\frac{3}{2}} \bigg[
    \frac{\zeta_{3}}{(4\pi)^{4}T^2} \Big(
        12 \lambda_h^{ } \lambda_m^{ } \mu_m^{ }
      + 2 \lambda_{m}^{2} (\mu_3^{ } + \mu_m^{ })
      + 24 \lambda_{h}^{2} \mu_m^{ }
      - \lambda_h^{ } \mu_m^{ } (3g^2 \xi_2 + \gp^2 \xi_1)
    \Big)
    \nn &\hp{=T^{\frac{3}{2}}\bigg[}
    - \frac{\zeta_{5}}{(4\pi)^{6}T^4} \frac{1}{2}\mu_{m}^{2} \Big(
        12 \lambda_h^{ } (2\mu_3^{ } + \mu_m^{ })
      + \lambda_m^{ } (8 \mu_3^{ } + 5 \mu_m^{ })
    \Big)
    \nn &\hp{=T^{\frac{3}{2}}\bigg[}
    + \frac{\zeta_{7}}{(4\pi)^{8}T^6} \frac{5}{8}\mu_{m}^{4} \Big(
        4\mu_{3}^{ }
      + \mu_{m}^{ }
    \Big)
  \bigg]\;.
\end{align}
Notably, all coefficients related to operators with Higgs field are {\em gauge dependent},
in analogy to the Higgs-singlet portal coupling in eq.~\eqref{eq:lamm}.
The class of topologies responsible for this uncancelled contribution
can be exemplified by the
Higgs-singlet portal interaction and
sextic Higgs marginal operator
\begin{align}
  \TopoVD(fex(\Lsc1,\Lcs1,\Lsr1,\Lsr1),\Agl2,\Asc1,\Asc1,\Asc1) &\simeq
  (3g^2\xi_2+\gp^2 \xi_1) \mu_{m}^{2}
  \;, \\[2mm]
  \TopoSD(fex(\Lcs1,\Lsc1,\Lsc1,\Lcs1,\Lsc1,\Lcs1),\Agl2,\Asc1,\Asc1,\Asc1) &\simeq
  (3g^2 \xi_2 + \gp^2 \xi_1)\lambda_{h}^{2}
  \;.
\end{align}
At zero external momenta the first diagram
vanishes only in Landau gauge due its transversality,
since it is proportional to
$\Tinti{P} P_\mu P_\nu D_{\mu\nu}(P)/P^{8} =
\xi \Tinti{P} 1/P^6$.
Where $D_{\mu\nu}(P)$ is the gauge field propagator in covariant gauge from
eq.~\eqref{eq:prop:gauge}.
Identically, the second diagram contributes to the Higgs 6-point correlator~\eqref{eq:c60}
where also a leftover gauge dependence remains.
Similarly, other marginal operators with external Higgs legs are $\xi$-dependent.
This leftover $\rmO(g^6)$ gauge dependence of the sextic correlator
was pointed out in ref.~\cite{Croon:2020cgk}.
Strikingly, the gauge dependence of
the Higgs-singlet portal coupling $\lambda_{m}$
arises already at $\rmO(g^4)$.
This underlines the subtlety of the power counting of the cubic portal coupling $\mu_{m}$
as discussed in sec.~\ref{se:dr:xsm}.

Higher dimensional operators can be used to estimate the accuracy of
the dimensional reduction by adding their effect at tree-level to 3d effective potential.
It remains to be understood how a gauge-invariant analysis is to be performed
as some of these operators are explicitly gauge-dependent.
We leave this endeavour as a future challenge, and note that at this stage
these operators can be used as mere numerical estimates of
the convergence of perturbation theory.

%
\subsection{Two-loop computation of correlators}
\label{se:cor:2l}

This appendix documents diagram-by-diagram the results for
the two-loop correlators used in sec.~\ref{se:hard:soft}
in terms of master sum-integrals of appendix~\ref{se:integrals}.
Despite being an algorithmic loop-diagrammatic exercise,
we believe that this explicit documentation can facilitate
future endeavours of dimensionally reduced high-$T$ effective theories.
Especially as one might find this to be
the non-trivial part of a dimensional reduction computation.
The computation was performed in general covariant gauge,
where gauge parameters enter via gauge field propagators
for SU(2)
\begin{align}
\label{eq:prop:gauge}
\langle A^{a}_{\mu}(P) A^{b}_{\nu}(-P) \rangle &=
  \delta^{ab} D_{\mu\nu}(P)
  \nn &=
  \frac{\delta^{ab}}{P^2} \Big( \mathcal{P}^T_{\mu\nu}(P) + \xi \frac{P_\mu P_\nu}{P^2} \Big)
\;,
\end{align}
and similarly for other gauge propagators of U(1) and SU(3).
The transverse projector is defined as
$\mathcal{P}^T_{\mu\nu}(P) \equiv \delta_{\mu\nu} - P_\mu P_\nu/P^2$.
This section displays results compactly employing Landau gauge ($\xi = 0$) where
gauge propagators are transversal.
This transversality decimates the number of integrals.
For the Higgs self-energy, we only list new contributions of the singlet scalar.

The corresponding Feynman rules and conventions (in Landau gauge) are outlined in
ref.~\cite{Brauner:2016fla}.
The pure singlet diagrams are listed above in sec.~\ref{se:cor:singlet},
wherefore here we only include diagrams wherein the singlet couples to the SM.
Note, that in this case also the pure singlet counterterm diagrams contains
SM contributions, and
similarly the Higgs counterterms have singlet contributions.
Again the correlator is minus sum of Feynman diagrams, and
at two-loop level (two-loop diagrams, one-loop counterterm diagrams)
we employ massless propagators sufficient for a NLO dimensional reduction;
see discussion at the end of sec.~\ref{se:cor:singlet}.

%
\subsubsection*{Singlet tadpole}

Diagrammatically, the renormalised singlet tadpole correlator $\hat\Gamma_{\sigma}$ in Landau gauge
composes of (excluding pure singlet terms
eqs.~\eqref{eq:sig:1pt:2l:a}--\eqref{eq:sig:1pt:2l:g})
\begin{multicols}{2}
\noindent
\begin{align}
\TopoOT(\Lsr1,\Acs1) &=
-\mu_m^{ } \Big( I^{4b}_1 - \mu_{h}^{2} I^{4b}_2 \Big)
\;, \\[3mm]
\ToptOE(\Lsr1,\Asc1,\Asc1,\Acs1) &=
6 \lambda_h^{ } \mu_m^{ } I^{4b}_2 I^{4b}_1
\;, \\[3mm]
\ToptOE(\Lsr1,\Asc1,\Asc1,\Asr1) &=
\frac{1}{2} \lambda_m^{ } \mu_m^{ } I^{4b}_2 I^{4b}_1
\;, \\[3mm]
\ToptOE(\Lsr1,\Asr1,\Asr1,\Acs1) &=
2\lambda_m^{ }\mu_3^{ } I^{4b}_2 I^{4b}_1
\;, \\[3mm]
\ToptOE(\Lsr1,\Acs1,\Acs1,\Agl1) &+
\ToptOE(\Lsr1,\Acs1,\Acs1,\Agl2) =
\nn &
+ \frac{1}{4} \mu_m^{ } (3g^3 + \gp^2) d I^{4b}_2 I^{4b}_1
\;, \\[3mm]
\ToptOM(\Lsr1,\Asc1,\Aqu,\Asc1,\Luqu) &=
-2\mu_m^{ } \Nc^{ } \gY^{2} F_5
\;, \\[3mm]
\ToptOM(\Lsr1,\Acs1,\Agl1,\Acs1,\Lcs1) &+
\ToptOM(\Lsr1,\Acs1,\Agl2,\Acs1,\Lcs1) =
\nn &
-\frac{1}{4} \mu_m^{ } (3g^3 + \gp^2) B_{4}
\;, \\[3mm]
\ToptOM(\Lsr1,\Asr1,\Asc1,\Asr1,\Lcs1) &=
-\frac{1}{2} \mu_{m}^{2} \mu_3^{ } S_4
\;, \\[3mm]
\ToptOM(\Lsr1,\Asc1,\Asc1,\Asc1,\Lsr1) &=
-\frac{1}{4} \mu_{m}^{3} S_4
\;, \\[3mm]
\ToptOS(\Lsr1,\Asc1,\Asc1,\Lsr1) &=
\mu_m^{ } \lambda_m^{ } S_3
\;, \\[3mm]
\TopoOTa(\Lsr1,\Asc1) &=
-\delta\mu_m^{ } I^{4b}_1
\;, \\[3mm]
\TopoOTx(\Lsr1,\Asc1) &=
\mu_m^{ } \Big(\delta Z_\phi I^{4b}_1 + \delta\mu_{h}^{2} I^{4b}_2 \Big)
\;.
\end{align}
\end{multicols}

%
\subsubsection*{Singlet self-energy}

The renormalised singlet self-energy $\hat\Pi_{\sigma\sigma}$
in Landau gauge composes of (excluding pure singlet terms
eqs.~\eqref{eq:sig:2pt:2l:a}--\eqref{eq:sig:2pt:2l:o})
\begin{multicols}{2}%
\noindent
\begin{align}
\TopoST(\Lsr1,\Acs1) &=
-2\lambda_m^{ } \Big( I^{4b}_1 - \mu_{h}^{2} I^{4b}_2 \Big)
\;, \\[3mm]
\TopoSB(\Lsr1,\Acs1,\Acs1) &=
\frac{1}{2}\mu_{m}^{2} \Big( I^{4b}_2 - \mu_{h}^{2} I^{4b}_3 \Big)
\;, \\[4mm]
\ToptSTT(\Lsr1,\Asc1,\Asc1,\Acs1) &=
12\lambda_h^{ } \lambda_m^{ } I^{4b}_2 I^{4b}_1
\;, \\[4mm]
\ToptSTT(\Lsr1,\Asc1,\Asc1,\Asr1) &=
\lambda_{m}^{2} I^{4b}_2 I^{4b}_1
\;, \\[4mm]
\ToptSTT(\Lsr1,\Asr1,\Asr1,\Acs1) &=
6 \lambda_\sigma^{ } \lambda_m^{ } I^{4b}_2 I^{4b}_1
\;, \\[4mm]
\ToptSTT(\Lsr1,\Acs1,\Acs1,\Agl1) &+
\ToptSTT(\Lsr1,\Acs1,\Acs1,\Agl2) =
\nn &
+ \frac{1}{2} \lambda_m^{ } (3g^2 + \gp^2) d\, I^{4b}_2 I^{4b}_1
\;, \\[3mm]
\ToptSBT(\Lsr1,\Asc1,\Asc1,\Asc1,\Acs1) &=
-6\mu_{m}^{2} \lambda_h^{ } I^{4b}_3 I^{4b}_1
\;, \\[4mm]
\ToptSBT(\Lsr1,\Asc1,\Asc1,\Asc1,\Asr1) &=
-\frac{1}{2} \mu_{m}^{2} \lambda_m^{ } I^{4b}_3 I^{4b}_1
\;, \\[4mm]
\ToptSBT(\Lsr1,\Asr1,\Asr1,\Asr1,\Acs1) &=
-8 \mu_{3}^{2} \lambda_m^{ } I^{4b}_3 I^{4b}_1
\;, \\[4mm]
\ToptSBT(\Lsr1,\Asc1,\Asc1,\Asc1,\Agl1) &+
\ToptSBT(\Lsr1,\Asc1,\Asc1,\Asc1,\Agl2) =
\nn &
-\frac{1}{4} \mu_{m}^{2} (3g^2 + \gp^2) d\, I^{4b}_3 I^{4b}_1
\;, \\[3mm]
\ToptSal(\Lsr1,\Asr1,\Asc1,\Asr1,\Asc1) &=
-4\mu_m^{ } \mu_3^{ } \lambda_m^{ } S_4
\;, \\[3mm]
\ToptSal(\Lsr1,\Asc1,\Asr1,\Asc1,\Acs1) &=
-2\mu_{m}^{2} \lambda_m^{ } S_4
\;, \\[3mm]
\ToptSM(\Lsr1,\Asc1,\Asc1,\Asc1,\Asc1,\Lsr1) &=
\frac{1}{8} \mu_{m}^{4} S_5
\;, \\[3mm]
\ToptSM(\Lsr1,\Asc1,\Asr1,\Asr1,\Asc1,\Lsc1) &=
\mu_{m}^{3} \mu_3^{ } S_5
\;, \\[3mm]
\ToptSM(\Lsr1,\Acs1,\Acs1,\Acs1,\Acs1,\Lgl1) &+
\ToptSM(\Lsr1,\Acs1,\Acs1,\Acs1,\Acs1,\Lgl2) =
\nn &
+ \frac{1}{8} \mu_{m}^{2} (3g^2 + \gp^2) B_{10}
\;, \\[3mm]
\ToptSBB(\Lsr1,\Acs1,\Acs1,\Acs1,\Aqu,\Aquu) &=
2 \Nc^{ } \mu_{m}^{2} \gY^{2} F_6
\;, \\[3mm]
\ToptSBB(\Lsr1,\Acs1,\Acs1,\Acs1,\Asr1,\Asc1) &=
\frac{1}{4} \mu_{m}^{4} S_6
\;, \\[3mm]
\ToptSBB(\Lsr1,\Asr1,\Asr1,\Asr1,\Asc1,\Asc1) &=
2 \mu_{m}^{2} \mu_{3}^{2} S_6
\;, \\[3mm]
\ToptSBB(\Lsr1,\Asc1,\Asc1,\Asc1,\Agl1,\Acs1) &+
\ToptSBB(\Lsr1,\Asc1,\Asc1,\Asc1,\Agl2,\Acs1) =
\nn &
+ \frac{1}{4} \mu_{m}^{2} (3g^2 + \gp^2) B_8
\;, \\[3mm]
\ToptSE(\Lsr1,\Asc1,\Acs1,\Acs1,\Asc1) &=
-3 \mu_{m}^{2} \lambda_h^{ } I^{4b}_2 I^{4b}_2
\;, \\[3mm]
\ToptSE(\Lsr1,\Asc1,\Asr1,\Asr1,\Asc1) &=
-2 \mu_m^{ } \mu_3^{ } \lambda_m^{ } I^{4b}_2 I^{4b}_2
\;, \\[3mm]
\ToptSTB(\Lsr1,\Asc1,\Asc1,\Auq,\Auqu) &=
-4 \Nc^{ } \lambda_m^{ } \gY^{2} F_5
\;, \\[3mm]
\ToptSTB(\Lsr1,\Asc1,\Asc1,\Asr1,\Acs1) &=
-\frac{1}{2} \mu_{m}^{2} \lambda_m^{ } S_4
\;, \\[3mm]
\ToptSTB(\Lsr1,\Asr1,\Asr1,\Asc1,\Asc1) &=
-\frac{3}{2} \mu_{m}^{2} \lambda_\sigma S_4
\;, \\[3mm]
\ToptSTB(\Lsr1,\Acs1,\Acs1,\Agl1,\Asc1) &+
\ToptSTB(\Lsr1,\Acs1,\Acs1,\Agl2,\Asc1) =
\nn &
-\frac{1}{2}\lambda_m^{ } (3g^2 + \gp^2) B_4
\;, \\[3mm]
\ToptSS(\Lsr1,\Asr1,\Asc1,\Lsc1) &=
2\lambda_{m}^{2} S_3
\;, \\[3mm]
\TopoSTx(\Lsr1,\Asc1) &=
2\lambda_m^{ } \Big( \delta Z_\phi I^{4b}_1 + \delta\mu_{h}^{2} I^{4b}_2 \Big)
\;, \\[3mm]
\TopoSTc(\Lsr1,\Asc1) &=
-2\delta \lambda I^{4b}_1
\;, \\[3mm]
\TopoSBx(\Lsr1,\Asc1,\Asc1) &=
-\mu_{m}^{2} \Big( \delta Z_\phi I^{4b}_2 + \delta\mu_{h}^{2} I^{4b}_3 \Big)
\;, \\[3mm]
\TopoSBa(\Lsr1,\Asc1,\Asc1) &=
\mu_m^{ } \delta\mu_m^{ } I^{4b}_2
\;.
\end{align}
\end{multicols}

%
\subsubsection*{Higgs self-energy}

The singlet contributions to the renormalised SM Higgs doublet self-energy
$\hat\Pi_{\phi^\dagger \phi}$
(in Landau gauge) read
\begin{multicols}{2}%
\noindent
\begin{align}
\TopoST(\Lsc1,\Asr1) &=
-\frac{1}{2}\lambda_m^{ } \Big( I^{4b}_1 - \mu_{\sigma}^{2} I^{4b}_2 \Big)
\;, \\[3mm]
\TopoSB(\Lsc1,\Asr1,\Asc1) &=
\frac{1}{4} \mu_{m}^{2} \Big( I^{4b}_2 - \mu_{\sigma}^{2} I^{4b}_3 \Big)
\;, \\[4mm]
\ToptSTT(\Lsc1,\Acs1,\Acs1,\Asr1) &=
3\lambda_h^{ } \lambda_m^{ } I^{4b}_2 I^{4b}_1
\;, \\[4mm]
\ToptSTT(\Lsc1,\Asr1,\Asr1,\Acs1) &=
\lambda_{m}^{2} I^{4b}_2 I^{4b}_1
\;, \\[4mm]
\ToptSTT(\Lsc1,\Asr1,\Asr1,\Asr1) &=
\frac{3}{2} \lambda_\sigma^{ } \lambda_m^{ } I^{4b}_2 I^{4b}_1
\;, \\[4mm]
\ToptSBT(\Lsc1,\Acs1,\Acs1,\Asr1,\Acs1) &=
-\frac{3}{2} \mu_{m}^{2} \lambda_h^{ } I^{4b}_3 I^{4b}_1
\;, \\[4mm]
\ToptSBT(\Lsc1,\Asr1,\Asr1,\Asc1,\Acs1) &=
-\frac{1}{2} \mu_{m}^{2} \lambda_m^{ } I^{4b}_3 I^{4b}_1
\;, \\[4mm]
\ToptSBT(\Lsc1,\Acs1,\Acs1,\Asr1,\Asr1) &=
-\frac{1}{8} \mu_{m}^{2} \lambda_m^{ } I^{4b}_3 I^{4b}_1
\;, \\[4mm]
\ToptSBT(\Lsc1,\Asr1,\Asr1,\Asc1,\Asr1) &=
-\frac{3}{4} \mu_{m}^{2} \lambda_\sigma^{ } I^{4b}_3 I^{4b}_1
\;, \\[4mm]
\ToptSBT(\Lsc1,\Acs1,\Acs1,\Asr1,\Agl1) &+
\ToptSBT(\Lsc1,\Acs1,\Acs1,\Asr1,\Agl2) =
\nn &
- \frac{1}{16} \mu_{m}^{2} (3g^2 + \gp^2) d\, I^{4b}_3 I^{4b}_1
\;, \\[3mm]
\ToptSal(\Lsc1,\Asr1,\Asc1,\Asc1,\Asc1) &=
-3\mu_{m}^{2} \lambda_h^{ } S_4
\;, \\[3mm]
\ToptSal(\Lsc1,\Acs1,\Asr1,\Asr1,\Asc1) &=
-\frac{1}{2} \mu_{m}^{2} \lambda_m^{ } S_4
\;, \\[3mm]
\ToptSal(\Lsc1,\Asr1,\Asr1,\Asc1,\Asr1) &=
-\mu_m^{ } \mu_3^{ } \lambda_m^{ } S_4
\;, \\[3mm]
\ToptSM(\Lsc1,\Asr1,\Acs1,\Asr1,\Asc1,\Lsc1) &=
\frac{1}{16} \mu_{m}^{4} S_5
\;, \\[3mm]
\ToptSM(\Lsc1,\Asr1,\Asr1,\Asc1,\Asc1,\Lsr1) &=
\frac{1}{4} \mu_{m}^{3} \mu_3^{ } S_5
\;, \\[3mm]
\ToptSBB(\Lsc1,\Acs1,\Acs1,\Asr1,\Aqu,\Aquu) &=
\frac{1}{2} \Nc^{ } \mu_{m}^{2} \gY^{2} F_6
\;, \\[3mm]
\ToptSBB(\Lsc1,\Asr1,\Asr1,\Asc1,\Asc1,\Asc1) &=
\frac{1}{8} \mu_{m}^{4} S_6
\;, \\[3mm]
\ToptSBB(\Lsc1,\Acs1,\Acs1,\Asr1,\Asr1,\Asc1) &=
\frac{1}{16} \mu_{m}^{4} S_6
\;, \\[3mm]
\ToptSBB(\Lsc1,\Asr1,\Asr1,\Asc1,\Asr1,\Asr1) &=
\frac{1}{2} \mu_{m}^{2} \mu_{3}^{2} S_6
\;, \\[3mm]
\ToptSBB(\Lsc1,\Acs1,\Acs1,\Asr1,\Agl1,\Asc1) &+
\ToptSBB(\Lsc1,\Acs1,\Acs1,\Asr1,\Agl2,\Asc1) =
\nn &
+ \frac{1}{16} \mu_{m}^{2} (3g^2 + \gp^2) B_8
\;, \\[3mm]
\ToptSE(\Lsc1,\Asr1,\Asr1,\Asc1,\Asc1) &=
-\frac{1}{4} \mu_{m}^{2} \lambda_m^{ } I^{4b}_2 I^{4b}_2
\;, \\[3mm]
\ToptSTB(\Lsc1,\Acs1,\Acs1,\Asr1,\Asc1) &=
-\frac{3}{2} \mu_{m}^{2} \lambda_h^{ } S_4
\;, \\[3mm]
\ToptSTB(\Lsc1,\Asr1,\Asr1,\Asc1,\Asc1) &=
-\frac{1}{4} \mu_{m}^{2} \lambda_m^{ } S_4
\;, \\[3mm]
\ToptSTB(\Lsc1,\Asr1,\Asr1,\Asr1,\Asr1) &=
-\mu_{3}^{2} \lambda_m^{ } S_4
\;, \\[3mm]
\ToptSS(\Lsc1,\Asr1,\Asr1,\Lcs1) &=
\frac{1}{2}\lambda_{m}^{2} S_3
\;, \\[3mm]
\TopoSTx(\Lsc1,\Asr1) &=
\frac{1}{2} \lambda_m^{ } \Big( \delta Z_\sigma I^{4b}_1 + \delta\mu_{\sigma}^{2} I^{4b}_2 \Big)
\;, \\[3mm]
\TopoSTc(\Lsc1,\Asr1) &=
-\frac{1}{2} \delta\lambda_m^{ } I^{4b}_1
\;, \\[3mm]
\TopoSBx(\Lsc1,\Asr1,\Asc1) &=
-\frac{1}{4} \mu_{m}^{2} \Big( \delta Z_\sigma I^{4b}_2 + \delta\mu_{\sigma}^{2} I^{4b}_3 \Big)
\;, \\[3mm]
\TopoSBx(\Lsc1,\Acs1,\Asr1) &=
-\frac{1}{4} \mu_{m}^{2} \Big( \delta Z_\phi I^{4b}_2 + \delta\mu_{h}^{2} I^{4b}_3 \Big)
\;, \\[3mm]
\TopoSBa(\Lsc1,\Asr1,\Asc1) &=
\frac{1}{2} \mu_m^{ } \delta\mu_m^{ } I^{4b}_2
\;.
\end{align}
\end{multicols}

%
\subsubsection*{Electroweak Debye masses and gauge couplings at two-loop}

The singlet contributions to the gauge field self-energies are
displayed in eqs.~\eqref{eq:mD1} and \eqref{eq:mD2}.
Diagrammatically the SU(2) gauge field self-energy composes of
\vspace{2mm}
\begin{align}
\Pi_{A_{\mu}^{a} A_{\nu}^{b}} &\supset
  \ToptSTT(\Lgl2,\Asc1,\Asc1,\Asr1)\;
  \ToptSTB(\Lgl2,\Asc1,\Asc1,\Asr1,\Acs1)\;
  \ToptSBT(\Lgl2,\Asc1,\Asc1,\Asc1,\Asr1)\;
  \ToptSM(\Lgl2,\Asc1,\Asc1,\Asc1,\Asc1,\Lsr1)\;
  \ToptSBB(\Lgl2,\Acs1,\Acs1,\Acs1,\Asr1,\Asc1)
  \;,
\end{align}
which is identical for U(1) when replacing the external legs:
$\Pi_{A_{\mu}^{a} A_{\nu}^{b}} \to \Pi_{B_{\mu}^{ } B_{\nu}^{ }}$.
Their corresponding SM contributions to the Debye mass align with
refs.~\cite{Gynther:2005dj,Gynther:2005av}:
\begin{align}
\label{eq:mD1:SM}
({\mD'^{2}})^{ }_{\rmii{SM}} &=
    T^{2}\gp^{2}\frac{1}{24}\Big(
        4\Ys^{2}
        + Y_{\rmi{2f}}^{ }\,\nf^{ }
      \Big)
    \nn &
    - \frac{T^{2}}{(4\pi)^2}\bigg[
        \bigg(
          \frac{2L_b - 5}{72}\Ys^{4}
        + \frac{9Y_{\rmi{4f}}^{ } + (L_b + 4L_f -2)Y_{\rmi{2f}}^{ }\Ys^{2}}{144}\nf^{ }
        + \frac{(L_f -1)}{144}(Y_{\rmi{2f}}^{ }\,\nf^{ })^{2}
      \bigg)\gp^{4}
    \nn &\hp{+\frac{T^{2}}{(4\pi)^2}\bigg[}
      - \frac{3}{8}\Big(\Ys^{2} - (\Yl^{2} + \Nc^{ }\Yq^{2})\nf\Big)\gp^{2}g^{2}
      - \gp^{2}\lambda_{h}^{ }\Ys^{2}
    \nn &\hp{+\frac{T^{2}}{(4\pi)^2}\bigg[}
      + \frac{2\Yq^{2}+\Yu^{2}+\Yd^{2}}{4}\CF^{ }\Nc^{ }\nf^{ }\,\gp^{2}\gs^{2}
      + \frac{6(\Yq\Yu-\Ys(\Yq - \Yu)) - 5\Ys^{2}}{12}\Nc^{ }\,\gp^{2}\gY^{2}
  \bigg]
  \;, \\[2mm]
\label{eq:mD2:SM}
({\mD^{2}})^{ }_{\rmii{SM}} &=
    T^{2}g^{2}\frac{1}{3}\Big(
        \frac{5}{2}
      + \frac{\Nc+1}{4}\,\nf
      \Big)
    \nn &
    + \frac{T^{2}}{(4\pi)^2}\bigg[
        \bigg(
          \frac{430L_b + 207}{72}
        + \frac{(\Nc+1)^{ }(43L_b - 20 L_f + 11)}{72}\nf^{ }
    \nn &\hp{+\frac{T^{2}}{(4\pi)^2}\bigg[}
        - \frac{(\Nc+1)^{2}(L_f -1)}{36}\nf^{2}
      \bigg)g^{4}
      + \frac{1}{8}\Big(\Ys^{2} - (\Yl^{2} + \Nc^{ }\Yq^{2})\nf\Big)\gp^{2}g^{2}
    \nn &\hp{+\frac{T^{2}}{(4\pi)^2}\bigg[\bigg(}
      + g^{2}\lambda_{h}^{ }\Ys^{2}
      - \frac{1}{2}\CF^{ }\Nc^{ }\nf^{ }\,g^{2}\gs^{2}
      - \frac{\Nc}{12}g^{2}\gY^{2}
  \bigg]
\;.
\end{align}
Additionally, we show the two-loop singlet contributions to the gauge couplings:
\begin{align}
\gp^2_{3} =
  \gp^4 T \frac{\zeta_{3}}{(4\pi)^{4}}\frac{1}{36}\bigg[
      \lambda_{m}
    - \frac{1}{(4\pi)^{2}T^{2}}\Big(6L_{b} + 11\Big)\mu_{m}^{2}
  \bigg]
  + (\text{SM terms})
\;,\\
g_{3}^2 =
  g^4 T \frac{\zeta_{3}}{(4\pi)^{4}}\frac{1}{36}\bigg[
      \lambda_{m}
    - \frac{1}{(4\pi)^{2}T^{2}}\Big(6L_{b} + 11\Big)\mu_{m}^{2}
  \bigg]
  + (\text{SM terms})
\;.
\end{align}
These contributions are formally of higher order, i.e. $\mathcal{O}(g^6)$ in our power counting.

%
\section{Collection of integrals}
\label{se:integrals}

This appendix collects definitions and results of sum-integrals
encountered in our computation.
Ref.~\cite{Osterman:2019xx} and references therein further showcase many explicit derivations.
We use dimensional regularisation in
$D = d+1 = 4-2\epsilon$ dimensions
in the \MSbar-scheme with renormalisation scales
$\Lambda$ in 4d and
$\Lamd$ in 3d.
Euclidean four-momenta are denoted as
$P \equiv (\omega_n,\vec{p})$
with the bosonic Matsubara frequency
$\omega_n = 2\pi n T$.
We define
the $d$-dimensional integral measure as
\begin{align}
\int_p \equiv
\Big( \frac{\Lambda^{2}e^\gammaE}{4\pi} \Big)^\epsilon
\int\frac{{\rm d}^{d}p}{(2\pi)^d}
\;,
\end{align}
and bosonic sum-integrals as
\begin{align}
\label{eq:Tint}
\Tint{P} &\equiv T \sum_{\omega_n} \int_p
\;, \quad\quad
\Tint{P}' \equiv T \sum_{\omega_n \neq 0} \int_p
\;,
\end{align}
where a primed integral denotes the absence of a zero mode.
For fermionic counterparts, we employ the definition of ref.~\cite{Brauner:2016fla}.

In pure 3d, we encounter the following integrals
\begin{align}
J_{\rmii{soft}} &\equiv \frac{1}{2} \int_p \ln(p^2 + m^2) =
  -\frac{1}{2}
  \Big( \frac{\Lambda^2_{\rmii{3d}}e^\gammaE}{4\pi} \Big)^\epsilon
  \frac{[m^2]^\frac{d}{2}}{(4\pi)^{\frac{d}{2}}} \frac{\Gamma(-\frac{d}{2})}{\Gamma(1)}
  \nn
  &= -\frac{(m^2)^{\frac{3}{2}}}{12 \pi} + \mathcal{O}(\epsilon)
  \;, \\
I^3_{\alpha}(m) &\equiv \int_p \frac{1}{[p^2+m^2]^\alpha} =
  \Big( \frac{\Lambda^2_{\rmii{3d}}e^\gammaE}{4\pi} \Big)^\epsilon
  \frac{[m^2]^{\frac{d}{2}-\alpha}}{(4\pi)^{\frac{d}{2}}} \frac{\Gamma(\alpha-\frac{d}{2})}{\Gamma(\alpha)}
  \;, \\
S^3_3(m_1, m_2, m_3) &\equiv
  \int_{p,q} \frac{1}{
    [p^2+m^2_1]
    [q^2+m^2_2]
    [(p+q)^2+m^2_3]}
  \nn &
  = \frac{1}{(4\pi)^2} \bigg(\frac{1}{4\epsilon} + \frac{1}{2} + \ln \Big( \frac{\Lambda_{\rmii{3d}}}{m_1 + m_2 + m_3} \Big) \bigg)
  + \mathcal{O}(\epsilon)
  \;,
\end{align}
where we define the shorthand notation
\begin{align}
\mathcal{D}^{\rmii{3d}}_{SS}(m_1,m_2) \equiv -I^3_1(m_1) I^3_1(m_2)
\;, \\
\mathcal{D}^{\rmii{3d}}_{SSS}(m_1,m_2,m_3) \equiv S^3_3(m_1, m_2, m_3)
\;.
\end{align}
In the 4d computation, we encounter sum-integrals parameterised by
\begin{align}
\label{eq:Tint:par}
  Z_{s_1;\sigma_1}^{\alpha_1} &\equiv
  \Tint{P\{\sigma_1\}} \frac{
    p_0^{\alpha_1}
    }{
    \bigl[ P^2 \bigr]^{s_1}}
  \;,\nn
  Z_{s_1 s_2 s_3;\sigma_1 \sigma_2}^{ } &\equiv
  \Tint{
    P\{\sigma_1\}
    Q\{\sigma_2\}}
    \frac{
    1
    }{
    \bigl[ P^{2} \bigr]^{s_1}
    \bigl[ Q^{2} \bigr]^{s_2}
    \bigl[ (P + Q)^{2} \bigr]^{s_3}}
  \;,
\end{align}
where the Matsubara four-momenta have implicit fermion signature
$P^2=\big[ (2n+\sigma_i)\pi T\big]^2 + \vec{p}^2$
with $\sigma_i=0(1)$ for bosons(fermions).
Below we list some of the recurring integrals in $d=3-2\epsilon$
\begin{align}
Z_{\alpha;0} = I^{4b}_{\alpha} &\equiv \Tint{P}' \frac{1}{[P^2]^\alpha} =
\bigg(\frac{\Lambda^{2}e^\gammaE}{4\pi}\bigg)^{\epsilon}
  2T\frac{[2\pi T]^{d-2\alpha}}{(4\pi)^{\frac{d}{2}}}
  \frac{\Gamma(\alpha-\frac{d}{2})}{\Gamma(\alpha)}
  \zeta_{2\alpha-d}
  \;, \\
Z_{\alpha;1} = I^{4f}_{\alpha} &\equiv \Tint{\{P\}} \frac{1}{[P^2]^\alpha} =
  \Big(2^{2\alpha-d}-1\Big) I^{4b}_{\alpha}
  \;, \\
\label{eq:Z111}
Z_{111;00} = S_3 &\equiv \Tint{P,Q} \frac{1}{P^2 Q^2 (P+Q)^2} = 0
  \;, \\
\label{eq:S4}
Z_{211;00} = S_4 &\equiv \Tint{P,Q} \frac{1}{P^4 Q^2 (P+Q)^2} =
  - \frac{1}{(d-5)(d-2)} I^{4b}_2 I^{4b}_2
  \;, \\
\label{eq:S5}
Z_{221;00} = S_5 &\equiv \Tint{P,Q} \frac{1}{P^4 Q^4 (P+Q)^2} = 0
  \;, \\
\label{eq:S6}
Z_{311;00} = S_6 &\equiv \Tint{P,Q} \frac{1}{P^6 Q^2 (P+Q)^2} =
  - \frac{4}{(d-7)(d-2)} I^{4b}_3 I^{4b}_2
  \;, \\
Z_{121;10} = \mathcal{F} &\equiv \Tint{\{P\},Q} \frac{1}{P^2 Q^4 (P+Q)^2} =
  \frac{1}{(d-5)(d-2)} \Big( I^{4f}_2 I^{4f}_2 - 2 I^{4f}_2 I^{4b}_2 \Big)
  \;, \\
F_5 &\equiv \Tint{\{P\},Q} \frac{P^2 + P \cdot Q}{P^2 Q^4 (P+Q)^2} =
  \Big(2^{2-d}-1\Big) I^{4b}_2 I^{4b}_1
  \;, \\
F_6 &\equiv \Tint{\{P\},Q} \frac{P^2 + P \cdot Q}{P^2 Q^6 (P+Q)^2} =
  I^{4f}_1 I^{4b}_3 - \frac{1}{2} \mathcal{F}
  \;, \\
B_4 &\equiv \Tint{P,Q} \frac{1}{P^4 (P+Q)^2}
  (2P+Q)_\mu
  (2P+Q)_\nu D_{\mu\nu}(Q) = 0
  \;, \\
B_8 &\equiv \Tint{P,Q} \frac{1}{P^6 Q^2}
  (P-Q)_\mu
  (P-Q)_\nu D_{\mu\nu}(P+Q) = I^{4b}_2 I^{4b}_2 + S_4
  \;, \\
B_{10} &\equiv \Tint{P,Q} \frac{1}{P^4 Q^4}
  (P-Q)_\mu
  (P-Q)_\nu D_{\mu\nu}(P+Q) = 4 S_4 - 3 I^{4b}_2 I^{4b}_2
  \;.
\end{align}
Here the power of integration-by-parts reduction
(cf.\ ref.~\cite{Nishimura:2012ee} and in particular
sec.~3.4 of ref.~\cite{Schicho:2020xaf}) is obvious.
All massless two-loop sum-integrals reduce to one-loop masters.
Similar sum-integral structures are listed in appendix~C of ref.~\cite{Gorda:2018hvi} and
can be used to compute pure SM contributions to the Higgs self energy at two-loop.
In broken phase computations,
we need massive sum-integrals (cf.\ ref.~\cite{Kajantie:1995dw}),
expanded at high-$T$ setting $m/T\ll 1$
\begin{align}
J_{b,\rmii{hard}}(m) &\equiv \frac{1}{2} \Tint{P}' \ln(P^2 + m^2)
  \nn &\simeq
      m^2 \frac{1}{2} I^{4b}_1
    - \frac{1}{4} m^4 I^{4b}_2
    + \frac{1}{6} m^6 I^{4b}_3
    - \frac{1}{8} m^8 I^{4b}_4
    + \frac{1}{10} m^{10} I^{4b}_5
    - \frac{1}{12} m^{12} I^{4b}_6
    \nn &=
      \frac{T^2}{24} m^2
    - \frac{1}{(4\pi)^2} \frac{m^4}{4} \Big(\frac{1}{\epsilon} + L_b \Big)
    + \frac{\zeta_{3}}{3(4\pi)^4} \frac{m^6}{T^2}
    - \frac{\zeta_{5}}{2(4\pi)^6} \frac{m^8}{T^4}
    \nn &
    + \frac{\zeta_{7}}{(4\pi)^8} \frac{m^{10}}{T^6}
    - \frac{7}{3} \frac{\zeta_{9}}{(4\pi)^{10}} \frac{m^{12}}{T^8}
    + \mathcal{O}(\epsilon)
    \;,\\[2mm]
J_{f}(m) &\equiv \frac{1}{2} \Tint{\{P\}}\!\! \ln(P^2 + m^2)
  \nn &\simeq
      \frac{1}{2} m^2 I^{4f}_1
    - \frac{1}{4} m^4 I^{4f}_2
    + \frac{1}{6} m^6 I^{4f}_3
    \nn &=
    - \frac{T^2}{48} m^2
    - \frac{1}{(4\pi)^2} \frac{m^4}{4} \Big(\frac{1}{\epsilon} + L_b \Big)
    + \frac{7}{3} \frac{\zeta_{3}}{(4\pi)^4} \frac{m^6}{T^2}
    + \mathcal{O}(\epsilon)
    \;, \\[2mm]
\mathcal{D}_{SS}(m_1,m_2) &\equiv
  - \Tint{P} \frac{1}{P^2+m^2_1} \Tint{Q} \frac{1}{Q^2+m^2_2}
  \nn &=
  - \bigg(
    T^2 I^3_1(m_1) I^3_1(m_2)
    + I^{4b}_1 I^{4b}_1
    \nn &
    + I^{4b}_1 \Big( -(m^2_1 + m^2_2) I^{4b}_2 + T I^3_1(m_1) + T I^3_1(m_2) \Big)
    \nn &
    - I^{4b}_2 T \Big(m^2_1 I^3_1(m_2) + m^2_2 I^3_1(m_1) \Big)
    + m^2_1 m^2_1 I^{4b}_2 I^{4b}_2
    + (m^4_1 + m^4_2) I^{4b}_1 I^{4b}_3
  \bigg)
  \nn &
  + \mathcal{O}(m^5)
  \;, \\
\label{eq:SSS}
\mathcal{D}_{SSS}(m_1,m_2,m_3) &\equiv
    \Tint{P,Q} \frac{1}{[P^2+m^2_1][Q^2+m^2_2][(P+Q)^2+m^2_3]}
  \nn &=
    T^2 S^3_3(m_1,m_2,m_3)
  + T \Big( I^3_1(m_1) + I^3_1(m_2) + I^3_1(m_3) \Big) I^{4b}_2
  \nn &
  - (m^2_1 + m^2_2 + m^2_3) S_4
  + (m^4_1 + m^4_2 + m^4_3) (S_5 + S_6)
  \nn &
  + \mathcal{O}(m^3)
  + \mathcal{O}(m^5)
  \;,\\[2mm]
\mathcal{D}_{S}(m,\delta m^2, \delta Z) &\equiv
    \Tint{P} \frac{ (\delta m^2 + P^2 \delta Z) }{P^2 + m^2}
    \nn &=
    \Big(\delta m^2 - m^2 \delta Z \Big)
    \Big( T I^3_1(m) + I^{4b}_1 - m^2 I^{4b}_2 \Big)
    + \mathcal{O}(m^4)
    \;.
\end{align}
Note that if one uses resummed propagators,
all 3d integrals in these expressions obtain a resummed 3d mass instead of a 4d mass.
In
$\mathcal{D}_{SS}$ and
$\mathcal{D}_{SSS}$
all mixed soft/hard terms (that are non-analytic in $m^2$) will be cancelled in
resummation at $\mathcal{O}(g^4)$.
For the latter integral, we do not write down terms of $\mathcal{O}(m^3)$ explicitly
since these are absent in resummation at $\mathcal{O}(g^4)$.
Whereas the $\mathcal{O}(m^4)$ hard contribution is required for the matching of $\mu_{3}^4$ terms.
Indeed,
in addition to
$\mathcal{O}(m^2)$ terms we need
$\mathcal{O}(m^4)$ terms due to cubic interaction $\mu_{3}$
which can be compared
eqs.~(81) and (88) of the classic ref.~\cite{Kajantie:1995dw}.

%
\subsection*{Sunset sum-integrals}

Finally, let us inspect the high-$T$ expansion of
the massive two-loop sunset sum-integral.
See also ref.~\cite{Ekstedt:2020qyp}.
The result given in (\ref{eq:SSS}) shows almost surprising cleanliness in
the terms of the mass expansion.
In particular, it is very pleasant that the coefficients of even powers of mass
are given in terms of full sum-integral structures,
as opposed to linear combinations of (massless) mixed soft/hard and hard sum-integral structures.
Essentially, this can be shown to take place by carefully expanding separately
the mixed modes and hard modes of the original massive sum-integral.
The arithmetic challenge arises from a proper treatment of the mixed modes,
as a {\em naive} mass expansion can only occur in propagators with
non-vanishing Matsubara index
(hard scale contribution).
Hence, the proper order of the mass expansion is chosen to be isolated with
an iterative approach.
This operation is explicitly described below for the special case with
three degenerate masses.
This simplifies the book keeping of the computation without any loss of information,
due to the obvious symmetry between the three masses.

It is well-motivated to symmetrise the computation as far as possible.
Thus, we choose to consider the hard modes by setting each propagator structure of
the sunset integral identical, with non-vanishing thermal component:
\begin{equation}
D_{SSS}^\text{hard} = \Tint{P,Q} \frac{(1-\delta_{P_0})(1-\delta_{Q_0})(1-\delta_{P_0+Q_0})}{(P^2+m^2)(Q^2+m^2)[(P+Q)^2+m^2]}
\;.
\end{equation}
We can expand this in a trivial manner up to $\mathcal{O}(m^2)$ to find
the following non-vanishing contributions:
\begin{align}
&-3 T \int_p \Tint{Q} \frac{1}{p^2 Q^2 (p+Q)^2}
-3 m^2 \bigg[
    \Tint{PQ} \frac{(1-\delta_{P_0})(1-\delta_{Q_0})}{P^4 Q^2(P+Q)^2}
  - T \int_p\Tint{Q} \frac{(1-\delta_{Q_0})}{p^2Q^4(p+Q)^2} \bigg]
\;,
\end{align}
where we took note that all scaleless spatial integrals vanish, and
more notably the massless sunset integral $S_{3}$~\eqref{eq:Z111}
vanishes~\cite{Arnold:1994ps,Arnold:1994eb}.
The first integral expression mixes soft and hard modes.
This naturally follows from the symmetrisation of the propagators,
which essentially adds contributions from the mixed part of the expansion.
Hence, we reinsert it in the second structure of interest.
\begin{align}
D_{SSS}^\rmii{mix} &=
    3 T \int_p \Tint{Q} \frac{1-\delta_{Q_0}}{(p^2+m^2)(Q^2+m^2)[(p+Q)^2+m^2]}
  - 3 T \int_p \Tint{Q} \frac{1}{p^2 Q^2 (p+Q)^2}
\end{align}
Let us start the expansion towards $\mathcal{O}(m^2)$ by solely considering
the propagators with non-vanishing thermal scale, which yields
\begin{align}
&
  3 T \int_p \Tint{Q} \frac{(1-\delta_{Q_0})}{(p^2+m^2)Q^2(p+Q)^2}
- 3 T \int_p \Tint{Q} \frac{1}{p^2 Q^2 (p+Q)^2}
- 6T m^2 \int_p \Tint{Q} \frac{1-\delta_{Q_0}}{(p^2+m^2)Q^4(p+Q)^2}
\nn &=
- 3T m^2 \int_p \Tint{Q} \bigg[
    \frac{1-\delta_{Q_0}}{p^2(p^2+m^2)Q^2 (p+Q)^2}
  + \frac{2(1-\delta_{Q_0})}{(p^2+m^2)Q^4(p+Q)^2}
  \bigg]
  \;.
\end{align}
In both of these sum-integral terms it suffices to remove
the mass terms inside the integral, as we only wish to find contributions exactly
at $\mathcal{O}(m^2)$.
Thus, we find the quadratic coefficient of the expansion as
\begin{align}
D_{SSS}^{m^2} &= - 3 \bigg[
    \Tint{PQ} \frac{(1-\delta_{P_0})(1-\delta_{Q_0})}{P^4 Q^2(P+Q)^2}
  + T\int_p \Tint{Q}\frac{1-\delta_{Q_0}}{p^4 Q^2 (p+Q)^2}
  + T \int_p \Tint{Q}\frac{2(1-\delta_{Q_0})}{[p^2+m^2]Q^4(p+Q)^2}
  \bigg]
  \nn[2mm] &\equiv
  -3 S_4
  \;,
\end{align}
where $S_4$ is defined by eq.~\eqref{eq:S4}.

We follow a similar procedure to find the coefficient of $\mathcal{O}(m^4)$,
with the slight difference of using the results found above as
the iterative subtraction element for the mixed element expansion.
This time, the hard expansion yields
\begin{align}
&
3 m^4 \Tint{PQ} \bigg[
    \frac{(1-\delta_{P_0})(1-\delta_{Q_0})(1-\delta_{P_0+Q_0})}{P^6 Q^2 (P+Q)^2}
  + \frac{(1-\delta_{P_0})(1-\delta_{Q_0})(1-\delta_{P_0+Q_0})}{P^4 Q^4 (P+Q)^2}
  \bigg]
  \\[2mm] &=
3 m^4 \bigg[
    \Tint{PQ} \frac{(1-\delta_{P_0})(1-\delta_{Q_0})}{P^6 Q^2 (P+Q)^2}
  - T \int_p \Tint{Q} \frac{1-\delta_{Q_0}}{p^2 Q^6 (p+Q)^2}
  \bigg]
  \\[2mm] &
+ 3 m^4 \bigg[
    \Tint{PQ} \frac{(1-\delta_{P_0})(1-\delta_{Q_0})}{P^4 Q^2 (P+Q)^4}
  - T \int_p \Tint{Q} \frac{1-\delta_{Q_0}}{p^4 Q^2 (p+Q)^4}
\bigg]
\;.
\end{align}
In order to fully extract the mixed contributions, we again expand
the suitable propagators and follow-up with a removal of the results of previous orders.
This results in three separate computations:
\begin{align}
&
  3 T \int_p \Tint{Q} \bigg[
      \frac{(1-\delta_{Q_0})}{(p^2+m^2)Q^2(p+Q)^2}
    - \frac{(1-\delta_{Q_0})}{p^2 Q^2(p+Q)^2}
    + \frac{m^2}{p^4 Q^2 (p+Q)^2}
\bigg] \nn[2mm] &\mapsto
  3 m^4 T \int_p \Tint{Q} \frac{(1-\delta_{Q_0})}{p^6 Q^2 (p+Q)^2}
  \;,
\\ &
- 6 T m^2 \int_p \Tint{Q} \bigg[
    \frac{(1-\delta_{Q_0})}{(p^2+m^2)Q^4(p+Q)^2}
  - \frac{(1-\delta_{Q_0})}{p^2 Q^4(p+Q)^2}
  \bigg]
  \nn[2mm] & \mapsto
  6T m^4 \int_p \Tint{Q} \frac{(1-\delta_{Q_0})}{p^4 Q^2 (p+Q)^4}
  \;,
\\ &
3 T m^4 \int_p \Tint{Q}\bigg[
    \frac{2(1-\delta_{Q_0})}{(p^2+m^2)Q^6 (p+Q)^2}
  + \frac{(1-\delta_{Q_0})}{(p^2+m^2)Q^4(p+Q)^4}
  \bigg]
  \nn[2mm] & \mapsto
  3m^4 T \int_p \Tint{Q}\bigg[
    \frac{2(1-\delta_{Q_0})}{p^2 Q^6 (p+Q)^2}
  + \frac{(1-\delta_{Q_0})}{p^2 Q^4(p+Q)^4}
  \bigg]
  \;.
\end{align}
These contributions combine to the $\mathcal{O}(m^4)$ coefficient of
the mass expansion as
\begin{equation}
D_{SSS}^{m^4} = 3 \bigg[
    \Tint{PQ} \frac{1}{P^6 Q^2(P+Q)^2}
  + \frac{1}{P^4 Q^2 (P+Q)^4} \bigg] \equiv 3 [S_5+S_6]
  \;,
\end{equation}
where
$S_5$ is defined in eq.~\eqref{eq:S5} and
$S_6$ in \eqref{eq:S6}.

{\small
%

}
\end{document}